\documentclass[a4paper,11pt]{article}
\pdfoutput=1 

\usepackage{jheppub} 

\usepackage[T1]{fontenc} 

\usepackage{amsmath}
\usepackage{dsfont}
\usepackage{slashed}
\usepackage{bbold}
\usepackage{psfrag}
\PassOptionsToPackage{caption=false}{subfig}
\usepackage{subfig}
\usepackage{multirow}
\usepackage{booktabs}
\usepackage{hyperref}
\usepackage{feynmp}
\usepackage{color}
\usepackage[normalem]{ulem} 

\usepackage[utf8]{inputenc}

\usepackage{enumitem}

\newcommand{\todo}[1]{{\color{red} \ifmmode\else[todo]\fi #1}}
\usepackage[usenames,dvipsnames]{xcolor}
     
\newcommand{\micron}{\mu \text{m}}

\def\beq{\begin{equation}}
\def\eeq{\end{equation}}

\graphicspath{{figures/}}

\begin{document}
\title{Low-Energy Radiative Backgrounds in CCD-Based Dark-Matter Detectors}

\author[a]{Peizhi Du,}
\author[b]{Daniel Ega\~na-Ugrinovic,}
\author[c]{Rouven Essig,}
\author[d]{Mukul Sholapurkar}

\affiliation[a]{New High Energy Theory Center, Rutgers University, Piscataway, NJ 08854, USA}
\affiliation[b]{Perimeter Institute for Theoretical Physics, Waterloo, ON N2L 2Y5}
\affiliation[c]{C.N.~Yang Institute for Theoretical Physics, Stony Brook University, Stony Brook, NY, 11794, USA}
\affiliation[d]{Department of Physics, University of California, San Diego, CA 92093, USA}

\preprint{YITP-SB-2023-12}

\date{\today}

\abstract{
The reach of sub-GeV dark-matter detectors is at present severely affected by low-energy events from various origins. We present the theoretical methods to compute the single- and few-electron events that arise from secondary radiation emitted by high-energy particles as they pass through detector materials and perform a detailed simulation to quantify them at (Skipper) CCD-based experiments, focusing on the SENSEI data collected at Fermilab near the MINOS cavern.  The simulations account for the generation of secondaries from Cherenkov and luminescent recombination radiation; photo-absorption in the bulk, backside layer, pitch adapter, and epoxy; the photon reflection and refraction at interfaces; thin-film interference; the roughness of the interfaces; the dynamics of charges produced in the highly doped CCD-backside-layers; and the partial charge collection on the CCD backside. We  consider several systematic uncertainties, notably those stemming from the backside modeling, which we estimate with a ``fiducial'' and an ``extreme'' charge-diffusion model, with the former model being preferred due to better agreement with partial-charge collection data.  We find that Cherenkov photons constitute about $40\%$ of the observed single-electron events for both diffusion models; radiative recombination contributes negligibly to the event rate for the fiducial model, although it can dominate over Cherenkov for the extreme model.  
We also estimate the fraction of 2-electron events that arise from 1-electron event coincidences in the same pixel, finding that the entire 2-electron rate can be explained by coincidences of radiative events and spurious charge.  Accounting for both radiative and non-radiative backgrounds, we project the sensitivity of  future Skipper-CCD-based experiments to different dark-matter models. 
For light-mediator models with dark-matter masses of $1$, $5$, and $10$\,MeV, 
we find that future experiments with 10-kg-year exposures and successful background mitigation could have a sensitivity that is larger by 9, 3, and 2 orders of magnitude, respectively, when compared to an experiment without background improvements.  
}

\maketitle

\section{Introduction}\label{sec:intro}

In recent years, significant theoretical and experimental efforts have been dedicated to the study of dark matter (DM) candidates with sub-GeV masses~(see \cite{Essig:2022dfa} and references therein). At present, some of the most sensitive detectors for DM with mass near the MeV-scale are solid-state devices that aim to find DM-induced ionization from DM-electron scattering~\cite{Essig:2011nj} or absorption~\cite{An:2014twa,Bloch:2016sjj,Hochberg:2016sqx}, or DM-nucleus scattering with the Migdal effect~\cite{Ibe:2017yqa}, but the reach of these detectors is severely limited by large backgrounds of unknown origin, especially towards lower energies~\cite{Fuss:2022fxe}. Identifying, modeling, and mitigating these backgrounds is critical for improving the DM discovery reach. The Charge-Coupled-Device (CCD)-based experiments such as SENSEI~\cite{Tiffenberg:2017aac,Crisler:2018gci,SENSEI:2019ibb,SENSEI:2020dpa,SENSEI:2023gie}, DAMIC~\cite{DAMIC:2016lrs,DAMIC:2020cut}, and DAMIC-M~\cite{Castello-Mor:2020jhd,DAMIC-M:2023gxo}, in particular, offer world-leading sensitivities for a variety of DM models across important regions of parameter space, but observe a significant number of low-energy ionization events that remain unexplained. The observed rates are significantly larger than those expected from thermal events, which are well described by the Shockley-Read-Hall theory~\cite{PhysRev.87.835}, and than those that can be obtained from Compton scattering.

A plausible hypothesis for the origin of at least some of these backgrounds, pointed out in~\cite{Du:2020ldo}, is that a sizable fraction of the observed rates could come from secondary photons emitted by energetic particles that pass through detector materials, either via the Cherenkov effect or luminescence, the latter of which in CCDs may occur due to electron-hole radiative recombination. The objective of this paper is to test this hypothesis by performing the first detailed computation of such backgrounds in CCD-based experiments. The computations depend sensitively on the particular detector details, so for concreteness we focus on the SENSEI detector located at Fermilab near the MINOS cavern and the data presented in~\cite{SENSEI:2020dpa}.  However, the methods that we present can be adapted to other existing and proposed DM and neutrino detectors, including SENSEI at SNOLAB, DAMIC, DAMIC-M, Oscura~\cite{Oscura:2022vmi,Oscura:2023qch,Oscura:2023qik}, CONNIE~\cite{CONNIE:2019xid}, $\nu$IOLETA~\cite{Fernandez-Moroni:2021nap}, DarkNESS~\cite{DarkNESS}, and CCDs used in astronomy \cite{howell2006handbook}. 

We build a complete detector simulation that starts from high-energy particles and calculates the observed rates after analysis cuts. 
Our simulation includes modules for simulating the high-energy tracks; 
calculating the Cherenkov and luminescent radiative recombination rates; propagating these photons across the bulk of the device and material interfaces; calculating the photo-ionization probabilities across different material layers; performing dedicated first-principles computations of charge diffusion and of the CCD's charge-collection efficiency; and analyzing the collected ionization events with the implementation of detector masks and cuts used by the SENSEI experiment. To provide precise results and evaluate systematic uncertainties, we meticulously survey the literature to characterize the properties of all the relevant materials used in the fabrication of the CCD and of auxiliary materials around it, down to sub-micron scales.

Our results show that Cherenkov radiation likely dominates the radiative rates at low energies, accounting for $\approx 40\%$ of the observed 1-electron event rates, and for the entire observed single-pixel 2-electron rate due to 1-electron coincidences. The contribution from radiative recombination of electron-hole-pairs created in the doped layer in the backside of the CCD is subject to large systematic uncertainties due to  bandgap narrowing effects in that region of the apparatus, which affects charge diffusion.  In our ``fiducial'' model for the bandgap narrowing in the backside, which reproduces well data on the partial charge collection~\cite{Moroni_2021} from the backside, radiative recombination contributes negligibly to the 1-electron rate.  However, with a more ``extreme'' model for the bandgap narrowing, the contribution to the 1-electron rate from radiative recombination can even dominate over that from Cherenkov; however, the extreme model does not reproduce well the partial charge collection data.  In our fiducial model, our simulations thus suggest that the remaining 1-electron events are not radiative and are instead possibly due to intrinsic detector dark counts; we thus characterize the spatial distribution for these remaining events and provide plausible hypotheses for their origin. 

Our background analyses allow us to perform informed projections regarding the future expected sensitivity of detectors as they collect more data; from such projections we find that improvements in the reach to discover DM over wide regions of parameter space will critically depend on background mitigation strategies.

\begin{figure}[t]
	\centering
	\includegraphics[width=0.96\textwidth]{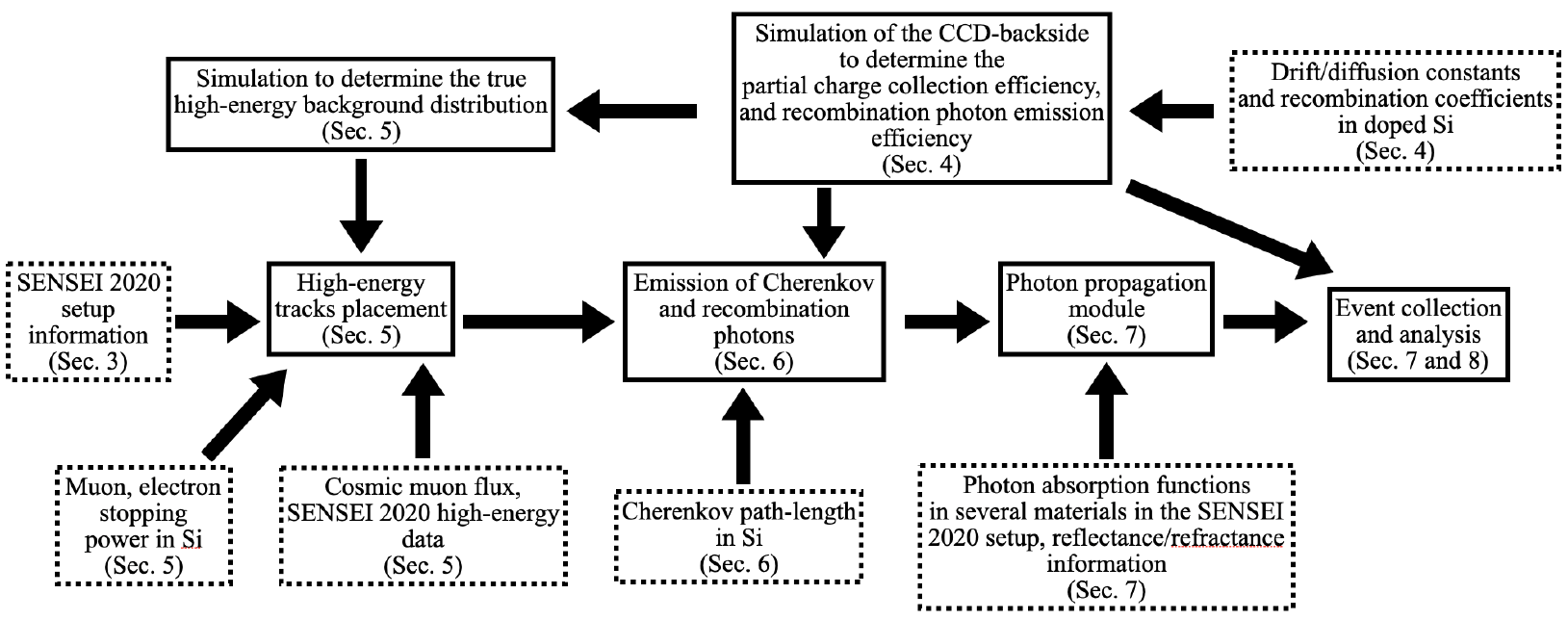}
	\caption{Schematic of the simulation performed in this work. Modules shown in solid boxes are different parts of the simulation and the information shown in dotted boxes are input data. } \label{fig:simulation_schematic}
\end{figure}

We organize this article as follows. We begin in Sec.~\ref{sec:theory} with a short theoretical overview of Cherenkov radiation and luminescence from radiative recombination. In Sec.~\ref{sec:structure}, we describe the details of the SENSEI detector located in a shallow underground site near the MINOS cavern at Fermilab. In Sec.~\ref{sec:backsidesim}, we present our module for analyzing the diffusion of charge on the CCD's backside (required to compute the charge collection efficiency of the device and radiative recombination rates).  Sec.~\ref{sec:cherenkovinsensei} describe the module for the simulation of high-energy events, while Sec.~\ref{sec:secondaryphotons} describes the module for the generation of Cherenkov and recombination photons.  In Sec.~\ref{sec:photon}, we discuss the propagation of the photons in the detector. We discuss the detector cuts (or ``masks'') in Sec.~\ref{sec:analysis}. We analyze the results in Sec.~\ref{sec:results}, which also includes a discussion of the systematic uncertainties and variations of the benchmark simulations to analyze the performance of the detector under diverse operating conditions.  We discuss simulations for a CCD that is ``thinned'' by removing the backside layer as well as a CCD in a low-background environment in Sec.~\ref{sec:thinned}.  Based on the understanding of the event rates resulting from our analysis, we speculate on the origin of the remaining backgrounds and perform DM-sensitivity projections for future CCD-based detectors in Sec.~\ref{sec:projections}. We conclude with a summary in Sec.~\ref{sec:conclusions}. We leave a discussion of epoxy-induced events to the appendix. We show a schematic of the  flow in  the simulation performed in this work in Fig.~\ref{fig:simulation_schematic}, which also shows the section in the draft where each module is discussed. 

\section{Radiative processes: theory}\label{sec:theory}
\subsection{Cherenkov radiation}
Energetic charged particles passing through dielectric materials emit secondary photons via the Cherenkov process. Emission of Cherenkov light with frequency $\omega$ occurs when the particle's velocity exceeds the condition \cite{budini1953energy}
\begin{equation}
v^2 \textrm{Re} \, \epsilon(\omega) > 1 \quad ,
\label{eq:cherenkovcond}
\end{equation}
where $\omega$ is the photon frequency, and $\epsilon(\omega)$ is the material's frequency-dependent dielectric function. When the condition Eq.~\eqref{eq:cherenkovcond} is met, the rate for emission of Cherenkov photons in the material is given by \cite{budini1953energy}
\begin{equation}
\label{eq:cherenkov}
\frac{d^2N_\gamma}{d\omega dx}=\alpha \left(1-\frac{\textrm{Re} \,\epsilon(\omega) }{v^2 |{\epsilon(\omega)|^2}}\right)  \quad ,
\end{equation}
where $dx$ is the path-length differential characterizing the unit-charge track trajectory. The emitted photons are sharply peaked at a polar angle with respect to a vector along the track's trajectory that is given by
\begin{equation}
\label{eq:angle}
\cos \theta_{\textrm{Ch}}= \frac{\sqrt{\textrm{Re}\, \epsilon(\omega)}}{v |\epsilon(\omega)|} \quad , 
\end{equation}
while the distribution around the azimuthal angle is isotropic. 

\subsection{Luminescence from radiative recombination}\label{sec:theoryLuminescence}
High-energy particles create electron-hole pairs through scattering, which  recombine radiatively via photon emission (this process is a type of luminescence), or non-radiatively by producing phonons. In this subsection, we briefly discuss  these processes.\\

The dynamics of electrons and holes is governed by the continuity equation, which reads 
\begin{align}\label{eq:continuity}
\frac{\partial n_{e,h}}{\partial t} = - \nabla \cdot \vec{j}_{e,h} - \Gamma_{e,h}\ ,
\end{align}
where $n_{e,h}$, $\vec{j}_{e,h}$ and $\Gamma_{e,h}$ are the electron or hole number density, current, and  loss rate from recombination processes, respectively. The current captures the motion of the charges, which can be due to a drift under the influence of an electric field, or because of diffusion. The electron and hole currents can thus be written as
\begin{align}
\vec{j}_e = -n_e \mu_e \vec{E}-D_e \nabla n_e \label{eq:currentelectron} ~,\\
\vec{j}_h = n_h \mu_h \vec{E}-D_h \nabla n_h \label{eq:currenthole} ~,
\end{align} 
where $\mu_{e,h}$ is the electron or hole mobility, $D_{e,h}$ is the electron or hole diffusion constant, and $\vec{E}$ is the electric field. The recombination term in Eq.~\eqref{eq:continuity}, on the other hand, can be written as a sum of three processes, 
\begin{align}\label{eq:recorates}
\Gamma_{e,h}= \Gamma^{\text{direct}}_{e,h}+\Gamma^{\text{Auger}}_{e,h}+\Gamma^{\text{trap}}_{e,h}\ , 
\end{align}
where $\Gamma^{\text{direct}}_{e,h}$, $\Gamma^{\text{Auger}}_{e,h}$ and $\Gamma^{\text{trap}}_{e,h}$ are the electron or hole rates of direct (or band-to-band) recombination,  Auger recombination and trap-assisted recombination. The direct or band-to-band recombination is radiative, and proceeds through an electron at the bottom of the conduction band recombining with a hole at the top of the valence band via the emission of a near-bandgap photon. The rate is given by 
\begin{align}
\Gamma^{\text{direct}}_e = \Gamma^{\text{direct}}_h = B(n_e n_h - \overline{n}_e \overline{n}_h)\ ,
\end{align}
where $\overline{n}_{e,h}$ is the background (equilibrium) electron or hole concentration, and $B$ is the direct recombination coefficient. The equilibrium electron and hole concentrations for an intrinsic non-degenerate semiconductor are
\begin{align} 
\overline{n}_e= \overline{n}_h= n_i \equiv \sqrt{N_V N_V} e^{-E_g/2T}\ ,
\end{align}
where $E_g$ is the direct bandgap and $N_C$ ($N_V$) is the effective density of states in the conduction (valence) band. For n- and p-type semiconductors, the background carrier concentrations are instead
\begin{align}
\overline{n}_e= N_D \quad , \quad  \overline{n}_h = \frac{n_i^2}{N_D} \quad \textrm{(n-type)} \\
\overline{n}_h= N_D \quad , \quad \overline{n}_e = \frac{n_i^2}{N_D}  \quad \textrm{(p-type)}
\end{align}
where $N_D$ is the n or p-type doping concentration. 

Auger recombination is a non-radiative process where an electron scatters with another electron and recombines with a hole, or a hole scatters with another hole and recombines with an electron. Effectively, an electron and a hole are lost in the process and the energy is emitted through phonons. The Auger rate is given by,
\begin{align}
\Gamma^{\text{Auger}}_e = \Gamma^{\text{Auger}}_h = (a_e n_e + a_h n_h)(n_e n_h - \overline{n}_e \overline{n}_h)\ ,
\end{align}
where $a_e$ and $a_h$ are the Auger recombination coefficients. 

Finally, trap-assisted recombination can be radiative or non-radiative, and proceeds through the capture of either an electron or a hole by traps that create energy levels between the valence and conduction bands. Unlike in the direct and Auger recombination processes, only an electron or a hole disappears in trap-assisted recombination. The corresponding trap-assisted recombination rates are,
\begin{align}
\Gamma^{\text{trap}}_{e} = c_e n_t [(1-f_t)n_e - f_t n_{et}]~,\\
\Gamma^{\text{trap}}_{h} = c_h n_t [f_t n_h - (1-f_t)n_{ht}]~,
\end{align}
where $c_{e,h}$ are trap-assisted recombination constants, $n_t$ is the density of impurities leading to traps, $f_t$ is the probability that a trap is occupied by an electron, and $n_{et}$, $n_{ht}$ are defined as,
\begin{align}
n_{et} \equiv N_C \exp{(E_T -E_C)/T}~,\\
n_{ht} \equiv N_V \exp{(E_V - E_T)/T}~,
\end{align}
where $N_C$ ($N_V$) is the conduction band (valence band) density of states, $E_T$ is the trap energy level, and $E_C$ ($E_V$) is the conduction (valence) band energy.

\section{The SENSEI experiment}\label{sec:structure}
\begin{figure}[t]
	\centering
	\includegraphics[width=0.48\textwidth]{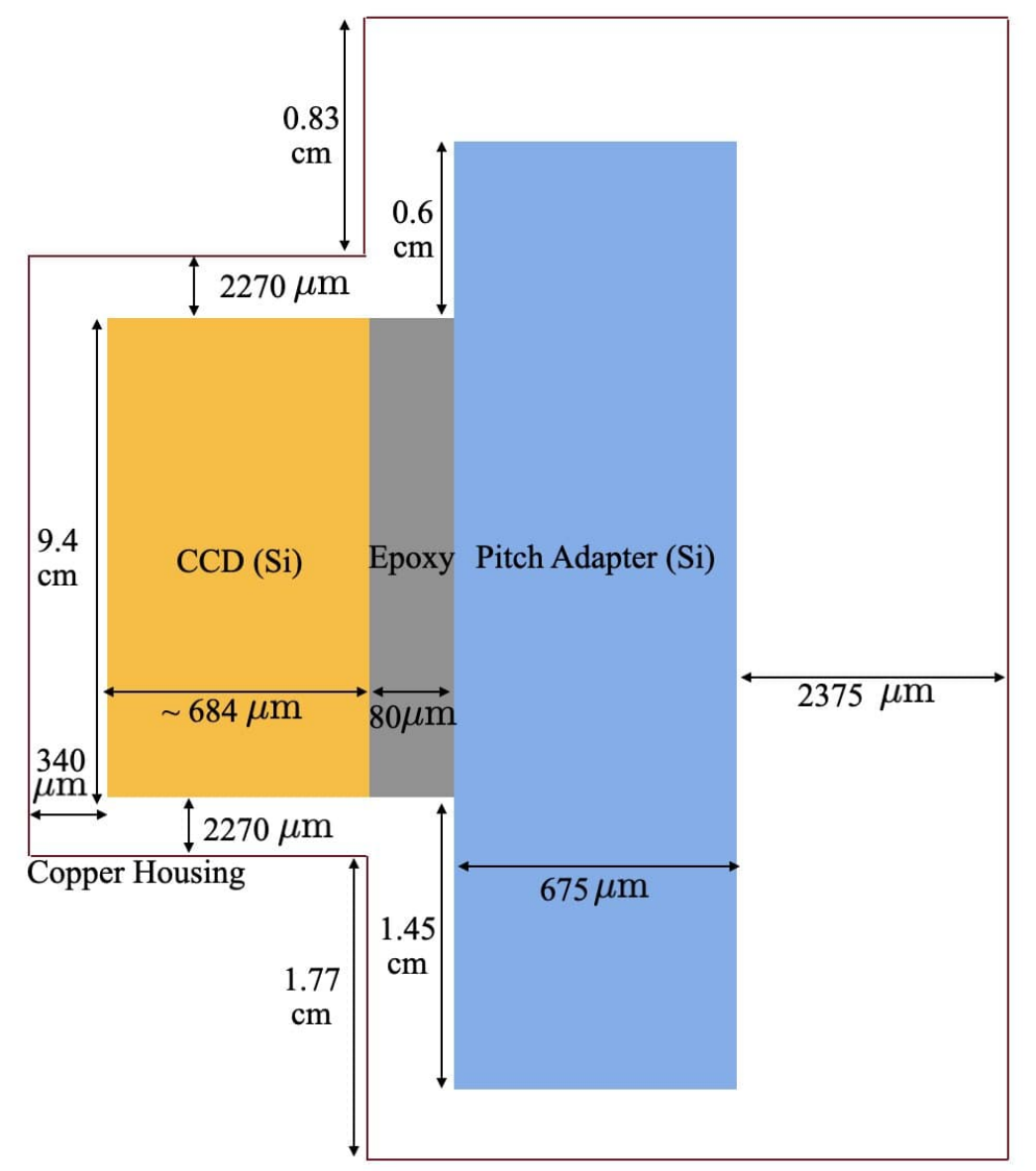}
	\includegraphics[width=0.48\textwidth]{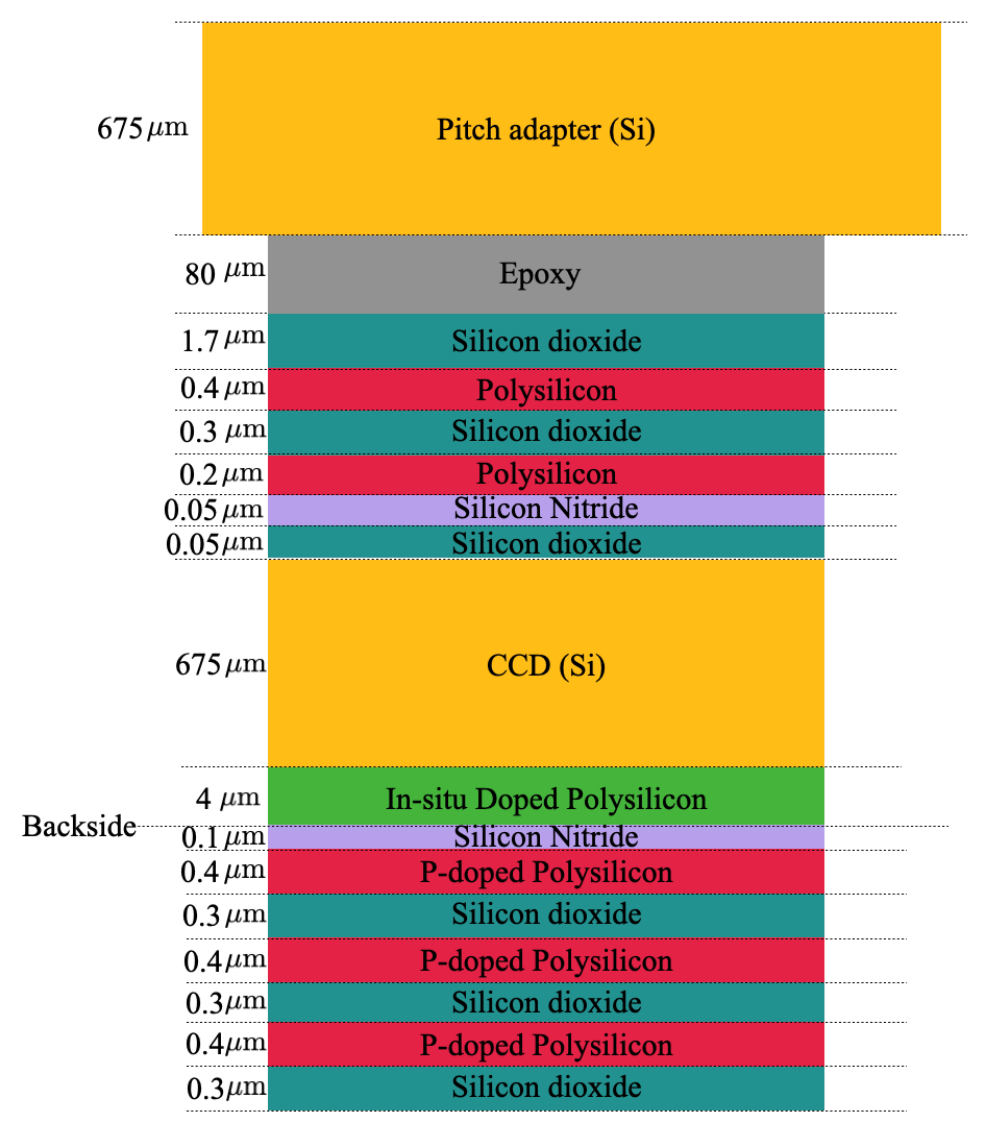}
	\caption{\textbf{Left:} Schematic of the SENSEI detector geometry. The width (horizontal direction in the figure) $\times$ length (vertical direction in the figure)  cross section of the detector geometry is shown. The backside of the CCD corresponds to the leftmost part of the device in the  figure, while its frontside corresponds to the part that interfaces with the epoxy. \textbf{Right:} Material layers on the backside and the frontside of the CCD~\cite{DAMIC:2021crr}. The length (horizontal direction in the figure) $\times$ width (vertical direction in the figure) cross section is shown. 
 We omit the ``buried channel'' layer~\cite{boyle1974buried} located on the CCD's frontside and made of p-type Si, as during detector operation this layer is depleted from holes and for the purposes of our simulations it behaves as pure Si. 
 The figures presented here are not to scale.} \label{fig:structure}
\end{figure}

\begin{figure}[t]
	\centering
	\includegraphics[width=0.48\textwidth]{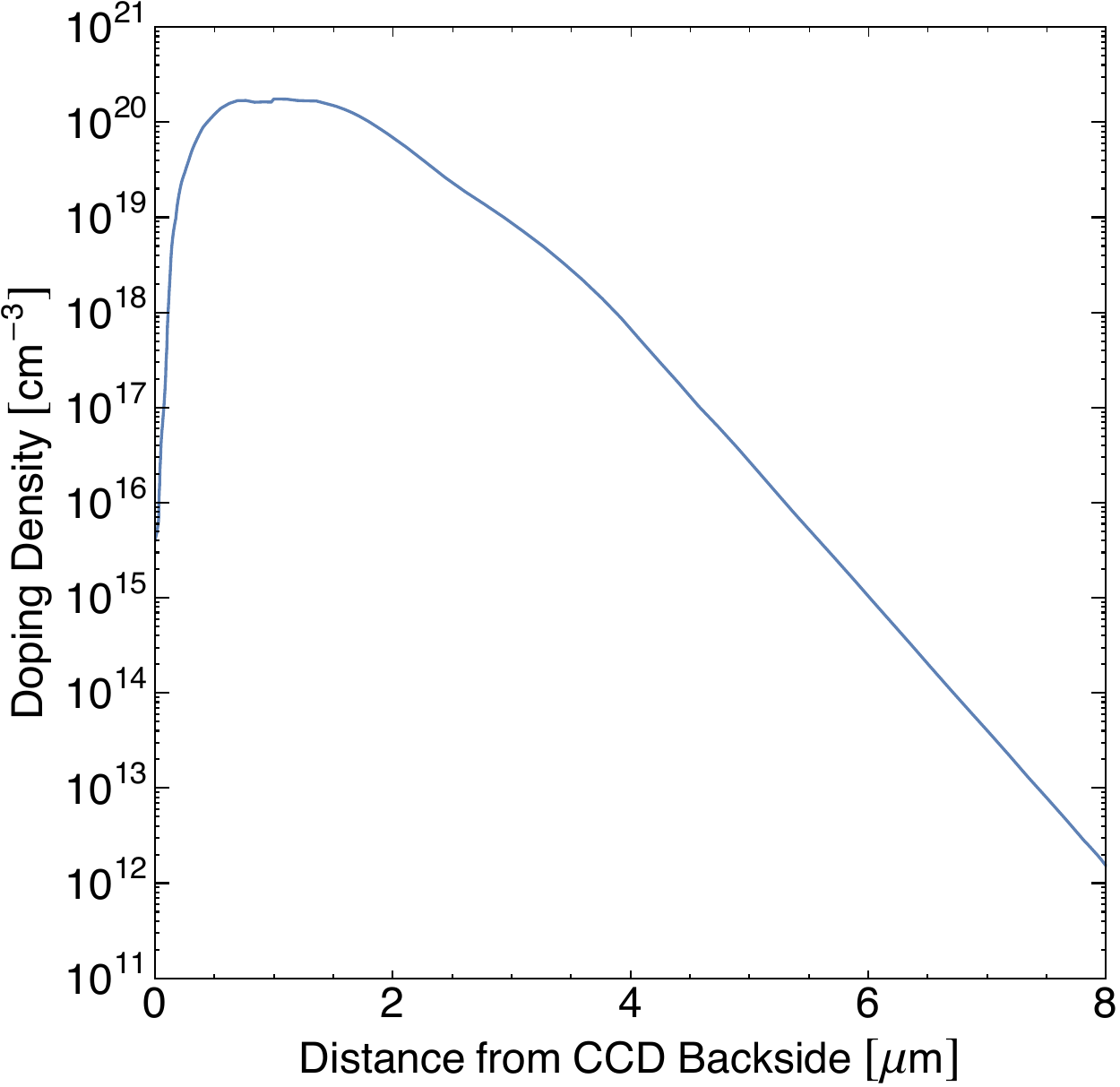}
	\caption{P-doping density versus the distance from the backside of the CCD,  taken from~\cite{DAMIC:2021crr}.} \label{fig:doping}
\end{figure}

SENSEI~\cite{SENSEI:2020dpa} uses silicon (Si) Skipper-CCDs to search for small ionization signals induced by DM interactions in the bulk of the CCD. A schematic of the detector is shown in the left panel of Fig.~\ref{fig:structure}. The CCD has a volume of $9.4~\text{cm}~\times1.6~\text{cm}~\times 684~\micron $, of which a central region of area $9.2~\text{cm}~\times1.3~\text{cm}$ times a width of $675~\micron$ is active. The CCD is held through epoxy glue ($\sim 80~\mu \text{m}$ wide) by a pitch adapter ($11.5~\text{cm}~\times1.2~\text{cm}~\times 675~\micron $) that is also made up of Si. When an ionization event occurs in the CCD, the charges are held by ``pixels'' of size $15~\micron~\times 15~\micron$ thanks to electrostatic potentials created by gates and channel stops located on the CCD's frontside (the side closest to the epoxy in Fig.~\ref{fig:structure})~\cite{janesick1987scientific}. Thus, the active area of the CCD is effectively pixelated in a plane that is normal to a vector pointing from left to right in Fig.~\ref{fig:structure} (left). This entire apparatus is placed inside copper shielding. The CCD-epoxy interface and the backside of the CCD contain several micrometer-scale layers of different materials, as shown in the right panel of Fig.~\ref{fig:structure}. The in-situ doped polysilicon on the backside is a P-doped Si layer with doping density as a function of distance from the backside that is shown in Fig.~\ref{fig:doping}.\\

The Skipper-CCD has an electric field applied across its width, which lies along the breadth of the CCD and points away from the backside (from left to right in Fig.~\ref{fig:structure}). When electron-hole pairs are created in the bulk of the CCD, the holes are quickly drifted to the frontside, where they are stored in a pixel until readout. The SENSEI CCD reads pixel charges by counting \textit{holes}. Electrons are drifted into the backside where they are eventually lost into contacts. The applied electric field does not completely penetrate the doped backside layer of the CCD (shown in green in the right panel of Fig.~\ref{fig:structure}), and hence any charges created in that layer are only partially collected.  This results in an active part of the CCD that is effectively smaller than its total volume, as noted above. 

To read out the charge of each pixel, gate voltages are changed to shift the holes into a readout stage. Due to the sub-electron noise achieved by the Skipper-CCD technology, SENSEI is sensitive to single-hole excitations in the CCD. We note that even though the CCD reads out holes, these signals are traditionally referred to as \textit{electron} excitations and denoted as ``electron''. Once the charges are collected, the resulting pixelated image is analyzed. 

In its recent run near the MINOS cavern~\cite{SENSEI:2020dpa}, SENSEI observed several high-energy events (pixels containing $>100$~electrons) as well as many low-energy events (pixels containing 1, 2, or a few electrons). In these data, events in only two quadrants of the CCD were considered. The high-energy events are typically tracks of either cosmic muons, or electrons that are ejected through Compton scattering or absorption by radiogenic photons in the detector. These high-energy particles create several secondary electron-hole pairs as they traverse the CCD, which are measured by SENSEI as high-energy events. To remove this high-energy background and any correlated low-energy events, SENSEI applies masks on the images based on the positions of these high-energy events. One of these masks is a ``halo-mask'' where all pixels within a circle of a certain radius, called the halo-mask radius, are removed around the pixels containing high-energy events. After applying a 60-pixel halo-mask in addition to other masks, SENSEI observes a 1-electron event rate of $\sim$450/(gram-day).\\   

It was suggested in~\cite{Du:2020ldo} that the background events observed by SENSEI could originate from high-energy tracks, either via secondary emission of Cherenkov photons or by photons generated  due to recombination of secondary electron-hole pairs generated by these tracks. In the bulk of the CCD, where an electric field is present, electron-hole pairs created by tracks are quickly drifted away and there is no recombination~\cite{janesick1987scientific,Du:2020ldo}. These events are vetoed by the halo mask, so these charge pairs do not represent backgrounds. In the doped backside of the CCD, however, the electric field does not penetrate completely as discussed above, and holes can also recombine with the background density of electrons supplied by the dopants. Thus, in the backside, a fraction of the generated holes are measured (and vetoed by the halo mask), and a fraction recombines radiatively. The resulting recombination photons can be reabsorbed in the active area of the CCD and far from their originating tracks (away from halo masks), in which case they represent a background. In order to simulate the recombination events generated by high-energy tracks, and also in order to compute the partial-charge collection in the CCD backside, which will be relevant to predict the observed events in SENSEI, we need to first understand in detail the dynamics of charges in this doped-backside layer. We do this in the following section.

\section{Dynamics of charges in the doped backside of the CCD}\label{sec:backsidesim}

In this section, we simulate the dynamics of charge carriers in the backside of the SENSEI CCD. Since the CCD operates by reading out holes, we will focus here on computing the dynamics of holes. 
As discussed earlier, our aim is to understand the fraction of holes that recombine radiatively, and the fraction of holes that get collected and measured in the CCD. The dynamics of holes are governed by the continuity equation, which is  given by Eq.~\eqref{eq:continuity}. For the particular case of the doped backside of SENSEI, we make two assumptions to simplify the transport equation:
\begin{itemize}
\item We take the electron density in the backside to be  $n_e = N_D$, where $N_D$ is the doping density. First, this amounts to assuming that the dopants are fully ionized, which at SENSEI's operating temperature of $T=135$~K is a good approximation given that the P-dopant ionization energy is $E_I \approx 45$~meV~\cite{shklovskii2013electronic}, which leads to a non-ionization fraction $e^{-E_I/T}\approx 2 \times 10^{-2}$. And second, when the dynamics of electrons and holes around high-energy tracks is being considered, this amounts to neglecting the secondary electrons created by the tracks. This is justified by the timescales of recombination, which are much slower than the timescales of diffusion~\cite{Du:2020ldo}. The secondary electron density may be higher than the doping density initially, but diffusion brings down the density quickly and drastically, so on the relevant timescales of recombination the secondary electron density is always negligible compared to the background doping density. This approximation simplifies the recombination rates in Eq.~\eqref{eq:recorates}. In particular, the direct band-to-band radiative recombination rate and the Auger recombination rate can be written as,
\begin{align}
\Gamma^{\text{direct}}_{h,\text{doped}} = B N_D (n_h - \overline{n}_h) \label{eq:directrecoDoped}~,\\
\Gamma^{\text{Auger}}_{h,\text{doped}} = N_D(a_e N_D + a_h n_h)(n_h - \overline{n}_h) \label{eq:AugerDoped}~ .
\end{align}   
\item The trap-assisted recombination timescale in the bulk of the CCD is $\sim 10^{-3}$ seconds~\cite{trapassistedref}. We assume that the timescale is similar in the doped backside of the CCD. As we will see later, this timescale is always much slower than other processes on the backside (see Fig.~\ref{fig:timescale}). Thus, we neglect trap-assisted recombination, and only consider direct and Auger recombination. 
\end{itemize}
With these simplifications, the continuity equation for holes in the doped backside reads 
\begin{align}\label{eq:dopedcontinuity}
\frac{\partial n_{h}}{\partial t} = - \nabla \cdot \vec{j}_{h} - \Gamma^{\text{direct}}_{h,\text{doped}} - \Gamma^{\text{Auger}}_{h,\text{doped}}~, 
\end{align}  
where $\Gamma^{\text{direct}}_{h,\text{doped}} $ and $\Gamma^{\text{Auger}}_{h,\text{doped}}$ are as given by Eqs.~\eqref{eq:directrecoDoped}, \eqref{eq:AugerDoped}, and $\vec{j}_{h}$ is as given by Eq.~\eqref{eq:currenthole}. To solve this equation, the drift-diffusion constants and the recombination constants need to be understood for Si at 135~K, and at various doping levels. In the following subsections, we discuss these constants in detail. We summarize the values we use in Table~\ref{tab:driftdiffusionvalues}.
\begin{table}[t]

\centering
\begin{tabular}{ |c|c| }
\hline
Constant & Value used / Figure \\
\hline
Hole mobility $\mu_h$ & Right panel of Fig.~\ref{fig:mobility} \\
\hline
Hole diffusion constant $D_h$ & Related to $\mu_h$ through Eq.~\eqref{eq:einstein}\\
\hline
\multirow{2}{*}{Direct-recombination coefficient $B$}  & $3.3 \times 10^{-15}~\text{cm}^3 \text{s}^{-1}$ for $N_D > 10^{18}~\text{cm}^{-3}$ \\ & $6.3 \times 10^{-14}~\text{cm}^3 \text{s}^{-1}$ for $N_D < 10^{18}~\text{cm}^{-3}$ \\
\hline
Auger recombination coefficient $a_e$ & $0.9 \times 10^{-31}~\text{cm}^6 \text{s}^{-1}$ \\
\hline
Auger recombination coefficient $a_h$ & $9.9 \times 10^{-32}~\text{cm}^6 \text{s}^{-1}$ \\
\hline
\end{tabular}
\caption{Values used, or reference to figure, for the drift-diffusion constants and recombination coefficients in this work.}
\label{tab:driftdiffusionvalues}
\end{table}

\subsection{Drift-diffusion constants}
The diffusion constant $D_h$ and the hole mobility $\mu_h$ in Eq.~\eqref{eq:currenthole} are related by the Einstein equation \cite{hu2010modern},
\begin{align}\label{eq:einstein}
D_h = \frac{\mu_h T}{e}~,
\end{align}
where $e$ is the electric charge, and $T$ is the temperature. For doping densities $N_D < 10^{17}~\text{cm}^{-3}$, we model the hole mobility using the analytic model in~\cite{reggiani2000analytical}. As shown in~\cite{reggiani2000analytical}, for higher doping densities this analytic model fails to reproduce the experimentally measured values of electron and hole mobilities in Si below room temperature, so  for $N_D \geq 10^{17}~\text{cm}^{-3}$ we instead use the mobility model of ~\cite{klaassen1992unified}, which leads to better agreement with data at the SENSEI operating temperature. This forms our base model for mobility, which we denote by $\mu_h^0$. 

\begin{figure}[t]
	\centering
	\includegraphics[width=0.48\textwidth]{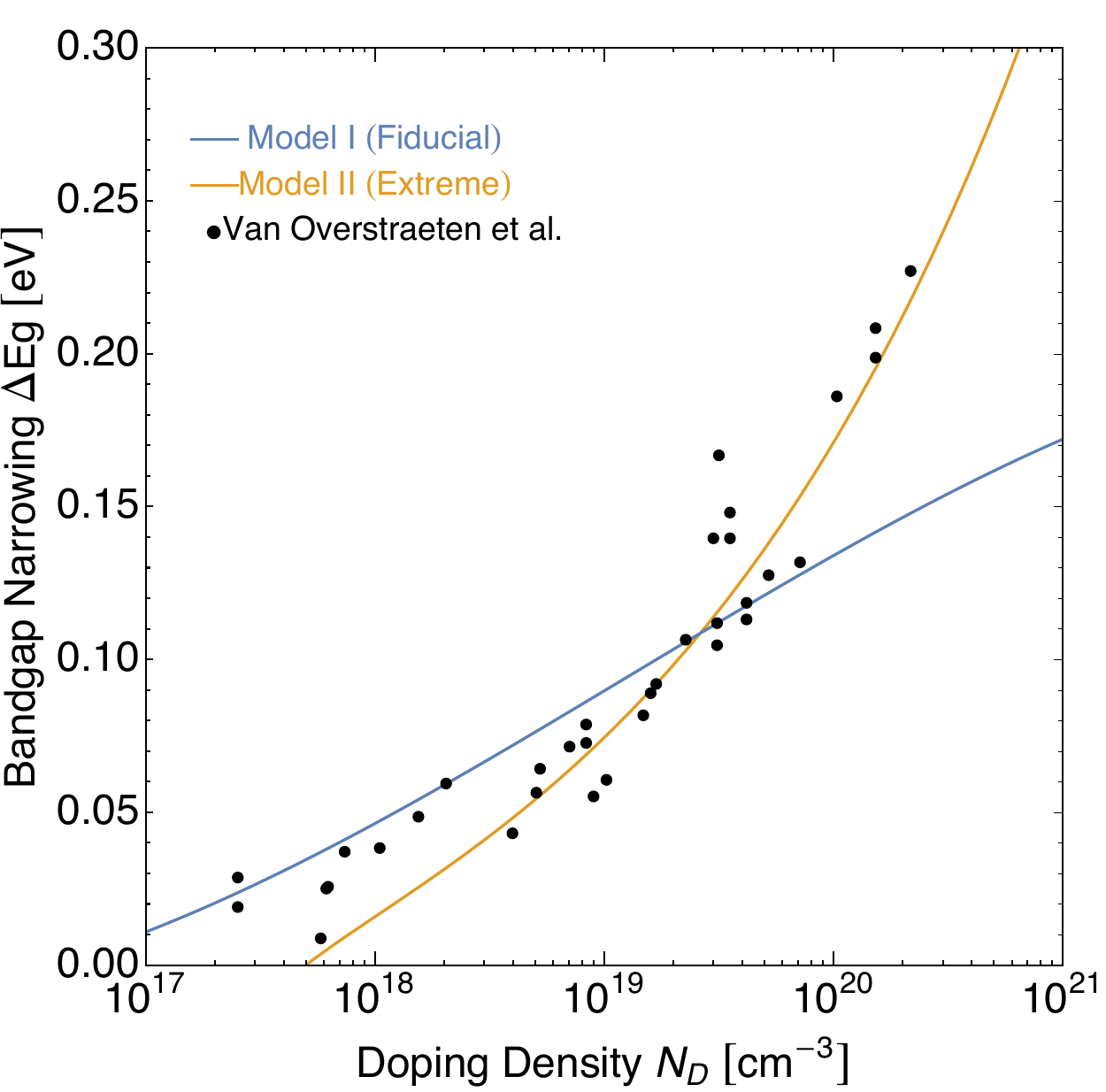}
	\includegraphics[width=0.48\textwidth]{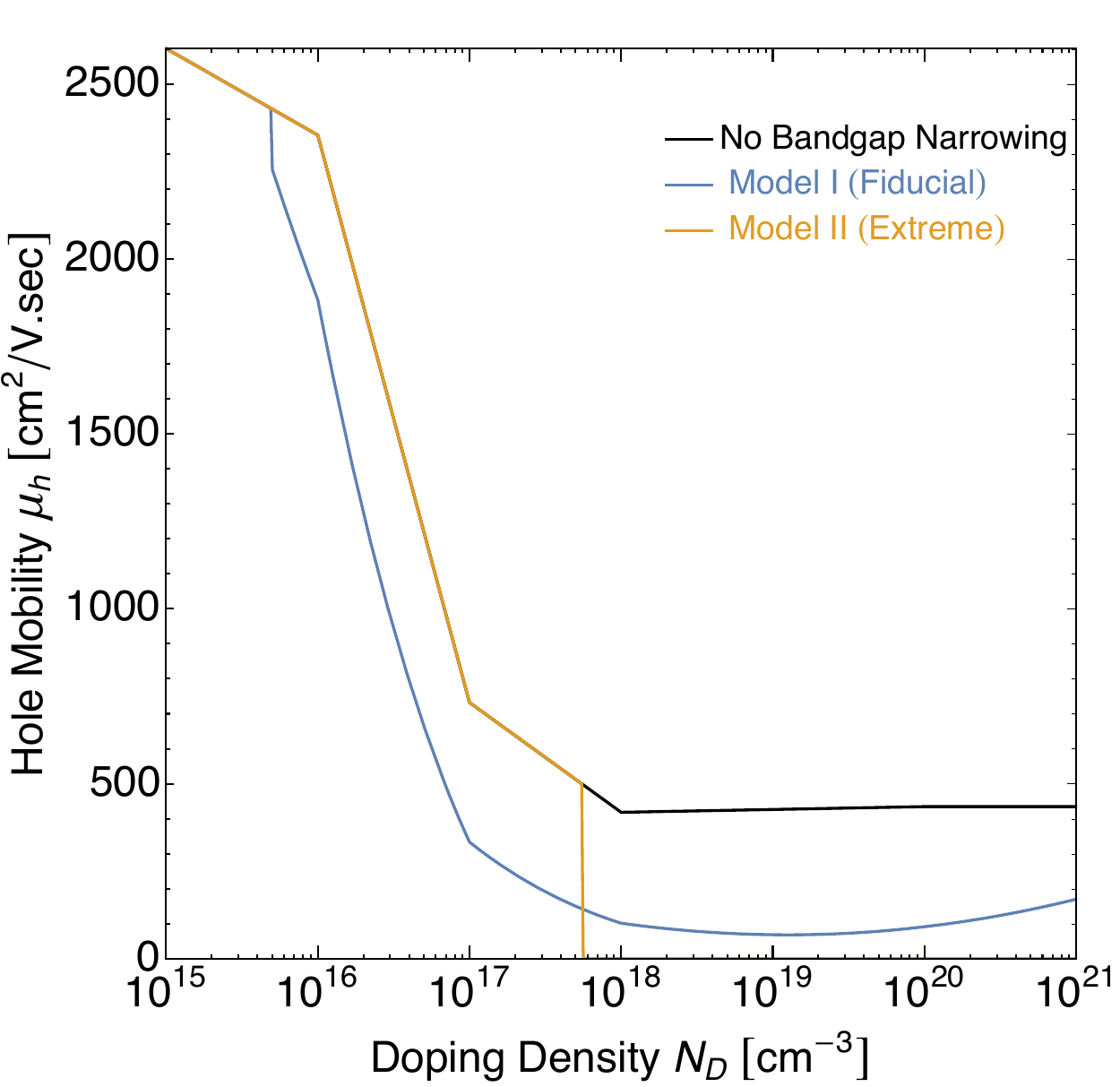}
	\caption{\textbf{Left:} The data for bandgap narrowing in Si as a function of the doping density. The black points represent the actual data~\cite{van1987heavy}, and the blue and the orange curves show the two fitting functions we consider as Model I and Model II, respectively. \textbf{Right:} Hole mobility in Si as a function of the doping density with Model I of bandgap narrowing (blue), Model II of bandgap narrowing (orange), and no bandgap narrowing (black).} \label{fig:mobility}
\end{figure}

We further improve the mobility model by accounting for bandgap narrowing due to the high doping concentrations, which leads to a correction that is given by~\cite{kane1993effect}
\begin{align}\label{eq:bandgapnarrowing}
\mu_h = \mu_h^0 \left(1- \frac{N_D}{2T}\frac{\delta \Delta E_g}{\delta N_D}\right)~,
\end{align} 
where $\Delta E_g$ is the bandgap narrowing due to the dopants.  Measurements of $\Delta E_g$ have been summarized in~\cite{van1987heavy}; we present them in the left panel of Fig.~\ref{fig:mobility}. Given the large scatter in the measured data, a systematic error arises from our inability to accurately determine $\Delta E_g$ as a function of the doping density $N_D$, which is required to compute the derivative in Eq.~\eqref{eq:bandgapnarrowing}. To account for this systematic, we consider two interpolations of the data, which we label as Model~I and Model~II, as shown in the left panel of Fig.~\ref{fig:mobility}. Model~I is our fiducial model and represents a mild bandgap narrowing effect. Model~II is an ``extreme'' case, which represents a more aggressive bandgap narrowing effect.
Using these two models, the mobility computed using Eq.~\eqref{eq:bandgapnarrowing} is shown in the right panel of Fig.~\ref{fig:mobility}.

While Model~II is a reasonable interpolation of the bandgap narrowing data, we refer to it as being ``extreme'' since it completely suppresses diffusion for doping densities above $\sim 5\times 10^{17}\,\mathrm{cm}^3$, as seen on the right panel of Fig.~\ref{fig:mobility}.  We will also see in Sec.~\ref{sec:chargecollection} that Model~I predicts  a partial charge collection in the backside that agrees well with data, while Model~II disagrees with measurements, so we will refer to Model I as being ``fiducial''. We also show our base model for mobility before including the bandgap narrowing effect. The diffusion constant is calculated from the mobility using Eq.~\eqref{eq:einstein}.\\
\subsection{Direct-recombination coefficient}
The direct-recombination coefficient $B$ in Eq.~\eqref{eq:directrecoDoped} sets the rate of radiative recombination in the doped backside. Direct recombination can proceed through the capture of an electron in the conduction band by a hole in the valence band, or through the annihilation of an electron-hole pair bound in an exciton. In doped Si, exciton formation is strongly suppressed for doping levels above a critical level referred to as the exciton Mott value. This critical doping density in Si is given as an analytic function in~\cite{kane1993effect}, and is estimated to be $\sim 5 \times 10^{17}~\text{cm}^{-3}$ at 135~K. In~\cite{liu2016quantifying}, the critical doping density is instead suggested to be $\sim 3 \times 10^{18}~\text{cm}^{-3}$.  
To account for the excitonic Mott transition, we only consider electron-hole recombination for values above the critical density, while for values below the Mott density we also account for the contribution of exciton annihilations. These two contributions are obtained from the theoretical model presented in~\cite{doi:10.1002/pssa.2210210140}. For the critical Mott density we simply take a value that is intermediate within the ones found in the literature, $N_D=10^{18}~\text{cm}^{-3}$. 
This results in a recombination coefficient
 $B=3.3 \times 10^{-15}~\text{cm}^3\text{s}^{-1}$  for $N_D > 10^{18}~\text{cm}^{-3}$, and $B= 6.3 \times 10^{-14}~\text{cm}^3\text{s}^{-1}$ for $N_D < 10^{18}~\text{cm}^{-3}$.

\subsection{Auger recombination coefficient}
The Auger recombination coefficients $a_e$ and $a_h$ in Eq.~\eqref{eq:AugerDoped} are set as follows. The coefficient $a_e$ in n-doped Si at 300~K is measured to be $1.6 \times 10^{-31}~\text{cm}^6\text{s}^{-1}$~\cite{wieder1980emitter}. It is expected to follow a temperature dependence of $T^{0.72}$~\cite{klaassen1992unified}. Using this dependence, the value of $a_e$ at 135~K is set to be $0.9 \times 10^{-31}~\text{cm}^6\text{s}^{-1}$. The value of $a_h$ is set to be $9.9 \times 10^{-32}~\text{cm}^6\text{s}^{-1}$~\cite{doi:10.1063/1.89694}. 
\subsection{Simulation of dynamics of charges in the doped backside of the CCD}
With the values of the drift-diffusion constants and the recombination coefficients discussed in the previous section, we now simulate the dynamics of charge carriers created in the doped backside layer of the SENSEI experiment. We consider a $15~\micron \times 15~\micron \times 15~\micron$ volume on the backside of the CCD, which roughly amounts to the volume of one CCD pixel across the whole depth of the backside, and discretize this volume to create a $20 \times 20 \times 20$ mesh such that the distance between consecutive mesh points (denoted by $\Delta l$) is $0.75~\micron$. Each point in this mesh can then be numbered $\{i,j,k\}$, with $i$ along the length of the CCD, $j$ along the breadth of the CCD, and $k$ along the depth of the CCD. We define the origin so that the points on the backside of the CCD have $k=0$, with $k>0$ points lying inside the CCD, and the $i=j=0$ line passes through the center of the volume under consideration. We then discretize Eq.~\eqref{eq:dopedcontinuity} in space and time, and denoting the over-density of holes at the position $\{i,j,k\}$ on the mesh and at the time $t_n$ by $\delta n_h(i,j,k,t_n)$, we get,
\begin{align}\label{eq:discrete}
\frac{ \delta n_h(i,j,k,t_{n+1})-  \delta n_h(i,j,k,t_{n})}{\Delta t} &= D_h(k) \Big( \frac{\Delta^2 \delta n_h(i,j,k,t_{n})}{\Delta l^2}\Big) + \Big(\frac{\Delta_z \delta n_h(i,j,k,t_{n})}{\Delta l} \times \frac{\Delta_z D_h (k)}{\Delta l} \Big) \nonumber \\& - \Big(\frac{\Delta_z (\mu_h(k)\delta n_h(i,j,k,t_{n})E(k))}{\Delta l} \Big) - \Gamma^{\text{direct}}_{h,\text{doped}} (i,j,k,t_{n})  \nonumber \\& - \Gamma^{\text{Auger}}_{h,\text{doped}}(i,j,k,t_{n})\ ,
\end{align} 
where, 
\begin{align}
\Delta^2 \delta n_h(i,j,k,t_{n}) &= (\delta n_h(i-1,j,k,t_{n}) + \delta n_h(i+1,j,k,t_{n}) - 2 \delta n_h(i,j,k,t_{n})) \nonumber \\&+  (\delta n_h(i,j+1,k,t_{n}) + \delta n_h(i,j-1,k,t_{n}) - 2 \delta n_h(i,j,k,t_{n})) \nonumber \\&+ (\delta n_h(i,j,k+1,t_{n}) + \delta n_h(i,j,k-1,t_{n}) - 2 \delta n_h(i,j,k,t_{n}))\ ,
\end{align}
\begin{align}
\Delta_z \delta n_h(i,j,k,t_{n}) = \delta n_h(i,j,k,t_{n}) - \delta n_h(i,j,k-1,t_{n})\ ,
\end{align}
\begin{align}
\Delta_z D_h (k) = D_h (k) - D_h (k-1),
\end{align}
\begin{align}
\Delta_z (\mu_h(k)\delta n_h(i,j,k,t_{n})E(k)) &= (\mu_h(k)\delta n_h(i,j,k,t_{n})E(k)) \nonumber \\& - (\mu_h(k-1)\delta n_h(i,j,k-1,t_{n})E(k-1))\ ,
\end{align}
\begin{align}\label{eq:directrecosim}
\Gamma^{\text{direct}}_{h,\text{doped}} (i,j,k,t_{n}) = B(k) N_D(k) \delta n_h(i,j,k,t_{n})\ ,
\end{align}
\begin{align}\label{eq:Augerrecosim}
\Gamma^{\text{Auger}}_{h,\text{doped}}(i,j,k,t_{n}) = N_D(k)(a_e N_D(k) + a_h (\delta n_h(i,j,k,t_{n})+ \overline{n}_h(k)))\delta n_h(i,j,k,t_{n}) ,
\end{align}
\begin{align}
\Delta t \equiv t_{n+1}-t_n\ .
\end{align}
Eq.~\eqref{eq:discrete} gives the change in $\delta n_h$ at all points of the mesh from $n^{\text{th}}$ time-step to $(n+1)^{\text{th}}$ time-step. We start from an initial condition where there is an over-density of $\delta n_h$ at one of the points in the mesh. Since the doping density, the drift-diffusion constants, and the electric field only depend on the distance from the backside, the dynamics only depend on the initial $k$ of the point where the over-density is placed. Hence, we choose $i = j = 0$ for the initial over-density, and vary the value of $k$. For a chosen value of $k=k_0$ for the over-density, the initial conditions for $\delta n_h$ are then
\begin{align}
\delta n_h (0,0,k_0,t_0)= 10^{18}~\text{cm}^{-3}\ ,\\
\delta n_h (i,j,k,t_0)= 0~\text{for}~\{i,j,k\}\neq\{0,0,k_0\}\ . 
\end{align}
We choose the value of $10^{18}~\text{cm}^{-3}$ for the initial over-density because it is the typical carrier density created by a crossing high-energy track \cite{Du:2020ldo}. Changing this initial value does not affect the outputs of the charge-diffusion simulation (fractional recombination rates and partial-charge collection), as these are insensitive to the normalization of the initial overdensity. From these initial conditions, we simulate time-steps with $\Delta t = 1~\mu \text{s}$, and use Eq.~\eqref{eq:discrete} to calculate how the hole density evolves on the mesh. We employ Neumann boundary conditions for the simulation, and set the first discrete derivative to zero at all boundary points. This ensures charge conservation if only diffusion is considered on the mesh. However, as we also have drift and recombination, the charge density drops with time. We run this simulation until the entire over-density of the holes on the mesh is low enough to have less than a single hole around every point within a distance of the mesh resolution $\Delta l$. In other words, we run the simulation for $n_\text{max}$ steps, where $n_{\text{max}}$ satisfies 
\begin{align}
\delta n_h(i,j,k,t_{n_{\text{max}}}) \leq 1/(\Delta l)^3\ ,
\end{align}  
for all $\{i,j,k\}$ on the mesh. The fraction of the hole density that undergoes direct recombination, which we denote by $f_{\text{direct}}$, is then given by 
\begin{align}
f_{\text{direct}} = \Big(\sum_{\{i,j,k\}\in\text{mesh}}\sum_{n=0}^{n_{\text{max}}} \Gamma^{\text{direct}}_{h,\text{doped}} (i,j,k,t_{n}) \Delta t \Big)/(10^{18}~\text{cm}^{-3})\ ,
\end{align}
where $\Gamma^{\text{direct}}_{h,\text{doped}} (i,j,k,t_{n})$ is given by Eq.~\eqref{eq:directrecosim}. The fraction of the hole density that undergoes Auger recombination, which we denote by $f_{\text{Auger}}$, is given by 
\begin{align}
f_{\text{Auger}} = \Big(\sum_{\{i,j,k\}\in\text{mesh}}\sum_{n=0}^{n_{\text{max}}} \Gamma^{\text{Auger}}_{h,\text{doped}} (i,j,k,t_{n}) \Delta t \Big)/(10^{18}~\text{cm}^{-3})\ ,
\end{align}
where $\Gamma^{\text{Auger}}_{h,\text{doped}} (i,j,k,t_{n})$ is given by Eq.~\eqref{eq:Augerrecosim}. As the only other mechanism of depletion of holes is the collection of holes through drift under the effect of the electric field, the fraction of the hole density that gets collected by the field, which we denote by $f_{\text{collected}}$, is given by 
 \begin{align}\label{Eq:fcollected}
f_{\text{collected}} = 1-f_{\text{direct}}-f_{\text{Auger}}\ .
\end{align}
We calculate the values of $f_{\text{collected}}$ and $f_{\text{direct}}$ by running simulations for various initial positions of over-densities. 

\begin{figure}[t]
	\includegraphics[width=0.48\textwidth]{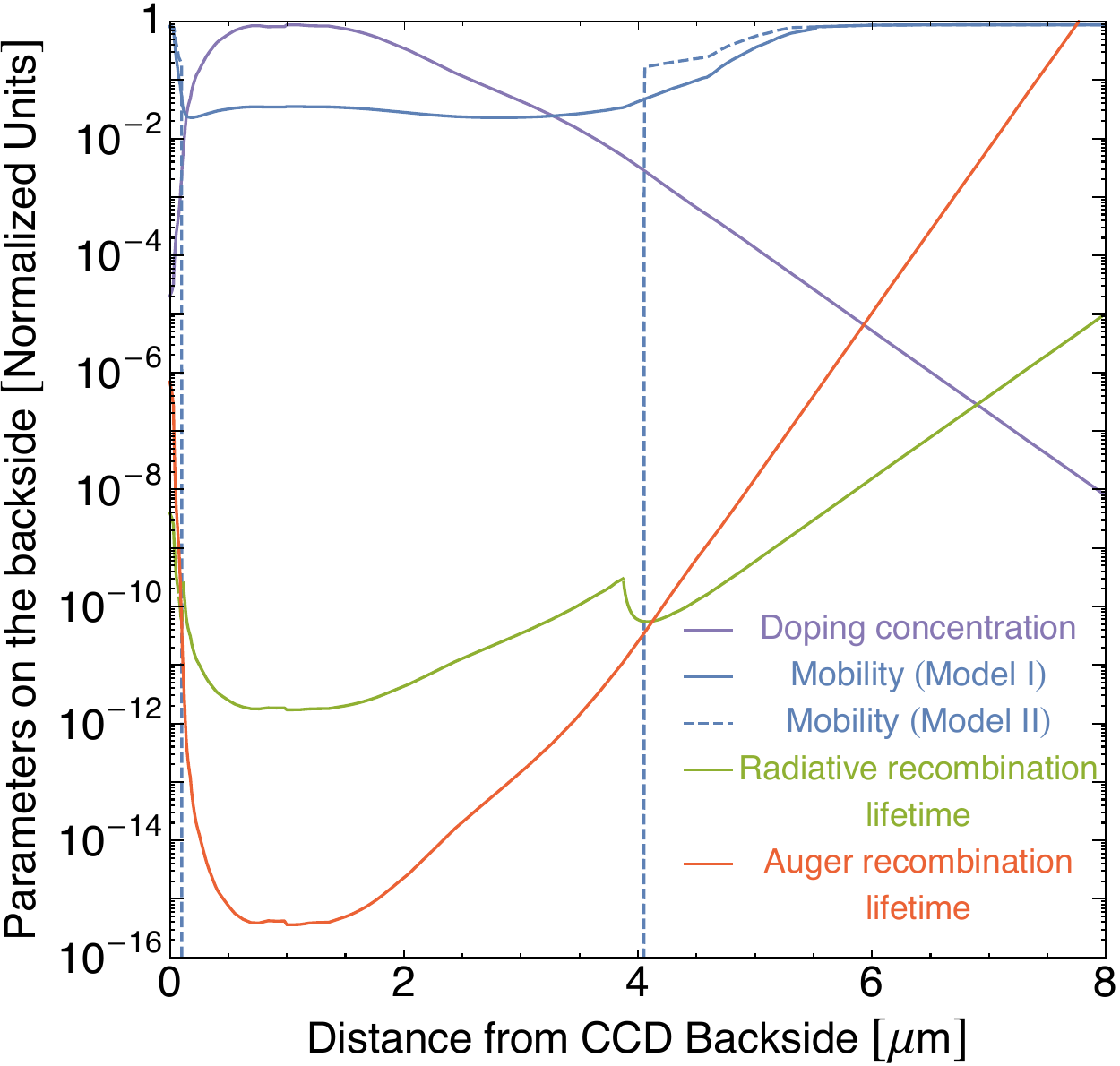}
	\includegraphics[width=0.48\textwidth]{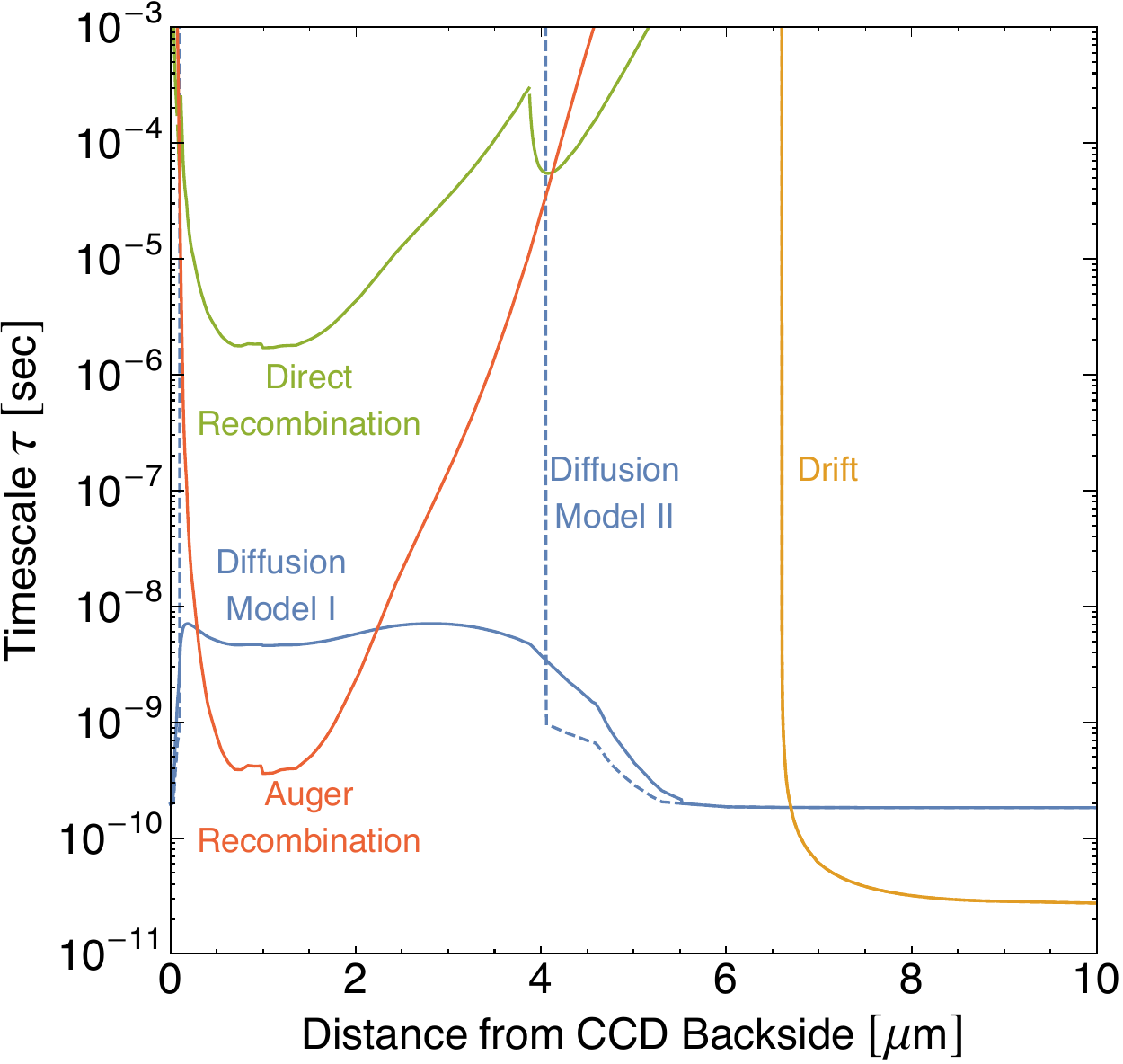}
	\caption{\textbf{Left:} Parameters on the backside of the CCD in normalized units of $2 \times 10^{20}~\text{cm}^{-3}$ for the doping concentration, $3000~\text{cm}^2/\text{V}/\text{sec}$ for the hole mobility, and $10^6~\text{sec}$ for the recombination lifetimes. \textbf{Right:} The timescales of various processes in the doped backside of the CCD.} \label{fig:timescale}
\end{figure}
To understand the results presented in the following subsections, it is convenient to first present the timescales of various processes in this discrete simulation. Diffusion attempts to even out the density of charges, and is the most active when there are large gradients in the densities. From Eq.~\eqref{eq:discrete}, the timescale for diffusion to smooth out densities on the length-scale corresponding to $\Delta l$ is roughly given by 
\begin{equation}
\tau_{\text{diffusion}}\sim \frac{\Delta l^2}{D_h}\ .
\end{equation} 
In the presence of an electric field $E$, the drift carries charges in the direction of the field. The timescale for drift to carry charges on the length-scale of $\Delta l$ is given by 
\begin{equation}
\tau_{\text{drift}}\sim \frac{\Delta l}{\mu_h E}\ .
\end{equation} 
The direct and Auger recombination timescales are given by,
\begin{align}
\tau_{\text{direct}}\sim (B N_D)^{-1}\\
\tau_{\text{Auger}}\sim (a_e N_D^2)^{-1}\ .
\end{align}
We show the parameters on the doped backside in the left panel of Fig.~\ref{fig:timescale}, and show the timescales of various processes as a function of the distance from the backside of the CCD in the right panel of Fig.~\ref{fig:timescale}. The extreme Model II has no diffusion/mobility above the doping levels of $\sim 5 \times 10^{17}~\text{cm}^{-3}$ due to the strong bandgap narrowing. This corresponds to the diverging diffusion timescale in the right panel of Fig.~\ref{fig:timescale} and zero mobility in the left panel of Fig.~\ref{fig:timescale} within $\sim 4~\micron$ of the distance from the backside of the CCD. The timescale of drift under the electric field diverges below the distance of $\sim 6.6~\micron$ as there is no electric field in that region. The sudden transition in the direct recombination timescale just below $\sim 4~\micron$ from the backside of the CCD arises from the activation of excitonic processes below the doping levels of $\sim 10^{18}~\text{cm}^{-3}$. 

\subsubsection{Partial charge collection in the doped backside of the CCD}\label{sec:chargecollection}

\begin{figure}[t]
	\centering
	\includegraphics[width=0.48\textwidth]{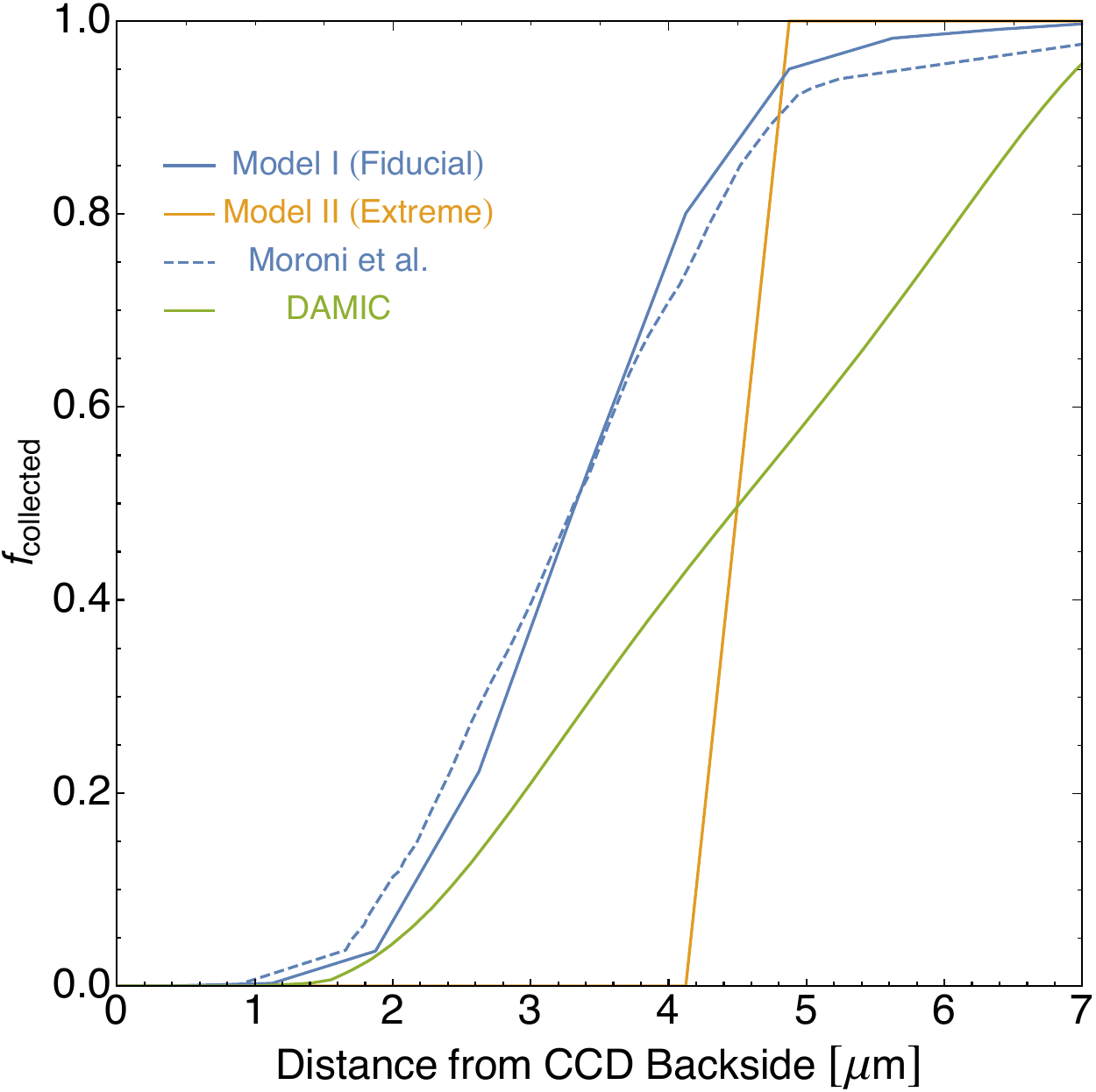} \quad 
	\includegraphics[width=0.48\textwidth]{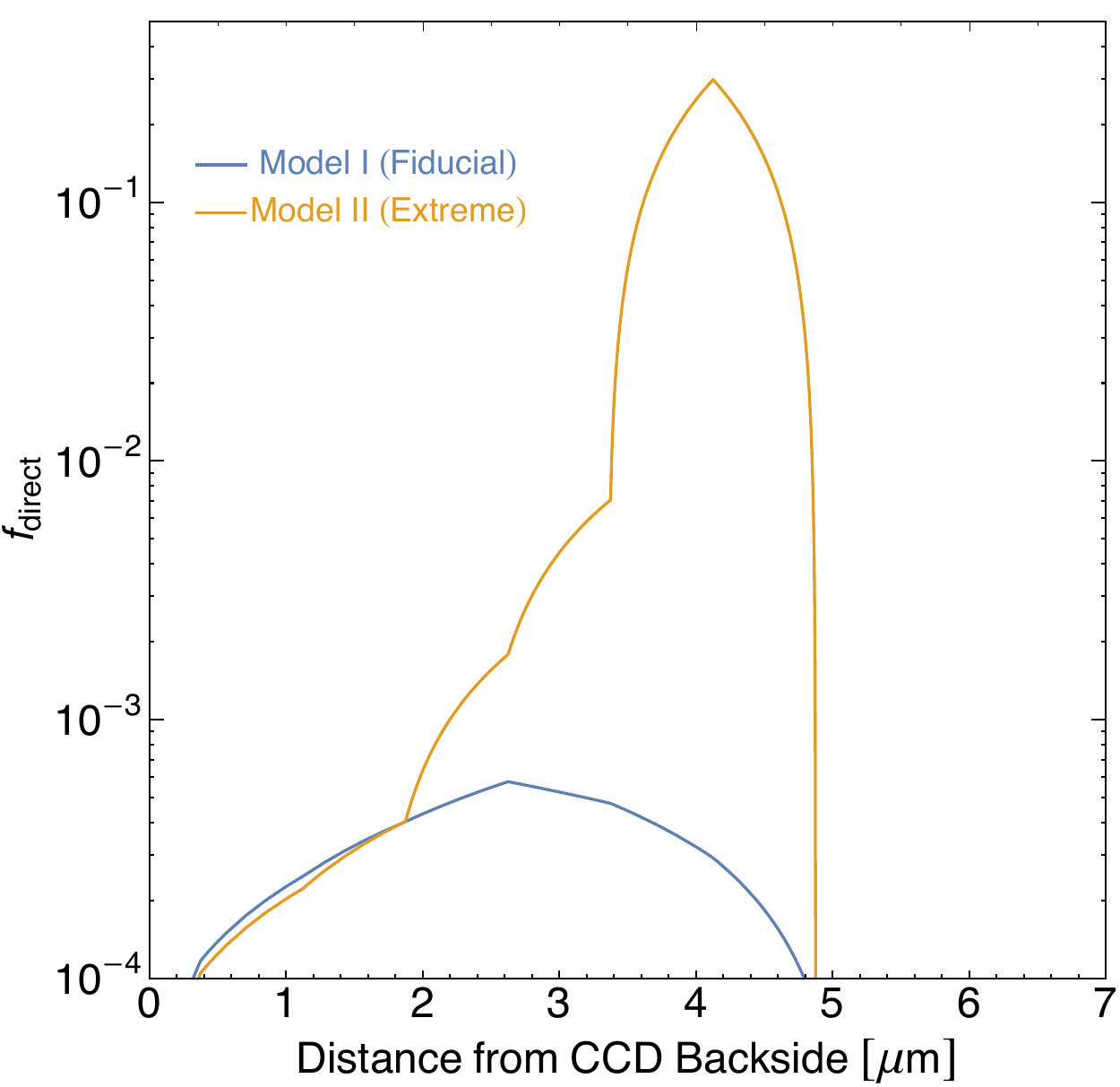}
	\caption{\textbf{Left:} The fraction of secondary charges that are drifted by the electric field towards the bulk and later collected by the detector, $f_{\text{collected}}$,  as a function of the initial position of the charge over-density. The blue (orange) solid line shows $f_{\text{collected}}$ for the fiducial Model I (extreme Model II) of transport parameters. The blue dashed line shows the measured $f_{\text{collected}}$ in~\cite{Moroni_2021}. The green line shows $f_{\text{collected}}$ considered in the DAMIC analysis in~\cite{DAMIC:2021crr}. \textbf{Right:} The fraction of secondary charges that undergo radiative recombination $f_{\text{direct}}$ as a function of the initial position of the charge over-density. The blue (orange) solid line shows $f_{\text{direct}}$ for the fiducial Model I (extreme Model II) of transport parameters.} \label{fig:fractions}
\end{figure} 

Only a fraction, $f_{\text{collected}}$ (see Eq.~\eqref{Eq:fcollected}), of secondary charges that are created on the CCD backside are drifted by the electric field towards the bulk and eventually measured by the detector.
The left panel of Fig.~\ref{fig:fractions} shows $f_{\text{collected}}$ as a function of the initial position of the overdensity. The results can be explained by comparing the timescales of various processes happening in the CCD backside as shown in Fig.~\ref{fig:timescale}. For the diffusion Model~I, the Auger recombination is the fastest process within $\sim 2~\micron$ from the backside of the CCD. The region between $\sim 2~\micron$ and $\sim6.6~\micron$ is diffusion dominated. The region beyond $\sim6.6~\micron$ is dominated by the drift due to the electric field. Thus, in Model I, the holes in the region between $\sim 2~\micron$ and $\sim6.6~\micron$ flow to the underdensities of holes created by the Auger recombination and the drift on either sides. This produces a gradually increasing charge collection fraction as a function of increasing distance from the CCD backside. In the extreme Model II, on the other hand, recombination processes dominate the region up to a distance of $\sim 4~\micron$ from the backside. The region beyond that point is completely dominated by diffusion and the drift created by the electric field. This generates a steeply rising charge collection fraction, which is non-zero only for distances of $\sim 4~\micron$ or more from the backside of the CCD. It is noteworthy that the charge collection fraction that we estimate in the case of the fiducial Model I matches well with the measurements of the same in~\cite{Moroni_2021}. We also show the charge collection fraction in the DAMIC analysis in ~\cite{DAMIC:2021crr} for comparison. We can then take this measurement as a preliminary \textit{validation} of Model~I and as a \textit{falsification} of Model~II, justifying that we refer to Model I as the fiducial model. This reduces the systematics on the diffusion parameters. We stress, however, that this validation should only be taken as preliminary, as measurements of partial charge collection themselves could have experimental errors that invalidate this conclusion.

\subsubsection{Recombination in the doped backside of the CCD}\label{sec:recofraction}
The right panel of Fig.~\ref{fig:fractions} shows the fraction of the hole density that undergoes direct recombination, $f_{\text{direct}}$, as a function of the initial position of the over-density. In the fiducial Model I, direct recombination is always suppressed by other processes resulting in a small $f_{\text{direct}}$. On the other hand, in the extreme Model II, as there is no diffusion of holes in the first $\sim 4~\micron$ from the backside, the value of $f_{\text{direct}}$ steadily grows as $N_D^{-1}$ as we move away from the backside from high doping densities to low doping densities due to the relative suppression of the Auger process. However, the collection due to electric field starts dominating at the point at which the diffusion and drift turn on, and the value of $f_{\text{direct}}$ falls sharply. In the end, this gives a sharply peaked profile of $f_{\text{direct}}$, which gives maximum values of $\sim 0.3$ at a distance of $\sim 4~\micron$ from the backside of the CCD. Thus, the transport parameter model has an enormous impact on the values of $f_{\text{direct}}$, and can change it by a few orders of magnitude. This further highlights the importance of validating the diffusion models against partial charge-collection data, as discussed previously. \\

After having now discussed the two models of transport parameters and $f_{\text{collected}}$ and $f_{\text{direct}}$, the measurement of low energy events in SENSEI originating from radiative processes can be simulated. We discuss the components of this simulation in the following sections. 
\section{High-energy particles in SENSEI} \label{sec:cherenkovinsensei}

High-energy particles that traverse the SENSEI detector emit Cherenkov radiation and can create electron-hole pairs that recombine radiatively. The resulting  photons can then be absorbed in the CCD to create low-energy events. The first step in simulating this process is the simulation of high-energy tracks in the detector. For this, we need to identify the sources of high-energy particles and their flux as a function of solid angle and energy. For the SENSEI data-run near the MINOS cavern, the two biggest sources of high-energy events are cosmic muons and high-energy electrons ejected by X-rays and gamma rays from radioactive impurities. 

\subsection{Cosmic muon tracks}\label{sec:muon-flux}
The differential of the intensity $I$ of cosmic muons as a function of the solid angle $\Omega$ and the muon momentum $p$ measured in GeV at the depth of SENSEI in the MINOS cavern is given by~\cite{Bogdanova:2006ex},
\begin{align}\label{eq:muonflux}
\frac{dI}{dp~d\Omega}= (4.3\times10^{-3})\frac{18}{p \cos{\theta}+145} (p+ 2.7 \sec{\theta})^{-2.7} \frac{p+5}{p+5\sec{\theta}}~\text{cm}^{-2}\text{sec}^{-1}\ ,
\end{align}
where $\theta$ is the polar angle with respect to the vertical, and the factor of $4.3\times10^{-3}$ is the suppression of the muon flux because of the depth below the sea-level. The muon momentum $p$ is assumed to be distributed between $1~\text{GeV}$ and $10^5~ \text{GeV}$, as the flux away from this range is subdominant. We have verified that our flux computations match the muon flux observed by SENSEI.\\

In the simulation, we consider muon tracks in the CCD and the pitch adapter. For simplicity, we assume no correlation between tracks in the CCD and tracks in the pitch adapter, and simulate them independently. For each surface of the CCD and the pitch adapter, we first integrate the differential intensity in Eq.~\eqref{eq:muonflux} over the solid angle and the total muon momentum range, and normalize the intensity to the area of that particular surface and the exposure time. This gives the total number of muons crossing that surface during the exposure time. We then uniformly distribute the points on the surface where the muons enter the detector, and assign their angle with respect to the vertical and a momentum based on the distribution in Eq.~\eqref{eq:muonflux}. We track every muon until it leaves the detector. The energy lost by muons inside the detector, which is relevant to compute the number of electron-hole pairs in the CCD backside, is also calculated using their stopping power in Si~\cite{Groom:2001kq}. 

\subsection{High-energy electron tracks}\label{sec:electroneventrate}
High-energy electrons (with energies $\sim$keV-MeV) are produced when X-rays or gamma rays from impurities in or around the detector are absorbed in the detector. Work is under progress to model these using the Geant4~\cite{GEANT4:2002zbu} simulation toolkit, but in this paper, we extract the information about the high-energy electron tracks from the data collected by SENSEI during its run at MINOS, which also allows us to normalize the background rate directly to the data. SENSEI measured several high-energy electron events in the detector and measured their energies. These data, however, cannot be used directly in our simulation as the true spectrum of energetic electrons, since there is partial-charge collection in the detector's backside. Thus, SENSEI only measures a fraction of the energy deposited by an electron track if it passes through the backside. For our background computations, this is especially problematic if electrons are created via the photoelectric effect; in this case the CCD could be subject to low-energy electrons from backside x-ray absorption that go undetected, or that have an underestimated energy. This can occur for photons with energies below $\sim 5\,$keV, for which the absorption range in Si is $\mathcal{O}(10 \, \micron)$, which is comparable with the width of the partial-charge collection region. While these electrons have energies that are too low to create secondary Cherenkov backgrounds, they can produce radiative recombination photons, especially since such electrons are concentrated in the luminescent CCD's backside. If low-energy electrons are instead due to Compton scattering of higher-energy photons (that have an absorption range much larger than the backside thickness), then such electrons are primarily distributed in the non-luminescent bulk of the CCD, and are thus ineffective in generating secondary photons.

To account for this potential source of events, we build a ``true'' electron distribution as follows. We   
 simulate X-rays and gamma-rays entering the CCD through the backside, assuming that electrons are created through photoabsorption. We simulate photons with a range of energies between the bandgap of Si ($\sim 1.1~\text{eV}$) and $\sim 20~\text{keV}$. The point on the backside surface where the high-energy photon enters is chosen uniformly on that surface, and its direction is assumed to be random. We choose its propagation length based on the average penetration length in Si. We track the X-rays and gamma-rays from their initial point on the backside in discrete steps with a resolution of $10^{-5}$ of the dimensions of the CCD in all three directions. For the photon energies considered, the interaction of photons in Si is dominated by the photoelectric effect~\cite{nist_photoabsorption}, so we assume that an electron is ejected at the end of the propagation length and that the electron has the entire energy of the photon. The absorption point of the photon sets the starting point of the track, while the orientation of the electron track is chosen randomly. The energy of the electron sets the ionization stopping power of that electron in Si~\cite{nist_estar}, which determines the length of the electron track. We then propagate the electron track in discrete steps, also with a resolution of $10^{-5}$ of the dimensions of the CCD in all three directions. We assume here that the entire initial energy of the electron, which is equal to the energy of the incoming photon, is dissipated in the CCD into ionization as expected for the electrons within our energy range \cite{PhysRevD.98.030001}. We thus ensure that the entire electron track obtained from the starting point, orientation, and the length of the track lies inside the CCD. If this is not the case, we choose the orientation of the track again so that this condition is satisfied, as otherwise, the simulated track does not deposit the entire energy of the incoming high-energy photon. For all tracks of the same energy, we use the partial charge collection models calculated in Sec.~\ref{sec:backsidesim}, and find the energy measured in the detector. We then take the average of the measured energies for these tracks. We repeat this procedure for different initial energies of photons, and find the average measured energy as a function of the total deposited energy. 

\begin{figure}[t]
	\centering
	\includegraphics[width=0.45\textwidth]{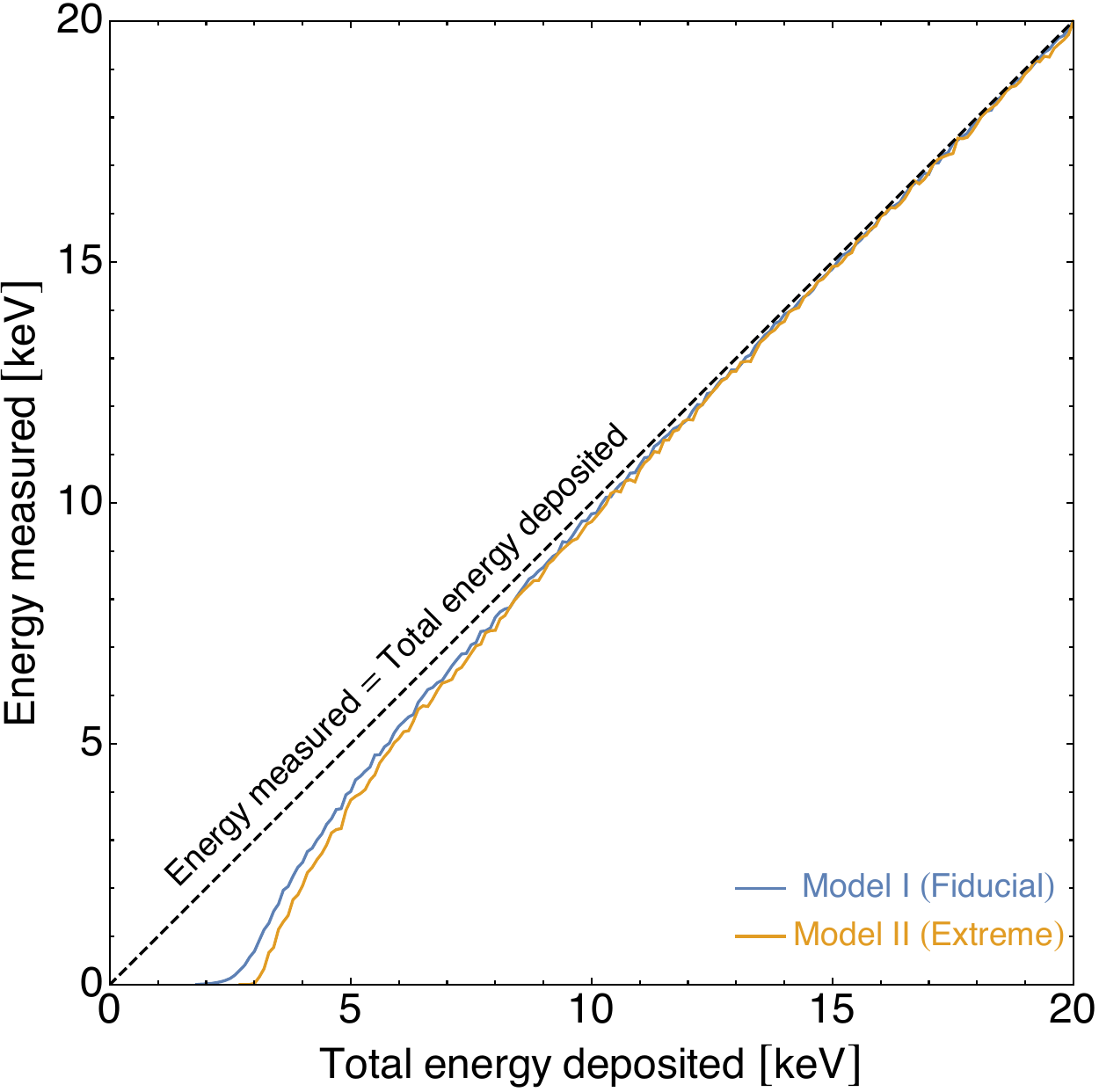}
	\includegraphics[width=0.48\textwidth]{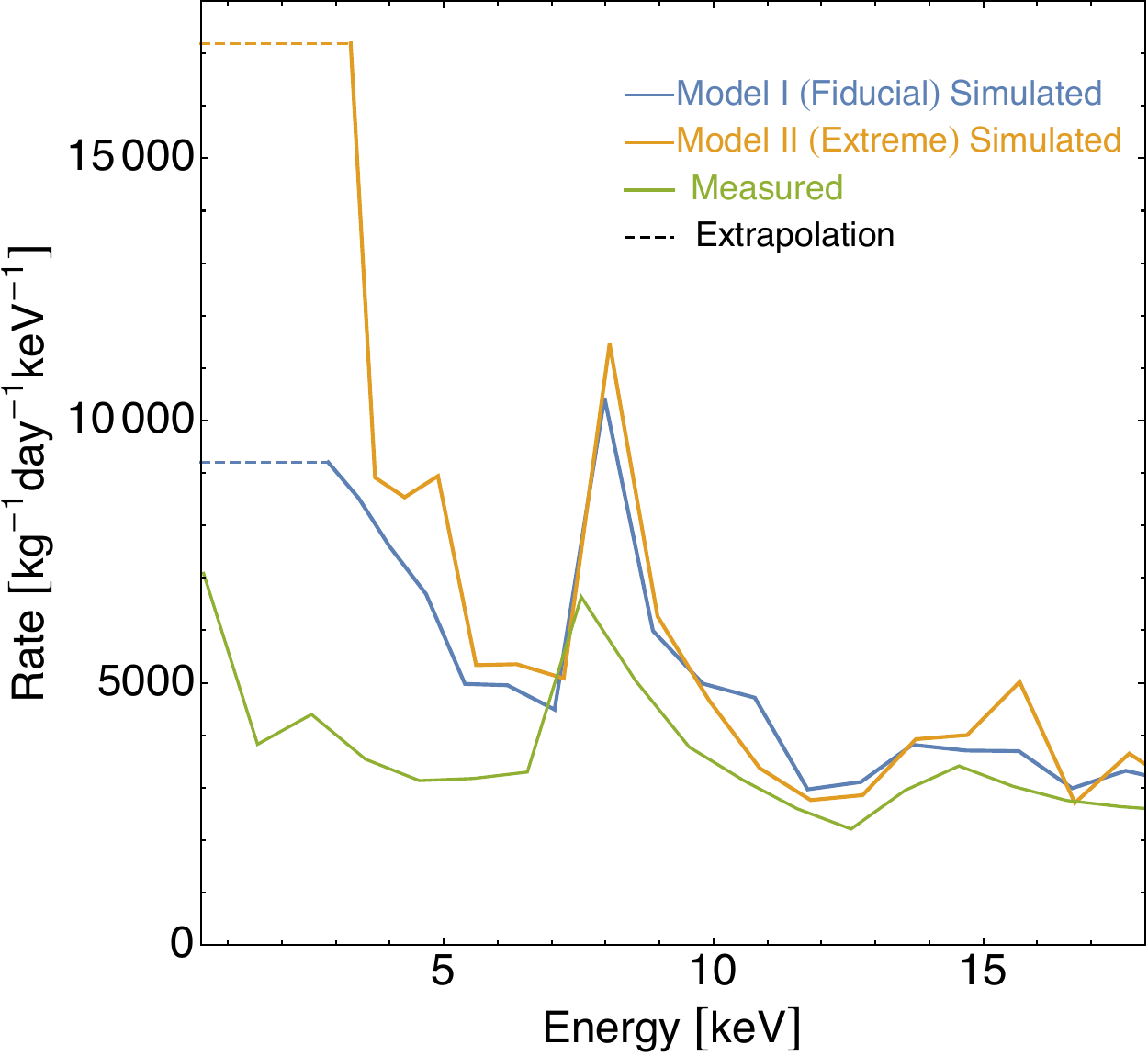}
	\caption{\textbf{Left:} Average energy measured through charge collection as a function of the total energy deposited by high-energy electron tracks.  \textbf{Right:} The estimated actual rate of high-energy electron tracks as a function of the total deposited energy. The blue (orange) line shows the rate for the fiducial Model~I (extreme Model~II) of transport parameters, with the dashed lines showing flat extrapolations at low energies. The green line shows the rate of high-energy electrons measured in the SENSEI CCD~\cite{SENSEI:2020dpa}, which is less than the actual rate due to partial charge collection. \label{fig:lowenergyelectronrate}}
\end{figure} 
The result of this simulation is shown in the left panel of Fig.~\ref{fig:lowenergyelectronrate}. The simulated measured energy as a function of the total deposited energy is used to convert the measured rate of electrons as a function of measured electron energy to the actual rate of high-energy electrons incident on the detector as a function of their total deposited energy. The actual rate as a function of the total deposited energy, computed for the two models of charge collection efficiency in Sec.~\ref{sec:chargecollection}, is shown in the right panel of Fig.~\ref{fig:lowenergyelectronrate}. We also show the measured rate for comparison. Note that we find that soft X-rays with energies less than $\sim 3$~keV get absorbed close to the backside of the CCD, and the resultant electron tracks are very short and are not measured. As a consequence, their rate cannot be inferred from the measurements of SENSEI. To overcome this limitation, we note that for the range of measured energies, and with the exception of the peaks in the spectrum that are likely associated to nuclear transitions (e.g., the peak at $\sim 8$~keV is likely from copper $K_\alpha$ fluorescence), the spectrum shows a moderate increase towards lower energies. In what follows we simply perform a flat extrapolation at energies lower than $\sim$ 3~keV, as shown by the dashed lines in the right panel of Fig.~\ref{fig:lowenergyelectronrate}. We also perform a simulation with a variation of this extrapolation, which we discuss in Sec.~\ref{sec:systematics}.\\ 

\begin{figure}[t]
	\centering
	\includegraphics[width=0.48\textwidth]{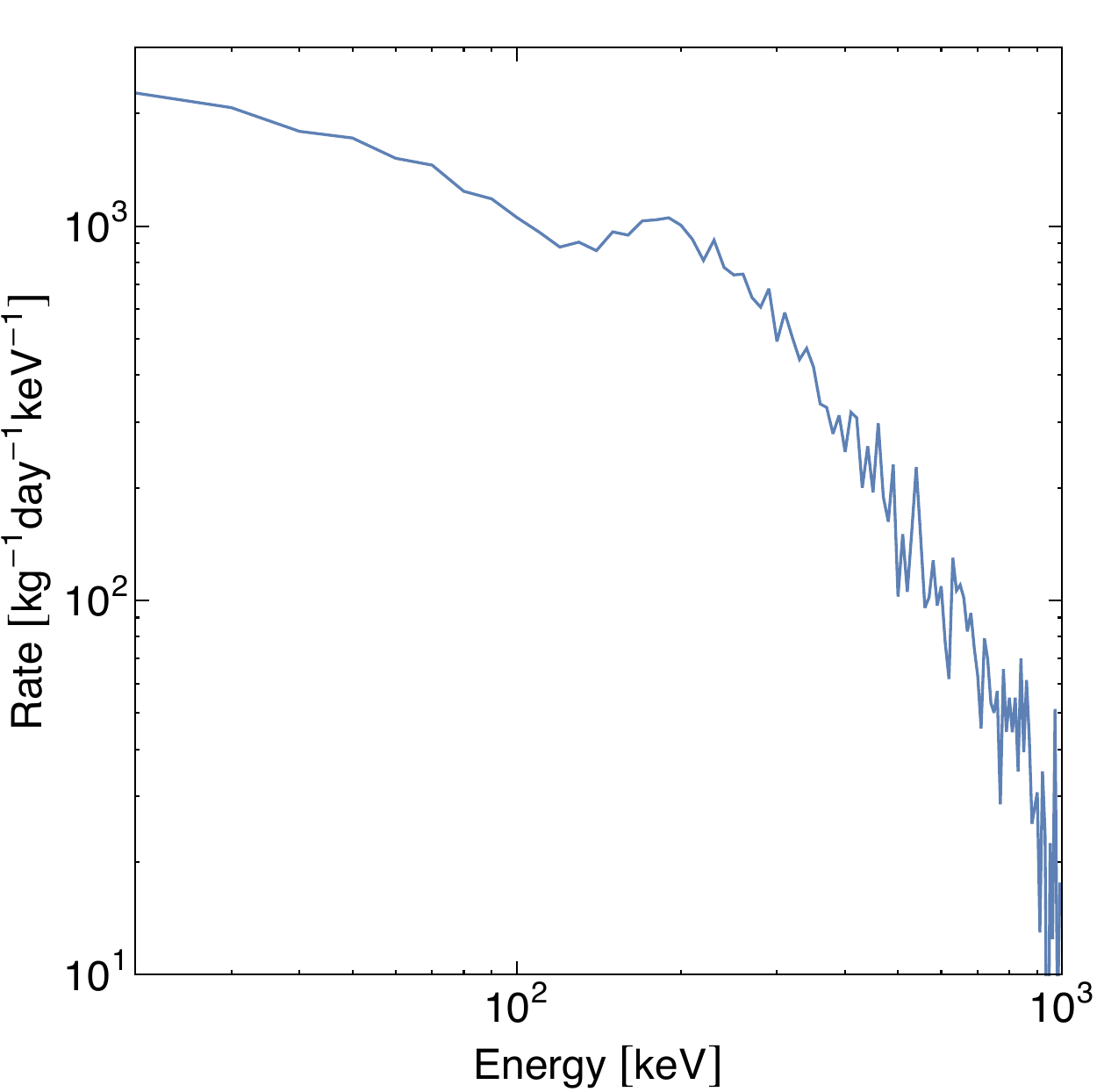}
	\caption{Rate of high-energy electron events in SENSEI with  energies above $20$\,keV~\cite{SENSEI:2020dpa}.} \label{fig:highenergyelectronrate}
\end{figure}

The actual rate and the measured rate of high-energy electrons ejected by X-rays match at energies at and above $\sim 20$~keV. Above this energy range, we thus consider the rate of high-energy electrons ejected by X-rays or gamma-rays to be equal to that observed by SENSEI, \textit{i.e.}, above these energies our  track reconstruction procedure is not used and instead we directly utilize the measured data. Since cosmic muons also create electrons as they cross the detector, we subtract the muon events from the measured distribution using the muon-energy loss in the detector discussed in the previous section. The resulting residual distribution, consisting purely of high-energy electron events ejected by photons, is shown in Fig.~\ref{fig:highenergyelectronrate}. \\ 

With the true high-energy electron spectrum at hand, we perform a simulation of electron tracks in the SENSEI CCD. To simulate the electron tracks, we again simulate the corresponding X-ray entering the CCD through the backside ejecting the electron through photoabsorption. Energetic photons entering the detector through the frontside of the detector either get absorbed in the pitch adapter, or are uniformly distributed in the CCD, as discussed below. The point where the photon is absorbed sets the starting point of the electron track. We first consider electrons with energies below $\sim 20$~keV, with energies picked from the distribution shown with the blue or orange lines (for Model~I or Model~II, respectively) in the right panel of Fig.~\ref{fig:lowenergyelectronrate}. These electrons are more concentrated near the backside of the CCD as the penetration length of soft X-rays is much smaller than the width of the CCD. In particular, note that these electrons cannot stem from the CCD frontside, as X-rays with energies below 20~keV get absorbed in the pitch adapter. We neglect low-energy electrons entering from the CCD's sides, as given the small thickness of the CCD they are a sub-leading contribution to the electron rate. For electron tracks ejected by photons with energies higher than 20 keV, the energy is picked from the distribution shown in Fig.~\ref{fig:highenergyelectronrate}. The penetration length of photons with such energies is larger than the width of the CCD, and hence we assume that the starting points of the corresponding electron tracks are uniformly distributed throughout the CCD. We also replicate the same process to simulate electron tracks in the pitch adapter. Similar to the simulation we perform to extract the true high-energy electron track distribution which is discussed earlier, the X-rays, gamma rays and electron tracks are propagated in discrete steps in the simulation with a resolution of $10^{-5}$ of the CCD dimensions in all directions. 

\section{Generation of secondary photons}\label{sec:secondaryphotons}

Using the high-energy muons and electron tracks simulated in the previous section, we now simulate secondary emission of low-energy photons by the Cherenkov process and radiative recombination.

\subsection{Generation of Cherenkov photons}\label{sec:Cherenkov_generation}
From Eq.~\eqref{eq:cherenkov} the mean path for a track to emit a Cherenkov photon with energy in the range $\omega^{\text{min}}\leq \omega\leq \omega^{\text{max}}$ is given by
\begin{align}
l_{\text{mfp}}=\left(\int_{\omega_{\text{min}}}^{\omega^{\text{max}}}d\omega~ \alpha \left(1-\frac{\text{Re}~\epsilon (\omega)}{v^2 |\epsilon(w)|^2}\right)\right)^{-1}\ ,
\label{eq:mfpcherenkov}
\end{align}
where $v$ is the speed of the track and $\epsilon(\omega)$ is the energy-dependent dielectric function of Si. In our simulation, we use $\omega_\text{min}=1.07~\text{eV}$ as photons with energy below this value are not absorbed in the device's active area (made of high-resistivity Si) so they do not constitute backgrounds.\footnote{Sub-eV photons can be absorbed by n or p-type dopants, which typically have ionization energies below $\approx 50$~meV \cite{shklovskii2013electronic}. In Skipper-CCDs, however, ionization of the n-type phosphorus dopants in the backside leads to the creation of a single electron without a hole, so it does not constitute a measured event. Ionization of the p-type boron dopants near the buried channel in the frontside (which would lead to a hole) is blocked, as that region of the detector is fully depleted. This means that the p-dopant impurity band is fully occupied, and ionization by electron photoabsorption from the valence band into the impurity band is thus not possible.} We use $\omega_{\text{max}}=2.2~\text{eV}$ for which the absorption length in undoped Si is a few micrometers. Photons with even higher energy are absorbed very close to the tracks and will be indistinguishable from them as the pixel-size of the CCD is $15~\micron$. 

The distance travelled by the track between emissions of two successive Cherenkov photons is sampled from an exponential distribution with mean equal to the mean free path Eq. \eqref{eq:mfpcherenkov}. If $\vec{x}_{\text{start}}$ is the position of the starting point and $\vec{x}_{\text{end}}$ is the end point of the track in the CCD or the pitch adapter, then the points $\vec{x}_i$ at which a Cherenkov photon is emitted are computed using 
\begin{align}
\vec{x}_{i+1}=\vec{x}_i + l  \vec{u}\ ,
\end{align}
where $\vec{u}$ is the unit vector along the direction of the track, $l$ is the distance sampled from an exponential distribution with mean $l_{\text{mfp}}$, and $\vec{x}_0 = \vec{x}_{\text{start}}$. In the case of muon tracks, starting from $\vec{x}_1$, all the points where a Cherenkov photon is emitted from the track are computed such that all $\vec{x}_i$ are between $\vec{x}_\text{start}$ and $\vec{x}_\text{end}$. For electron tracks, $\vec{x}_i$ are computed such that $|\vec{x}_i - \vec{x}_{\text{start}}| < l_{\text{threshold}}$, where $l_{\text{threshold}}$ is the length travelled by the track before its energy decreases below $20~\text{keV}$, which is the threshold energy for electrons to emit Cherenkov photons in Si. As the dielectric function $\epsilon(\omega)$ of Si is approximately flat between the chosen $\omega_{\text{min}}$ and $\omega_{\text{max}}$ \cite{1985hocs.book.....P}, the energy of every emitted Cherenkov photon is randomly chosen between $\omega_{\text{min}}$ and $\omega_{\text{max}}$. The direction of propagation of the photon is chosen such that the polar angle of the direction of the photon with respect to the direction of the track is the characteristic Cherenkov angle given by Eq.~\eqref{eq:angle}. The azimuthal angle around the direction of the track is chosen randomly.

\subsection{Generation of recombination photons}
The high-energy muon and electron tracks in the CCD generate secondary electron-hole pairs through scattering. Recall that a fraction of these secondary electron-hole pairs recombine radiatively to emit photons. The electron-hole pairs created in the bulk of the CCD are drifted by the electric field before they can recombine, so we focus on  computing the photons that are generated by charges that recombine in the backside where collection is inefficient. We estimated  the fraction of charges emitting recombination photons as a function of the distance from the backside of the CCD in Sec.~\ref{sec:recofraction}. Here we use the results of that section to simulate the backside's recombination photons. \\

The number of secondary charges created for every high-energy track as a function of the track penetration in the CCD is computed using the electronic stopping power of high-energy muons and electrons in Si. The number of recombination photons $N_\gamma$ generated at every position $(x,y,z)$ is then given by
\begin{align}\label{eq:noofrecophotons}
N_{\gamma} (x,y,z) = N_{\text{e-h}}(x,y,z) f_{\text{direct}}(z)\ ,
\end{align}
where $N_{\text{e-h}}(x,y,z)$ is the number of secondary charges created by high-energy tracks at the position $(x,y,z)$, and $f_{\text{direct}}$ is the direct recombination fraction computed in Sec.~\ref{sec:recofraction}. The energy spectrum of these photons is given by~\cite{5244084} 
\begin{align}
\frac{dN_{\gamma}}{d\omega} (x,y,z)  = B_0 \frac{\alpha (\omega,T,N_D(z)) n(\omega,T,N_D (z))^2 \omega^2}{\exp{(\omega /T)}-1}\ ,
\end{align} 
where $\alpha$ is the light absorption coefficient, $n$ is the refractive index, $\omega$ is the photon frequency, $N_D$ is the doping density, and $T$ is the temperature, which is set to 135~K for SENSEI. The normalization coefficient $B_0$ is defined such that 
\begin{align}
\int_{0}^{\infty} d\omega \frac{dN_{\gamma}}{d\omega} (x,y,z) = N_{\gamma}(x,y,z)\ ,
\end{align}
where $N_{\gamma}(x,y,z)$ is as calculated in Eq.~\eqref{eq:noofrecophotons}. Recombination photons are assumed to be emitted isotropically from the position where charges recombine.
\section{Propagation of secondary photons}\label{sec:photon}
The position of the emission of the photons, their direction, and their energy, discussed in the previous section, set their initial parameters. Photons then travel in the detector before getting absorbed. In this subsection, we explain the details of the photon propagation module in our simulation. We track the photons by propagating them in $1~\mu\textrm{m}$ steps, and we take the following effects into account: 
\begin{figure}[t]
	\centering
	\includegraphics[width=0.48\textwidth]{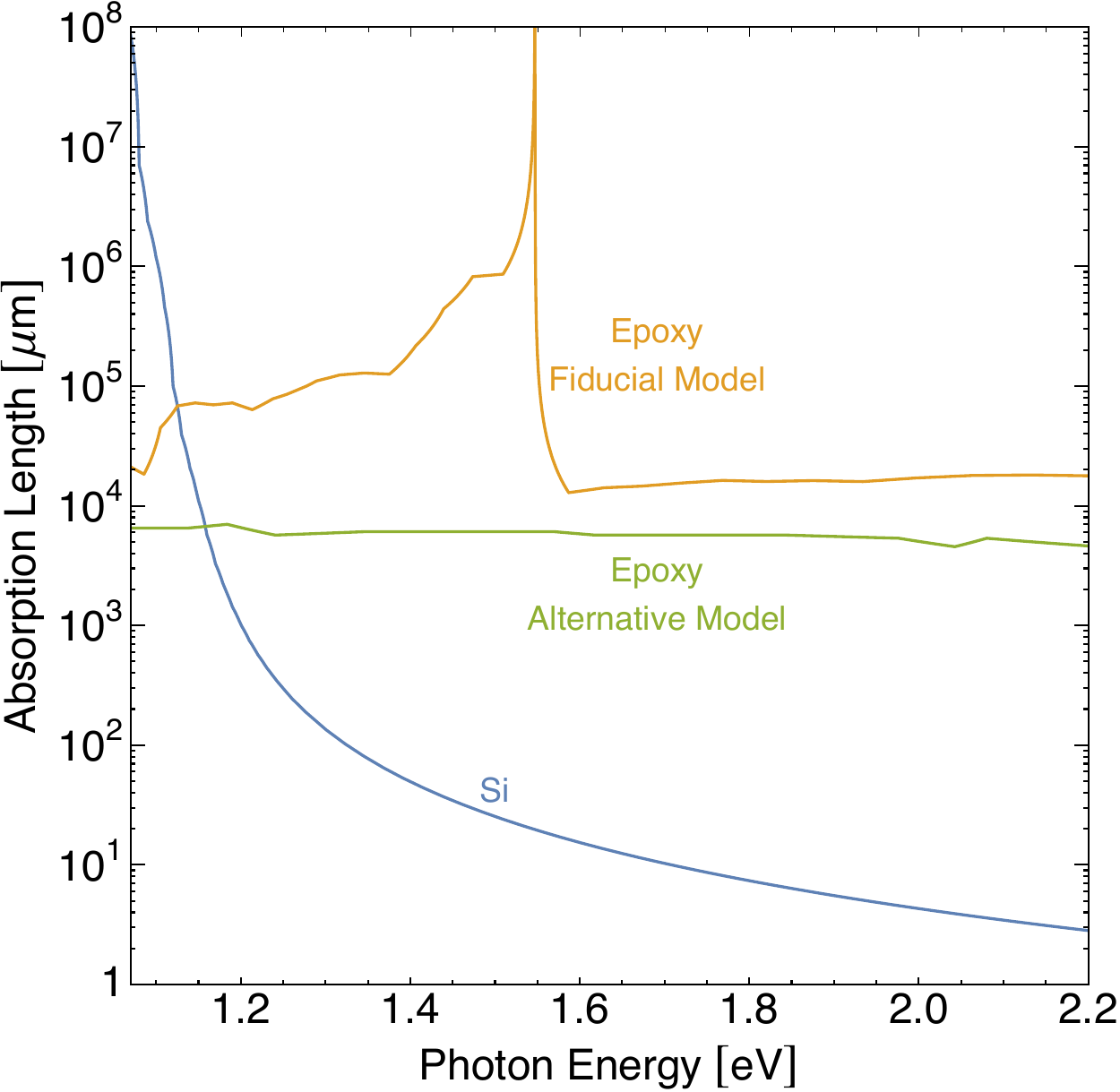}
	\caption{The absorption length of photons in Si~\cite{Stanford:2020xli} at 135~K and epoxy~\cite{epoxy,s2014study} as a function of the energy of the photon. The jump in the absorption coefficient for our fiducial model from~\cite{epoxy} is likely related to transmittance measurement uncertainties in the manufacturer's data.
 } \label{fig:absorptionlength}
\end{figure}

\begin{itemize}
\item \textbf{Absorption in the detector's bulk or pitch adapter:} the photon can get absorbed while travelling in the CCD and the pitch adapter. The pitch adapter and the bulk of the CCD are made of high-resistivity Si. In these regions of the detector, we sample the photon's absorption length from an exponential distribution with mean equal to the mean free path for a photon of that energy in undoped Si, which is taken from \cite{Stanford:2020xli} and is shown in Fig.~\ref{fig:absorptionlength}. For the lifetime of every secondary photon emitted, we keep a counter of the distance travelled by that photon. For every step the photon takes in the high-resistivity Si, we add the size of the step to the counter and check if the total distance exceeds the chosen absorption length in the beginning. If this is the case, the photon is considered to be absorbed at its current position. That position is stored and the propagation module is stopped.
\item \textbf{Absorption in epoxy:} the photon can also be absorbed in the epoxy between the pitch adapter and the CCD (see Fig.~\ref{fig:structure}). We take this into account by choosing an absorption length from an exponential distribution with mean equal to the mean free path for a photon of its energy in epoxy, which is shown in Fig.~\ref{fig:absorptionlength}. Due to uncertainties on the epoxy photo-absorption, we consider two models for the absorption length. Our fiducial model is from~\cite{epoxy}, and the alternative model is from~\cite{s2014study}. 
We then track the photon's absorption by keeping a counter of its travelled distance, as done for the undoped Si above.  
\begin{figure}[t]
	\centering
	\includegraphics[width=0.48\textwidth]{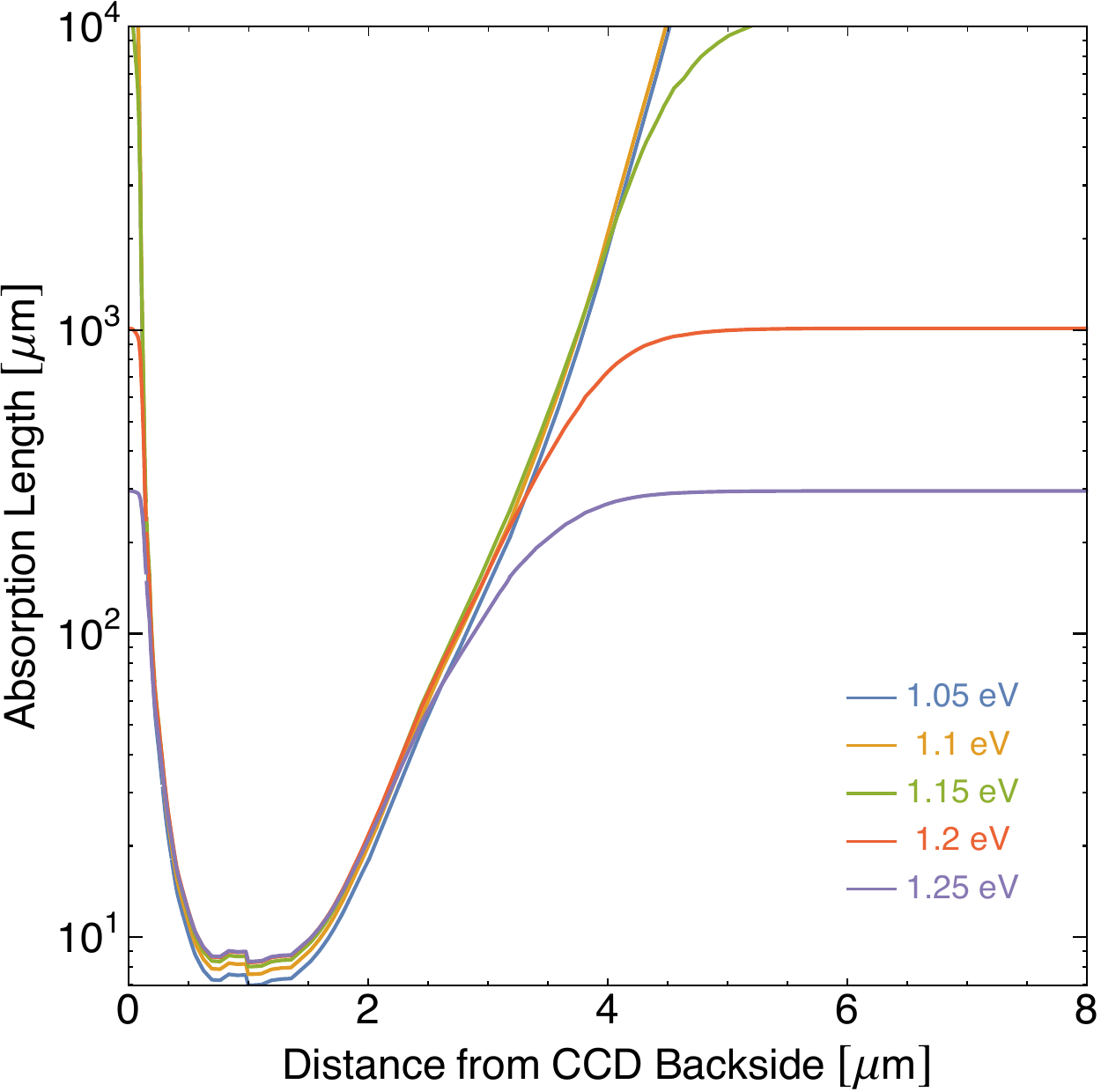}
	\includegraphics[width=0.48\textwidth]{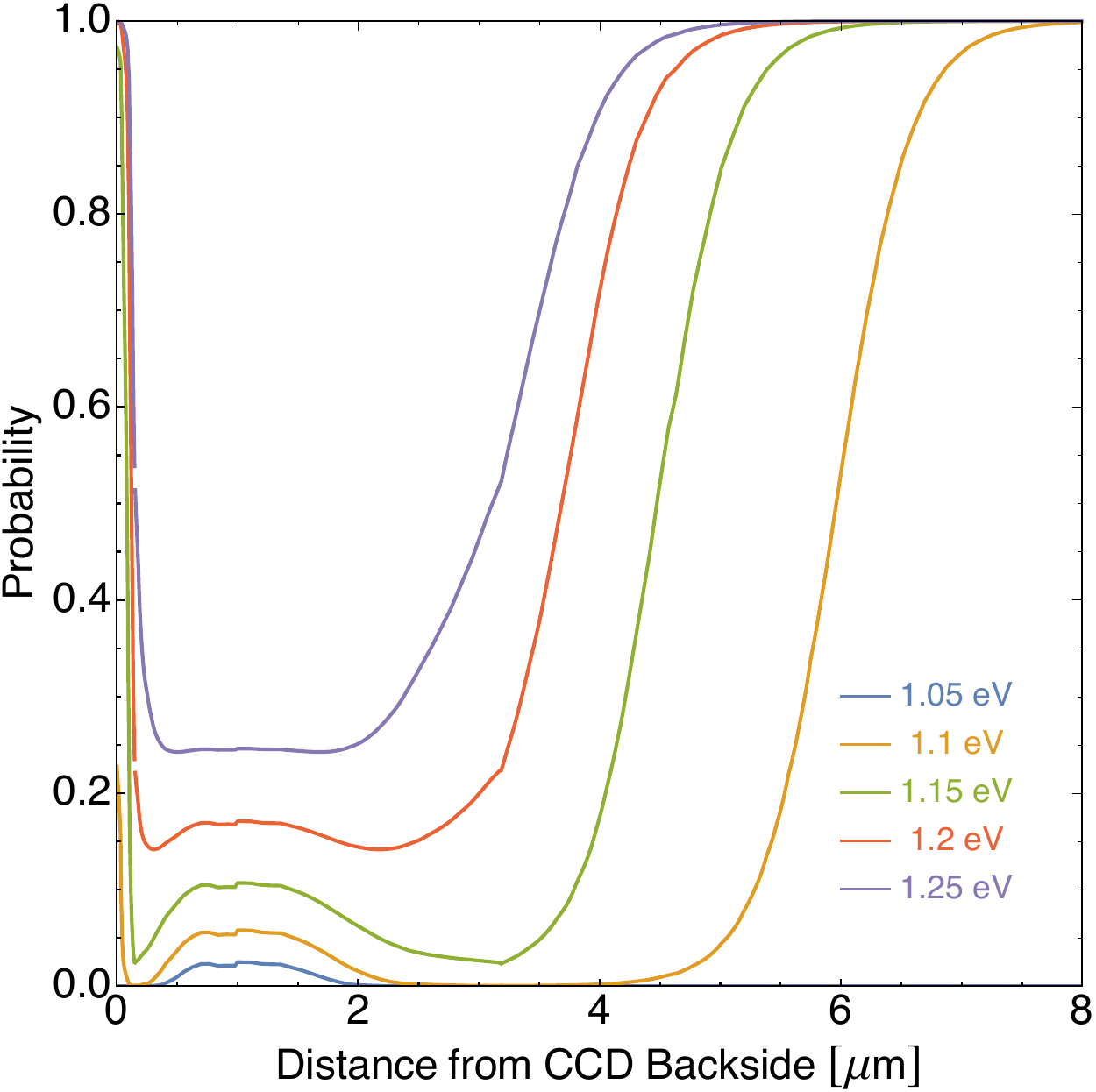}
	\caption{\textbf{Left:} Absorption length of photons in the doped backside layer as a function of distance from the backside of the CCD at 135 K \cite{Stanford:2020xli,2014JAP...116f3106B}. Different colored lines correspond to different photon energies as shown. \textbf{Right:} Probability that an absorbed photon will create an electron-hole pair in the doped backside layer as a function of the distance of the point where it is absorbed from the backside of the CCD. Different colored lines correspond to different photon energies as shown.} \label{fig:absorptionlengthdoped}
\end{figure} 
\item \textbf{Absorption in the doped Si on the backside of the CCD:} as shown in Fig.~\ref{fig:structure}, the backside of the CCD contains doped Si. The absorption length of photons depends on the doping concentration, and decreases sharply in the detector's backside due to the high levels of doping as shown in the left panel of Fig.~\ref{fig:absorptionlengthdoped}, where we present the absorption length taken from~\cite{2014JAP...116f3106B}. Dopants decrease the photon lifetime in the material, as they allow for free-carrier absorption (absorption of photons by the ionized dopant electrons in the conduction band).  Given the enhanced absorption, to accurately track the photon propagation in the backside we must increase our simulation's resolution in this region. We do so by reducing the maximum step size along the width by a factor of 10 (i.e., to $\sim 0.1~\micron$) when a photon falls within a distance of $9~\micron$ from the CCD's backside.  In every small step in this layer, an absorption length is picked from an exponential distribution with mean given by the mean absorption length as a function of the doping concentration at the current position of the photon. For a particular position in this backside layer, we use the absorption length presented in Fig.~\ref{fig:absorptionlengthdoped} to calculate the probability for a photon to get absorbed at that position. If the chosen absorption length is smaller than the step size in the layer, then the photon is considered to be absorbed at that position. At  high doping, photoabsorption results in the excitation of free carriers (followed by relaxation via phonon emission), instead of the creation of an electron-hole pair by ionization. Here we are only interested in the photons that create ionization upon being absorbed. To compute the ionization rate, we take the ratio of inter-band absorption length to the total absorption length at that position, and use this fraction to determine the ionization fraction. The probability that a photon absorbed in the doped backside layer will create ionization is shown in the right panel of Fig.~\ref{fig:absorptionlengthdoped}.

\item \textbf{Reflection/Refraction at interfaces:} a photon travels along a straight line until it reaches an interface between two materials, where it can either reflect or refract. In the case where the two materials have thickness larger than the typical wavelengths of the photons considered ($\sim \micron$), we treat the two materials as semi-infinite media, and ignore thin-film effects. In this case we simply use the Fresnel equations to determine the probability of the photon to reflect or refract based on the angle of incidence with respect to the normal of the interface, and the refractive indices of the two media. The reflection and refraction probabilities depend on the photon's polarization. For simplicity, we assume that whenever a photon encounters an interface, the polarization vector lies parallel to the plane of the interface (s-polarization); we have checked that performing our simulations with other polarizations does not significantly alter the results.  For this polarization state of the photon, the probability for the photon to get reflected, denoted by $P_r$ is given by 
\begin{align}\label{eq:Fresnel}
P_r = \Big( \frac{n_1 \cos{\theta_i} - n_2 \cos{\theta_t}}{n_1 \cos{\theta_i} +n_2 \cos{\theta_t}} \Big)^2 \ ,
\end{align}
where $n_1$ is the refractive index of the medium from which the photon is incident, $n_2$ is the refractive index of the second medium, $\theta_i$ is the angle of incidence with respect to the normal to the plane of the interface, and $\theta_t$ is the angle of refraction into the second medium. These quantities are  related by Snell's law,
\begin{align}\label{eq:Snell}
n_1 \sin{\theta_i} = n_2 \sin{\theta_t}\ .
\end{align}
Thus, for known $n_1$ and $n_2$, and a given angle of incidence $\theta_i$, we can calculate $\theta_t$ using Eq.~\eqref{eq:Snell}, and use that to calculate $P_r$ as given by Eq.~\eqref{eq:Fresnel}. The probability of the photon to refract into the second medium is simply $1-P_r$. For the special cases where there is no $\theta_t$ that satisfy Eq.~\eqref{eq:Snell}, the photon is always reflected back into the first medium (total internal reflection).\\

We use this method whenever a photon comes in contact with an interface where the two sides of the interface can be treated as semi-infinite media as compared to the photon wavelength: the CCD-vacuum interface, the epoxy-vacuum interfaces, the epoxy-pitch  adapter interfaces, and the vacuum-copper interfaces. The CCD frontside and CCD backside, i.e., interfaces along the length and the breadth of the CCD, are treated as discussed next.
 
\item \textbf{Thin-film interference:} As shown in Fig.~\ref{fig:structure}, there are thin films at the backside of the CCD, and also between the CCD and the epoxy. The thickness of these films is much smaller than the typical wavelengths of the photons in our simulation, so thin-film effects must be taken into account. When a photon crosses the thin films, it can get reflected, get absorbed, or can pass through the films. The probabilities for these three processes to happen as a function of the angle of incidence of the photon, its energy, and its polarization are called reflectance, absorptance, and transmittance, respectively. We have three sets of thin films in the SENSEI apparatus that we consider in our simulation. The first is the set of thin films at the backside of the CCD. These are placed between the Si in the CCD and the vacuum on the other side. The second set of films are at the frontside of the CCD and are sandwiched between the CCD and the $1.7~\micron$ thick $\text{SiO}_2$ layer. The third set is a $0.5~\micron$ thick $\text{SiO}_2$ coating between the vacuum and the pitch adapter. In all cases we input the geometry and composition of our thin-film layers into the complex-matrix Fresnel equation calculator~\cite{filmetrics} to get the reflectance, absorptance, and transmittance as a function of photon energy and the angle of incidence for the three sets of films, and for photons coming from either direction of the films.

\item \textbf{Roughness of the interfaces:} Interfaces and edges in the detector are not perfectly flat, and have some roughness. There is no exact data available regarding the surface roughness, so to estimate the impact of this effect we follow a simplified prescription. For $50\%$ of the photons impinging on a rough surface, we ignore the roughness and assume that the photons encounter a perfectly flat surface according to the geometry of the detector. For the other $50\%$ of the photons, we assume that the photons encounter an edge that is not oriented according to the geometry of the detector. For every such case, we choose a random orientation of the edge. This modifies the normal to the surface that the photon encounters and thus also changes the incident angle. The Fresnel equations or the thin-film calculations mentioned above are then applied with the new incident angle. We have checked that considering a perfectly flat device does not significantly alter our results.
\end{itemize}
We do not keep track of photons with energies less than 1.07 eV, as they have an absorption length  in the undoped bulk of the CCD that significantly exceeds the detector's size, and can only be absorbed in the part of the backside where there is no charge collection to create events. Note that our longest lived photons can get reflected several times by the copper housing around the SENSEI apparatus before being absorbed. The photons that get absorbed in the bulk of the CCD are assumed to have created an event measured by the SENSEI detector. If the photon gets absorbed in the doped backside of the CCD, we use the charge collection model estimated in Sec.~\ref{sec:chargecollection} to compute the fraction of these that generate measured events.

\section{Analysis}\label{sec:analysis}
Once the Cherenkov and recombination photoabsorption positions are known in the simulation, we project these on the $\text{length}\times\text{breadth}$ plane of the CCD, and bin them in $15~\micron \times 15~\micron$ pixels to create a 2-D image as observed by SENSEI. Note that the Cherenkov and recombination photons considered in the simulation can only lead to single electron-hole pairs in Si given their energies (2-electron events require photons with energies above at least $\simeq 3$~eV~\cite{Rodrigues:2020xpt,Ramanathan:2020fwm}). However, if several photons are coincidentally absorbed in the same pixel (or in contiguous pixels for 3~electrons or more), that is recorded as a multi-electron event (in~\cite{SENSEI:2020dpa}, only single pixels are considered for the 2-electron event analysis). Multi-electron events can also occur from coincidences of photoabsorption and spurious-charge events, which occur as the charge is moved towards readout. Events containing several electrons require several absorption coincidences, and hence are increasingly unlikely. In what follows, we perform the analysis of 1-electron to 4-electron bins, following the analysis procedure performed by the SENSEI collaboration in~\cite{SENSEI:2020dpa}. 

We also project the positions of the high-energy tracks on the 2-D pixelated image, and label the pixels that contain a high-energy event. As the charges are drifted to the surface, they can diffuse away from their original position on the 2-D plane of the image. As in the SENSEI analysis, we do not consider this effect for single-electron events, or for events made up of coincidences of single-ionization events. For high-energy tracks, we consider all the pixels in a 2-pixel radius around the original track on the 2-D image to be a part of the diffused track. 

In the following subsections, we describe the analysis of 1-electron to 4-electron events.
\subsection{1-electron events analysis}\label{subsec:1e}
For analyzing 1-electron events, we apply the following masks:
\begin{itemize}
\item \textbf{Edge Mask:} we divide the image into 4 quadrants and mask 60 pixels around the boundaries of the quadrants.
\item \textbf{Bleeding Zone Mask:} we mask 100 pixels in the horizontal and vertical direction starting from every pixel that is a part of a high-energy track and going away from the nearest corner of the image to that pixel. 
\item \textbf{Halo Mask:} we mask the pixels that lie within circles that surround any high-energy tracks. The circle's radius is referred to as the ``halo-mask radius.'' We  analyze the remaining 1-electron events for halo mask radii starting from 0 pixels to 60 pixels in increments of 10 pixels. 
\end{itemize}
\subsection{2-electron events analysis} 
2-electron events are obtained by identifying pixels with two coincident 1-electron events in our simulation. To include the possibility of coincidences with spurious charge,
we add a 1-electron component uniformly distributed on the CCD on each image obtained from the simulation, at a level of $1.6\times10^{-4}~\text{e}^{-}/\text{pixel}$, which is consistent with the measurements in~\cite{SENSEI:2020dpa}. Pixels with two coincident 1-electron events are considered to contain a 2-electron event. On these events, we then apply the following masks:
\begin{itemize}
\item \textbf{Edge Mask:} we divide the image into 4 quadrants and mask 60 pixels around the boundaries of the quadrants.
\item \textbf{Bleeding Zone Mask:} we mask 50 pixels in the horizontal and vertical direction starting from every pixel that is a part of a high-energy track and going away from the nearest corner of the image to that pixel.
\item \textbf{Halo Mask:} same mask as for the 1-electron events described in Sec.~\ref{subsec:1e}. 
\item \textbf{Loose Cluster Mask:} for all pixels containing a 1-electron event that is not still masked by the bleeding-zone mask, if any two such pixels are within a 20-pixel radius of one another, all pixels within a 20-pixel radius of those two pixels are masked. 
\end{itemize}
\subsection{3-electron and 4-electron events analysis} 
To identify 3-electron and 4-electron events, we again add a uniform spurious charge of $1.6\times10^{-4}~\text{e}^{-}/\text{pixel}$ on each image obtained from the simulation. Then contiguous pixels with 3- or 4-electron events are identified as 3-electron and 4-electron events, respectively. On these events, we apply the same masks as for the 2-electron events as mentioned above.
We present the results of the 1-electron to 4-electron analysis in the following section.

\section{Results}\label{sec:results}
In this section, we present the results of the analysis of the 1-electron and 2-electron events as seen in the simulation of radiative processes in SENSEI. As discussed in Sec.~\ref{sec:backsidesim}, we have two models for the transport parameters on the doped backside of the CCD. These two models differ in the charge collection efficiency and radiative recombination fraction, and also impact the high-energy-electron event rate as discussed in Sec.~\ref{sec:electroneventrate}. Thus, we run the simulation separately for both models and present the results for both. We simulate 200 images to reduce the statistical errors in the simulation. For each image, we simulate the high-energy tracks crossing the SENSEI detector for an exposure time of 22.5 hours, track the Cherenkov and radiative recombination photons, and analyze the ionization events measured in the CCD. We use Mathematica to simulate the tracks and the propagation of secondary photons, perform the analysis on resulting images using Python, and run the simulations on the Symmetry Cluster at Perimeter Institute. 

\begin{figure}[t]
	\centering
	\includegraphics[width=0.48\textwidth]{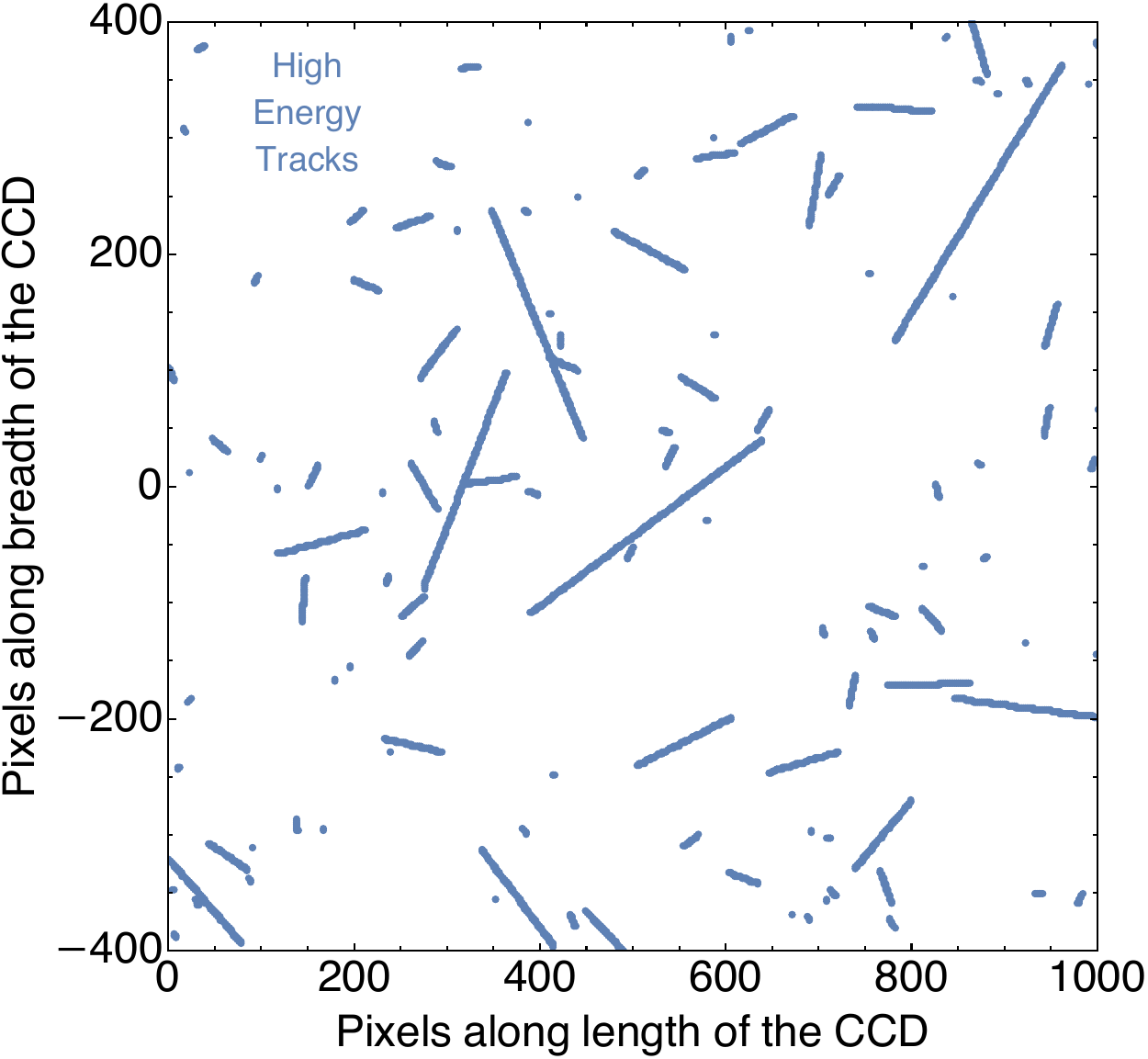}
	\includegraphics[width=0.48\textwidth]{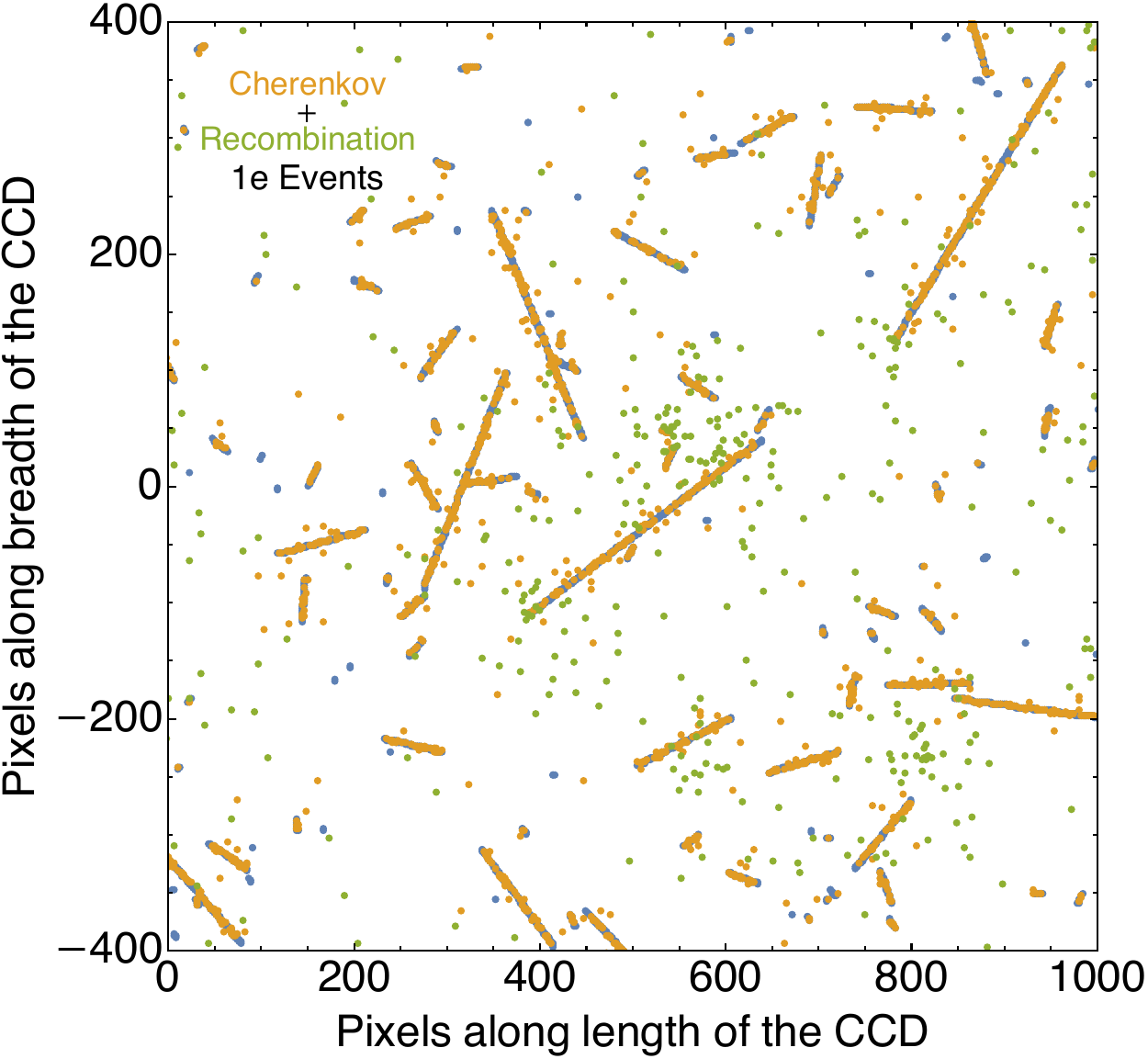}
	\caption{Example of a simulated image corresponding to the exposure of a SENSEI image. We use the extreme Model~II for this image; for Model~I, the recombination events will be negligible, while the Cherenkov events are very similar.   \textbf{Left:} Pixels containing high-energy tracks (in blue). \textbf{Right:} Pixels containing high-energy tracks (in blue) overlaid with pixels containing 1-electron events created by Cherenkov (orange) and radiative recombination (green) photons.} \label{fig:events}
\end{figure}

In Fig.~\ref{fig:events}, we show an example of a simulated image assuming the extreme transport parameter model (Model II). In the left panel, we show the pixels containing the high-energy tracks, and in the right panel, we overlay the tracks with the pixels containing 1-electron events generated by Cherenkov and radiative recombination photons.
\subsection{1-electron events in SENSEI}  
\begin{figure}[t]
	\centering
	\includegraphics[width=0.48\textwidth]{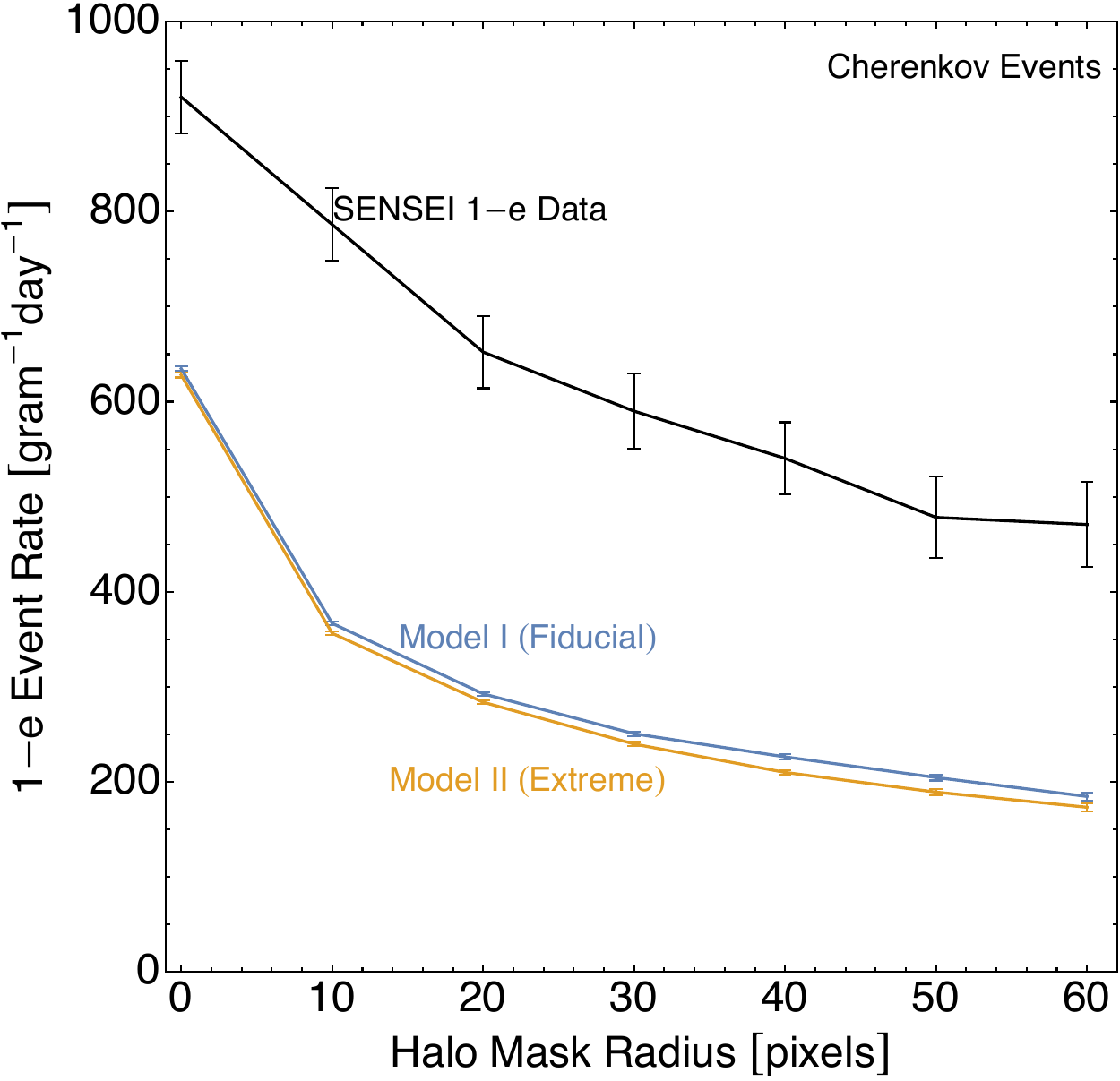}
~~\includegraphics[width=0.48\textwidth]{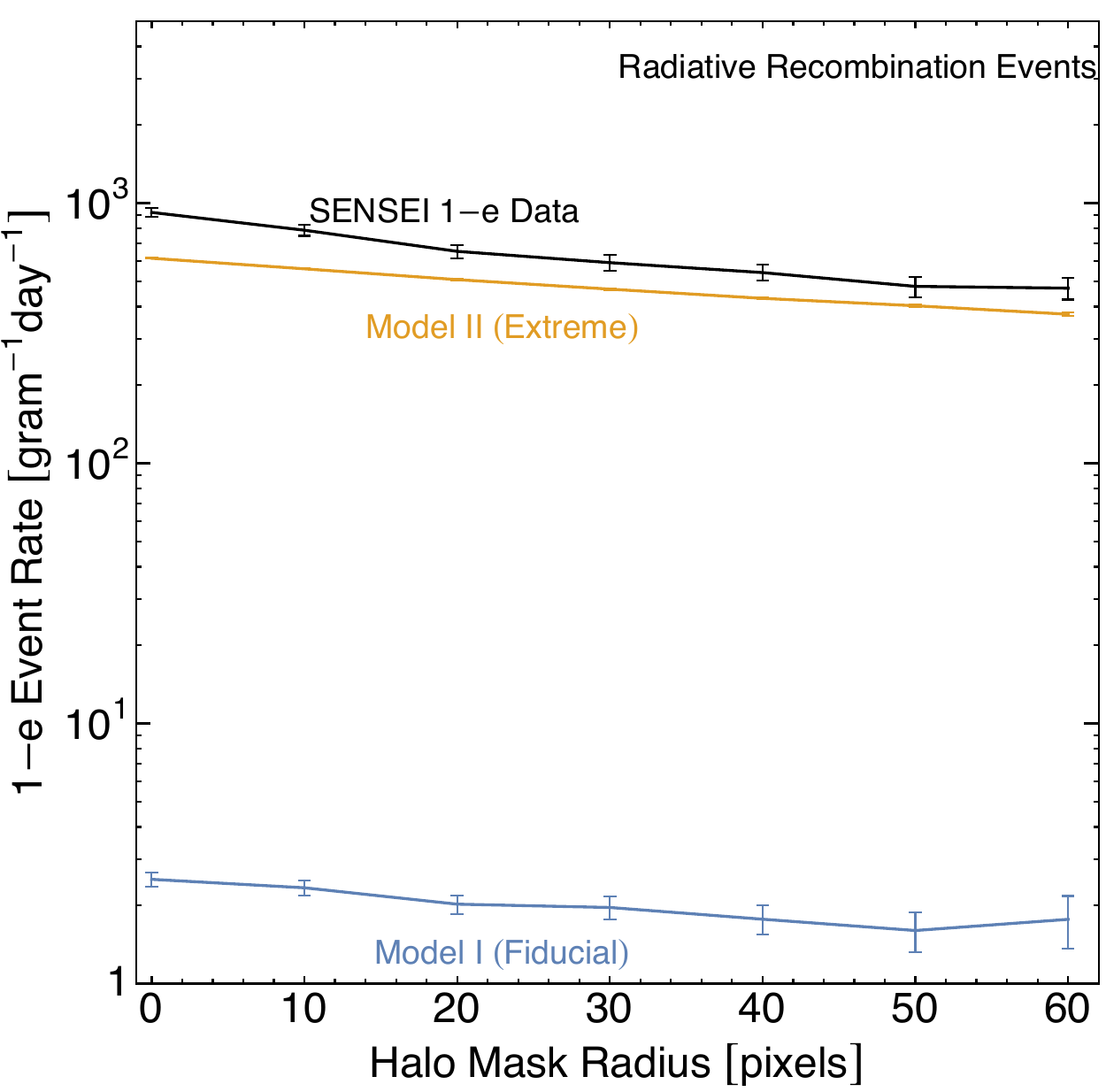}\\
\vskip 5mm
\includegraphics[width=0.48\textwidth]{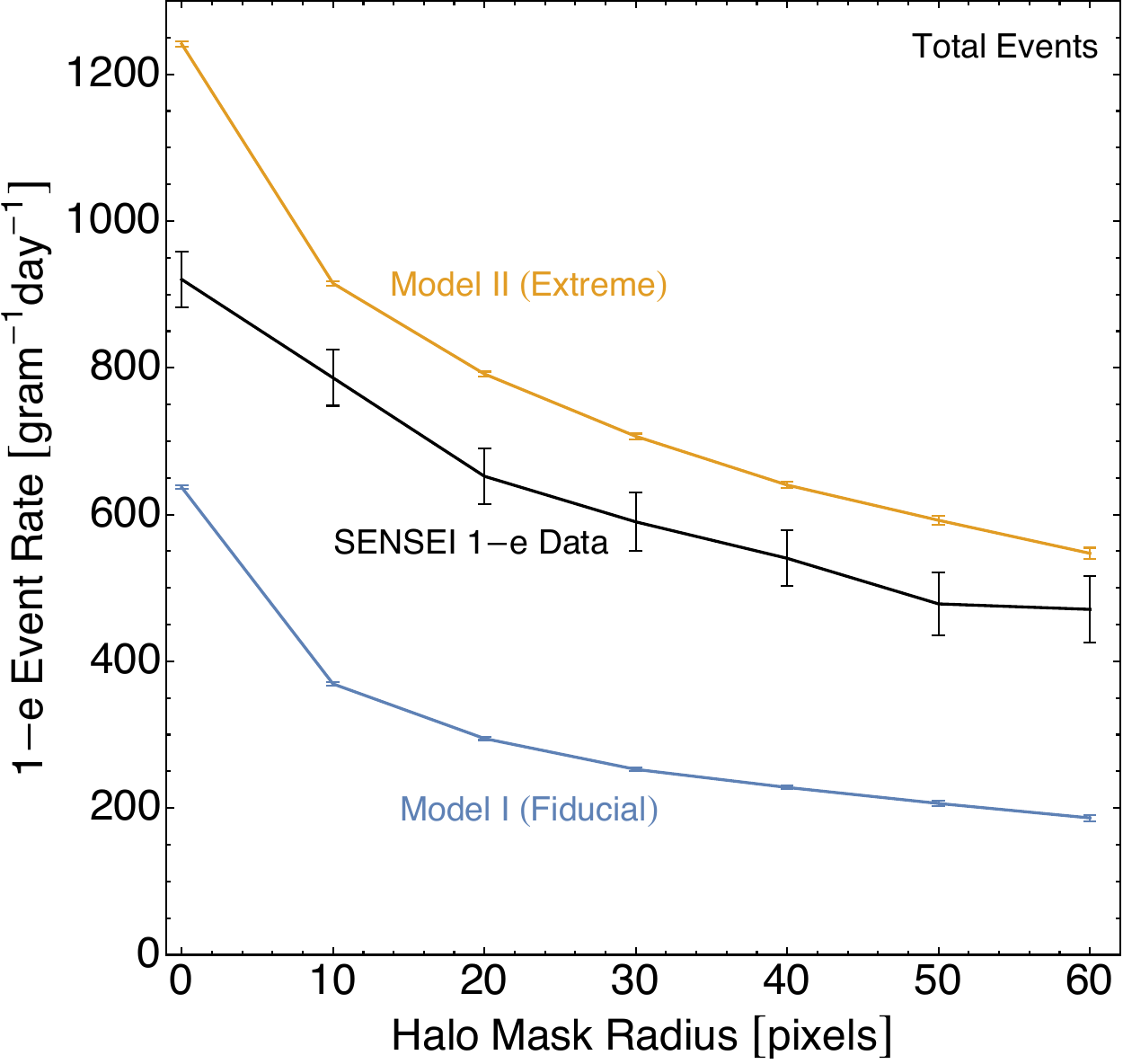}
	\caption{\textbf{Top left:} Results of the analysis of 1-electron events produced by Cherenkov photons in SENSEI.  \textbf{Top right:} Results of the analysis of 1-electron events produced by radiative recombination photons in SENSEI. \textbf{Bottom:} Results of the analysis of total 1-electron events produced by radiative processes in SENSEI. We compare with the SENSEI data in~\cite{SENSEI:2020dpa}.} \label{fig:results1e}
\end{figure}

Fig.~\ref{fig:results1e} shows the results of the analysis of 1-electron events averaged across 200 images. The top-left, top-right, and lower panels show the results for 1-electron events produced by Cherenkov photons, radiative recombination photons, and the sum of these two event-rate components, respectively. The event rate shown is after applying all the masks discussed in Sec.~\ref{sec:analysis}, and is shown as a function of varying halo mask radius.\\
   
In our fiducial transport parameter model (Model I), Cherenkov (radiative recombination) photons generate $\sim 1451~(3)$ 1-electron events per gram-day, at least 3 pixels away from the center of tracks, before applying any masks for the energy range of interest to us,  $1.07\leq \omega \leq 2.2$\,eV (see Sec.~\ref{eq:cherenkov}). Table~\ref{tab:1ecutflow} shows the fraction of exposure (geometric efficiency) and the fraction of events surviving after each successive cut. We also show the geometric efficiency of the cuts observed in the SENSEI data. After all cuts, including a 60-pixel halo mask cut, Cherenkov (radiative recombination) photons contribute $185 \pm 4$ ($2.0 \pm 0.17$) events per gram-day, where the quoted errors correspond to the simulation's statistical errors. Comparing the fraction of events surviving after the bleeding zone mask and the halo mask for the two processes, we see that the 1-electron events created by Cherenkov photons are much more correlated with the high-energy tracks compared to the 1-electron events created by radiative recombination photons. This is expected from the energy of these photons, as Cherenkov photons are almost uniformly distributed in energy whereas the radiative recombination photons are near-bandgap. Fig.~\ref{fig:absorptionlength} shows that near-bandgap photons have a larger absorption length, and thus travel farther away from tracks before getting absorbed. This makes 1-electron events created by radiative recombination photons less correlated with tracks than those created by Cherenkov photons. \\

\begin{table}[t!]

\begin{tabular}{ |p{2cm}|p{5cm}|p{2.5cm}|p{2.5cm}| }\hline
 Mask& Geometric efficiency in the simulation (SENSEI data) & Cherenkov 1-e  surviving & Recombination 1-e surviving\\
\hline
Edge Mask&  0.7 (0.7) &   69\% &   71\% \\
\hline
Bleeding Zone&  0.6 (0.6)&  27\% &  54\% \\
\hline
60-pixel Halo Mask&   0.14 (0.18)& 4\% & 10\% \\
\hline
\end{tabular}
\caption{Fraction of exposure (geometric efficiency) and the fraction of 1-electron events that survive after each successive cut. \label{tab:1ecutflow}}
\end{table}

In our extreme transport parameter model (Model II), we predict approximately the same number of Cherenkov photons as in Model I as seen in Fig.~\ref{fig:results1e}, but a much larger 1-electron radiative recombination event rate of $374 \pm 6$ events per gram-day after all cuts, for the 60-pixel halo mask. The large difference in the recombination rate between Models~I and~II for the CCD backside charge-transport parameters indicates that our simulation has a large systematic uncertainty for the 1-electron events due to these parameters. This uncertainty stems from the difference in the fraction of charges undergoing radiative recombination in the two models, as shown in Sec.~\ref{sec:recofraction}.  

For Model I, we predict a total number of $187 \pm 4$ 1-electron events per gram-day from radiative processes for the standard 60-pixel halo mask, most of them coming from the Cherenkov process. For Model II, we predict a total number of $547 \pm 8$ 1-electron events per gram-day from radiative processes for the same 60-pixel halo mask, with the majority arising from radiative recombination.  This can be compared with the observed rate at SENSEI, namely $\sim 450 \pm 45$ 1-electron events per gram-day.  We see that with Model II we can explain the entire 1-electron event rate at SENSEI with radiative processes; however, as discussed in Sec.~\ref{sec:chargecollection}, measurements of the partial charge collection efficiency in the detector's backside indicate a preference in data for Model I, which is thus taken as our fiducial model for our results.  With Model I we can only attribute a fraction of the observed 1-electron events to radiative processes. We discuss possible explanations for the remaining events in Sec.~\ref{sec:projections}.\\

\subsection{2-electron events in SENSEI}
\begin{figure}[t]
	\centering
	\includegraphics[width=0.48\textwidth]{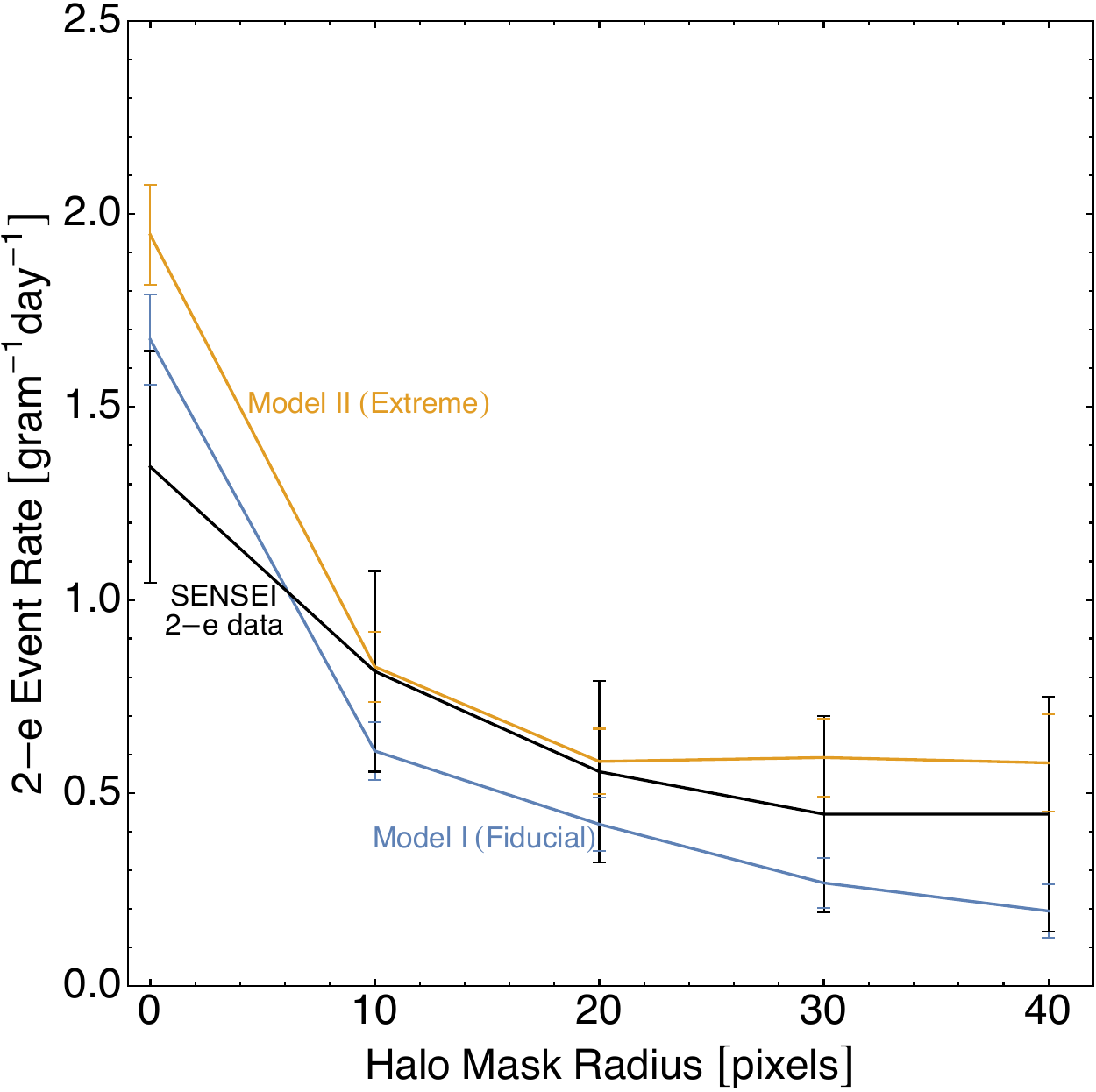}
	\caption{Results of the analysis of 2-electron events produced by radiative processes in SENSEI compared with the SENSEI data from~\cite{SENSEI:2020dpa}.\label{fig:results2e}}
\end{figure}

Fig.~\ref{fig:results2e} shows the results of the analysis of 2-electron events averaged across 200 images. The event rate shown is after applying all the masks discussed in Sec.~\ref{sec:analysis}, and is shown as a function of varying halo mask radius. We find that the large increase in the number of 1-electron events created by radiative recombination photons in Model II does not lead to a proportional increase in the 2-electron events found in Model II as compared to Model I. This indicates that the 2-electron event rate has a sizable component of self-coincidences of the 1-electron events created by Cherenkov photons. \\

In our fiducial Model I, 1-electron event coincidences generate $\sim 25$ 2-electron events per gram-day before applying any masks. Table~\ref{tab:2ecutflow} shows the fraction of exposure (geometric efficiency) and the fraction of events surviving after each successive cut. After all cuts, including a 20-pixel halo mask cut, the above rate is reduced to $0.42 \pm 0.07$ 2-electron events per gram-day. Comparing the geometric efficiency and the events surviving for each mask, we note that the loose cluster cut is extremely efficient in removing 2-electron events from Cherenkov photons. This indicates that the 2-electron events from radiative processes are correlated with the 1-electron event clusters.\\ 

SENSEI observed $\sim 0.55 \pm 0.23$ 2-electron events per gram-day after applying all masks, including a 20-pixel halo mask. Our estimate of $0.42 \pm 0.07$ 2-electron events per gram-day from radiative processes thus explains the rate observed by SENSEI within statistical error bars. If the 1-electron event rate from Cherenkov photons outside a 20-pixel halo radius was uniformly distributed, then a probabilistic estimate indicates that the expected 2-electron event rate from the coincidences of the Cherenkov 1-electron events and the uniformly distributed spurious charge should be $\sim 0.08~\text{gram}^{-1}\text{day}^{-1}$. The fact that the 2-electron event rate observed in the simulation is a factor of six more than this estimate suggests that the correlation of Cherenkov 1-electron events outside a halo-radius of 20 pixels is in large part responsible for the observed 2-electron events. \\ 

\subsection{3-electron and 4-electron events in SENSEI}
Our simulation results show a 3-electron event rate of $0.022 \pm 0.016~\text{gram}^{-1}\text{day}^{-1}$ averaged across 200 images and at a halo-radius of 20 pixels. This small event rate is consistent with the fact that SENSEI did not observe any 3-electron events in the 22 images analyzed. We also searched for 4-electron events in 200 images but did not find any in the simulations. The SENSEI data also does not contain any 4-electron events.

\begin{table}[t]
\begin{center}

\begin{tabular}{ |c|p{5cm}|c| }\hline
 Mask& Geometric efficiency in the simulation (SENSEI data) & Cherenkov 2-e  surviving \\
\hline
Edge Mask& 0.89 (0.90) & 83\%\\
\hline
Bleeding Zone& 0.78 (0.79)& 16\%\\
\hline
20-pixel Halo Mask& 0.76 (0.81)& 34\% \\
\hline
Loose Cluster& 0.94 (0.89)& 21\%\\
\hline
\end{tabular}
\caption{Fraction of exposure (geometric efficiency) and the fraction of 2-electron events that survive after each successive cut. \label{tab:2ecutflow}}
\end{center}
\end{table}

\subsection{Systematic errors in the simulation}\label{sec:systematics}

In addition to the systematics due to the CCD-backside transport parameters, our computations have a variety of other systematic errors, which we estimate by performing variations of our fiducial simulation. In particular, we consider the following additional sources of errors: 
\begin{itemize}
\item \textbf{Variations of the absorption parameters in doped Si:} The systematic uncertainty on the intra-band absorption of photons in doped Si is quantified  in~\cite{2014JAP...116f3106B}. To estimate the effect of the systematic uncertainty in the photon absorption length in the doped backside of the CCD, we perform a simulation with lower values of the intra-band absorption coefficient in~\cite{2014JAP...116f3106B}. We get the 1-electron event rate to be $207 \pm 14 $ events per gram-day, which represents a $\approx 10\%$ increase in the radiative background rate with respect to our fiducial simulation. 
\item \textbf{Variations of the recombination parameters in doped Si:} Experimental measurements of radiative and non-radiative recombination parameters in the literature have sizable variations~\cite{fossum1983carrier}. The radiative recombination coefficient that we use from~\cite{doi:10.1002/pssa.2210210140} is roughly a factor of four larger than the one in~\cite{nguyen2014temperature} at the SENSEI operating temperature. We expect that taking the smaller radiative recombination coefficient of~\cite{nguyen2014temperature} would thus reduce the recombination backgrounds roughly by a factor of four for both Models~I and~II of charge diffusion on the backside. Note also  that the radiative recombination coefficients in both~\cite{doi:10.1002/pssa.2210210140} and~\cite{nguyen2014temperature} grow exponentially towards lower temperatures, which indicates that if the detector were to be operated colder we would expect a \textit{larger} background from radiative recombination. 
\item \textbf{Variations of the absorption parameters for the epoxy:} If we use the higher absorption model in epoxy from~\cite{kufa}, we get the 1-electron event rate to be $164 \pm 8 $ events per gram-day, which represents a $\sim 10\%$ decrease in the radiative background rate with respect to our fiducial simulation.
\item \textbf{Extrapolation of the high-energy track spectrum to energies below 3 keV:} As discussed in Sec.~\ref{sec:electroneventrate}, high-energy electron tracks with energies below $\sim$ 3 keV are not measured, and the distribution in energies for these tracks needs to be determined by an extrapolation, or with a detailed Geant4 simulation, which is beyond the scope of this paper (and which would require precise knowledge of the radioactive contaminants in detector materials). While we use a flat extrapolation as discussed in Sec.~\ref{sec:electroneventrate}, we check the effect of a different extrapolation by first considering an extreme case of extrapolation, which has a bigger high-energy electron rate below 3~keV by a factor of 10. Even with this large enhancement of the rate of high-energy tracks below 3 keV, we find that the 1-electron event rate at a 60-pixel halo radius is 161 $\pm$ 11 per gram-day, which is slightly smaller than the fiducial result. This shows that the high-energy electron tracks below 3 keV do not contribute significantly to the 1-electron rate in our fiducial simulation. Since the fiducial rate is dominated by Cherenkov radiation and the threshold to generate Cherenkov is $\sim 20$~keV, this is expected. With high statistics of tracks below 3~keV, a very small fraction may be measured, which can lead to slightly larger halo masks, leading to smaller event rates. 
\item \textbf{Surface roughness:} In all our simulations, we use a model of surface roughness as explained in Sec.~\ref{sec:photon}. To check for systematics, we also run a simulation without taking any roughness into account and do not find significant differences in the final results.
\end{itemize}

\section{Simulations for thinned and low-background CCDs}\label{sec:thinned}
In addition to providing a detailed first-principles estimation of radiative backgrounds at SENSEI, our simulation setup can also be used to discuss the impact of possible future modifications to the detector setup. We consider two variations. First, we study the impact of thinning the CCD's backside in order to remove the regions of high-phosphorus doping, which is an area of the detector that plays an important role in the generation and absorption of secondary background photons, but that is not required for detector operation. From a technological perspective, thinning of the CCD's backside is a standard procedure, which is widely applied in backside-illuminated CCDs to increase their quantum efficiency~\cite{janesick1987scientific}. And second, we run a simulation to study the effects of reducing the high-energy track rate by improving the detector shielding. 

\subsection{Thinned CCD}
\begin{figure}[t]
	\centering
	\includegraphics[width=0.48\textwidth]{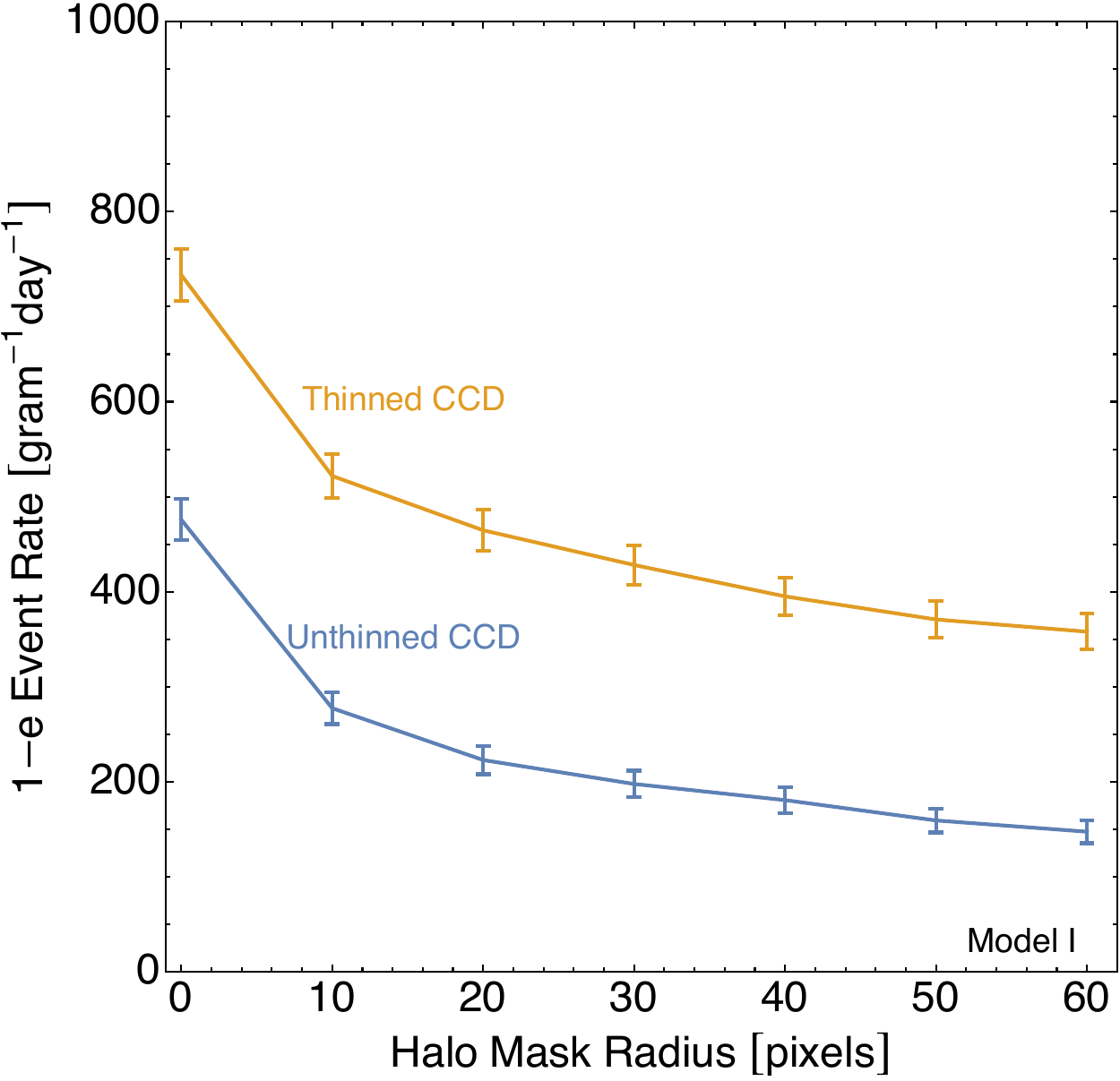}
	\caption{Results of the analysis of 1-electron events from radiative processes in a thinned CCD.} \label{fig:resultsthinned}
\end{figure}

We run a simulation with a thinned CCD for which the doped backside of the CCD, \textit{i.e.} all the region with high phosphorus doping,  is removed. We do not change any other parameters from the SENSEI simulation, and assume  the fiducial model for the transport parameters. We analyze the 1-electron events in this simulation, and show the results in Fig.~\ref{fig:resultsthinned}. 

We see that for a thinned-CCD, the 1-electron event-rate after applying all the masks \textit{increases} by $\sim 200$ events per gram-day for all halo-mask radii compared to the actual SENSEI simulation with an unthinned CCD. This clearly shows the effect of high absorptance of photons in the doped backside, which coincidentally acts as a sink for secondary Cherenkov photons that would otherwise constitute a background.  As shown in the left panel of Fig.~\ref{fig:absorptionlength}, the absorption length of near-bandgap photons is much smaller in the doped backside of the CCD, making the doped layer a trap for such photons. The right panel of Fig.~\ref{fig:absorptionlength} also shows that the photons absorbed deep in the doped backside have a small probability of being absorbed via ionization.
Furthermore, even if an ionization event is created deep in the doped layer, the left panel of Fig.~\ref{fig:fractions} shows that the corresponding charge is unlikely to be collected. The doped backside layer leads then to an overall reduction of the 1-electron events created by photons. In the thinned CCD case, the photons do not have a trap and can get absorbed anywhere in the CCD, increasing their chance to create events.  

Note that thinning of the CCD \textit{removes} backgrounds coming from radiative recombination, so naively and as pointed out in \cite{Du:2020ldo}, thinning could have resulted in an overall suppression of backgrounds. In our fiducial simulation this is not the case and instead backgrounds increase, since for our fiducial charge transport Model~I, radiative recombination backgrounds from an unthinned CCD are a subleading component of the total radiative background. A \textit{decrease}  of the background rate could occur with a thinned CCD, however, if \textit{e.g.}~Model~II of the diffusion parameters is correct, or in a CCD with different operating parameters with respect to the ones used by SENSEI at MINOS, as varying the operating conditions (such as the detector temperature) could lead to a higher radiative recombination rate in the backside. 

\begin{figure}[t]
	\centering
	\includegraphics[width=0.48\textwidth]{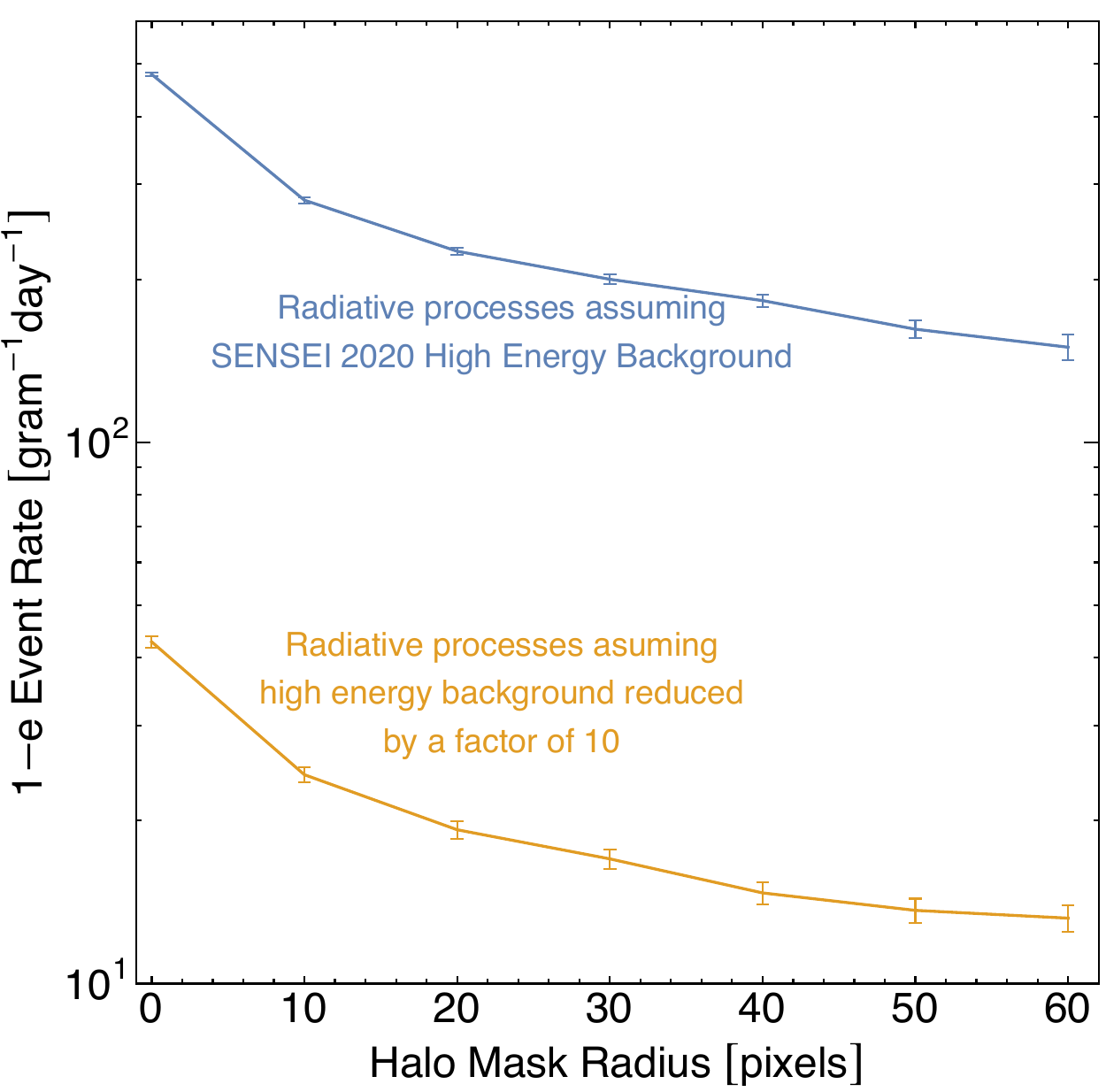}
	\caption{Results of the analysis of 1-electron events from radiative processes with the high-energy background reduced by a factor of 10 compared to the SENSEI data~\cite{SENSEI:2020dpa}.} \label{fig:resultsbackgroundreduced}
\end{figure} 

\subsection{Low-background simulation}
We run a simulation with the SENSEI CCD setup but assuming an improved shielding and a lower high-energy background rate. In this simulation, we simply assume that the detector is underground and shielded enough so that no muons pass through when it is taking data, and consider a high-energy electron rate that is smaller by a factor of 10 than the one considered for the SENSEI data-run simulation (see Fig.~\ref{fig:lowenergyelectronrate}). We do not change any other parameters and take the  fiducial Model I for the transport parameters. We analyze the 1-electron events in this simulation, and show the results in Fig.~\ref{fig:resultsbackgroundreduced}. We see that the rate of 1-electron events after masks also decreases by a factor of $\sim$10 as the high-energy background is reduced by an order of magnitude. This confirms that radiative backgrounds scale down as naively expected from the reduction of high-energy events, even if such reduction leads to  modifications in the geometric efficiencies of the detector due to the applied track-related masks.

\section{Projections for future detectors}
\label{sec:projections}

\begin{table}[b]
\centering
\begin{tabular}{ |c|c| }\hline
 &Coincident event rate \\
\hline
2-electron& $0.51~\text{gram}^{-1}\text{day}^{-1}$  \\
\hline
3-electron& $0.035~\text{gram}^{-1}\text{day}^{-1}$ \\
\hline
4-electron& $4.22 \times10^{-6}~\text{gram}^{-1}\text{day}^{-1}$ \\
\hline
\end{tabular}
\caption{The 2-, 3-, and 4-electron event rates from coincident 1-electron events, after applying all masks discussed in~\cite{SENSEI:2020dpa}, including a 20-pixel halo radius. The events arise from the radiative backgrounds as estimated in this work, combined with an additional (unspecified) uniform component to match the total 1-electron event rate measured in the SENSEI data~\cite{SENSEI:2020dpa}; we have also included a (known) uniform spurious charge component as measured in~\cite{SENSEI:2020dpa}. \label{tab:coincidencewithcherenkov}} 
\end{table}

The simulations performed in this work were done to mimic the analysis of the SENSEI run with an effective exposure of $\sim$10~gram-day~\cite{SENSEI:2020dpa}. Future planned CCD experiments will have  exposures that are larger by several orders of magnitude as well as much lower environmental background radiation. In this section, we estimate the sensitivities of large-exposure experiments assuming different background hypotheses. 
\subsection{Background hypothesis: radiative processes + uniform component}
We first consider a background hypothesis consistent with the SENSEI data~\cite{SENSEI:2020dpa}. We assume that the background events arise from a combination of two components: radiative processes, which we estimated to produce  a 1-electron event rate of $\sim 187~\text{gram}^{-1}\text{day}^{-1}$ (Model I), and an additional 1-electron component  uniformly distributed across the device with a rate of $\sim 300~\text{gram}^{-1}\text{day}^{-1}$. This rate is chosen so that the sum of the two background components roughly matches the 1-electron event rate in the SENSEI data for the 60-pixel halo mask, while the assumption of spatial homogeneity is motivated by noting that the difference between the SENSEI data and our radiative background simulations is almost constant across different halo-mask radii, as seen on the left panel of Fig.~\ref{fig:results1e}. 

The origin of this uniform extra component is unknown, but it may arise by a variety of processes. A plausible hypothesis is that these are surface events due to traps on the interface between the p-type Si that forms the buried channel and the insulating layers (such as divacancies~\cite{spratt1997effects,hopkinson1999proton}). To suppress such  dark counts these traps are filled with electrons by driving the CCD into ``inversion mode'' \cite{saks1980technique,hynecek1981virtual} before operation. Afterwards, however, the CCD is operated in non-inverted mode to take data, in which case some traps could empty out thermally or due to processes assisted by electric fields, and hence produce dark counts. Another possibility is that charge leakage occurs due to electric-field assisted tunnelling of valence-band electrons from the p-type Si into traps in the insulating Si$\mathrm{O}_2$ or Si$_3\mathrm{N}_4$ layers (rather than traps located at the interface), or into the gates, leading to holes that are collected in the buried channel. In fact, a wide variety of tunnelling processes are known to occur in MOS devices that are similar to the CCD's front-side layers (for a review we refer the reader to~\cite{ranuarez2006review,schenk1992model}), and similar processes have been hypothesized to be a source of dark counts in CCDs~\cite{janesick1987scientific}. We have checked that the electric fields across the insulating layers during non-inverted operation are  large (of the order of $\sim 10^{5}-10^{6}$~V/cm), and are likely sufficient to induce charge leakage. Note that for both aforementioned hypotheses the dark counts would occur near the CCD's frontside surface, in which case they could be mitigated by the device proposed in~\cite{Tiffenberg:2023vil}. In what follows, we remain agnostic regarding the physical process behind the extra uniform background, and as pointed out above, simply add this component with a rate chosen appropriately to match the observed 1-electron rate. 

\begin{figure}[t]
	\centering
	\includegraphics[width=0.48\textwidth]{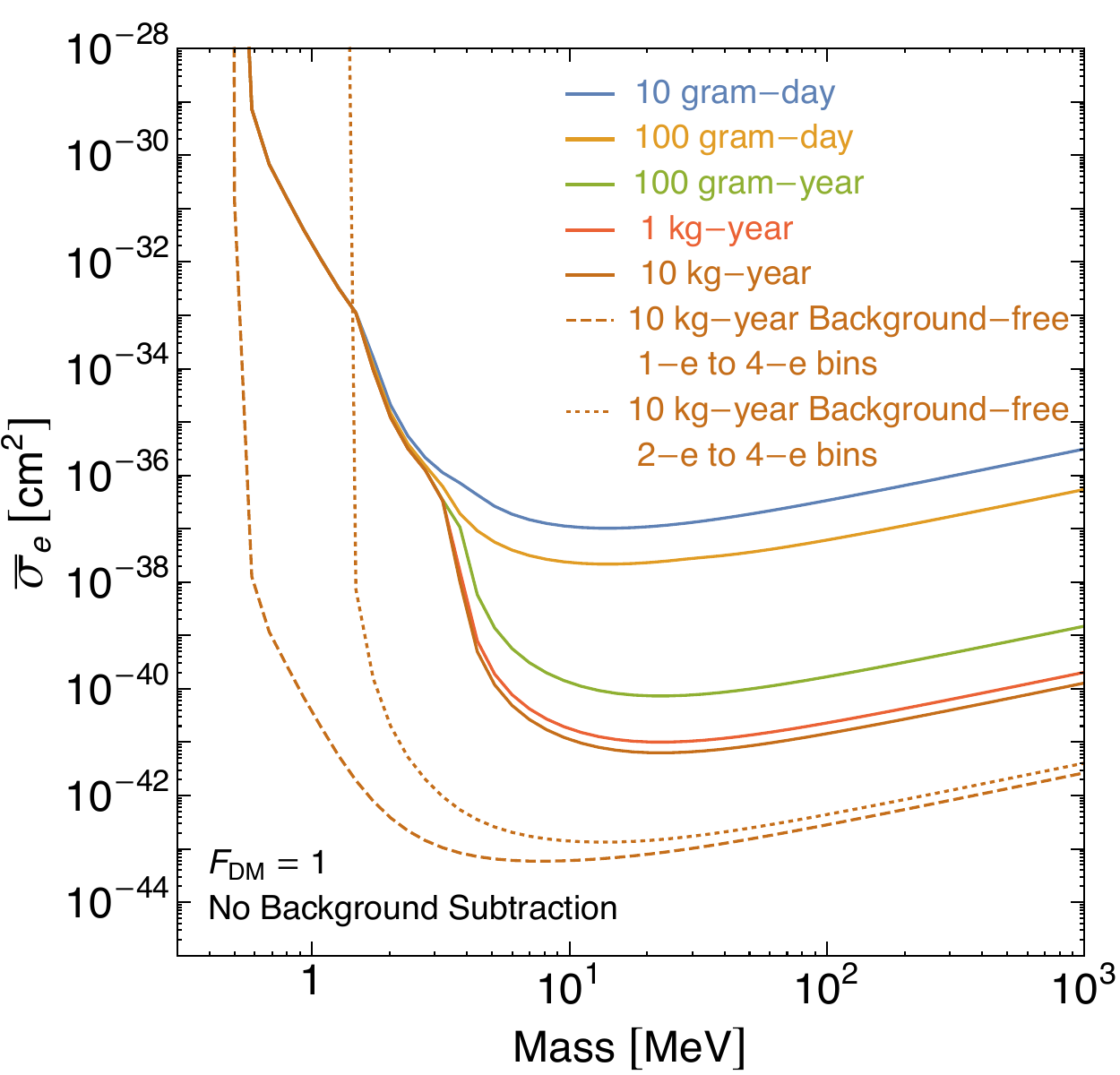}
	\includegraphics[width=0.48\textwidth]{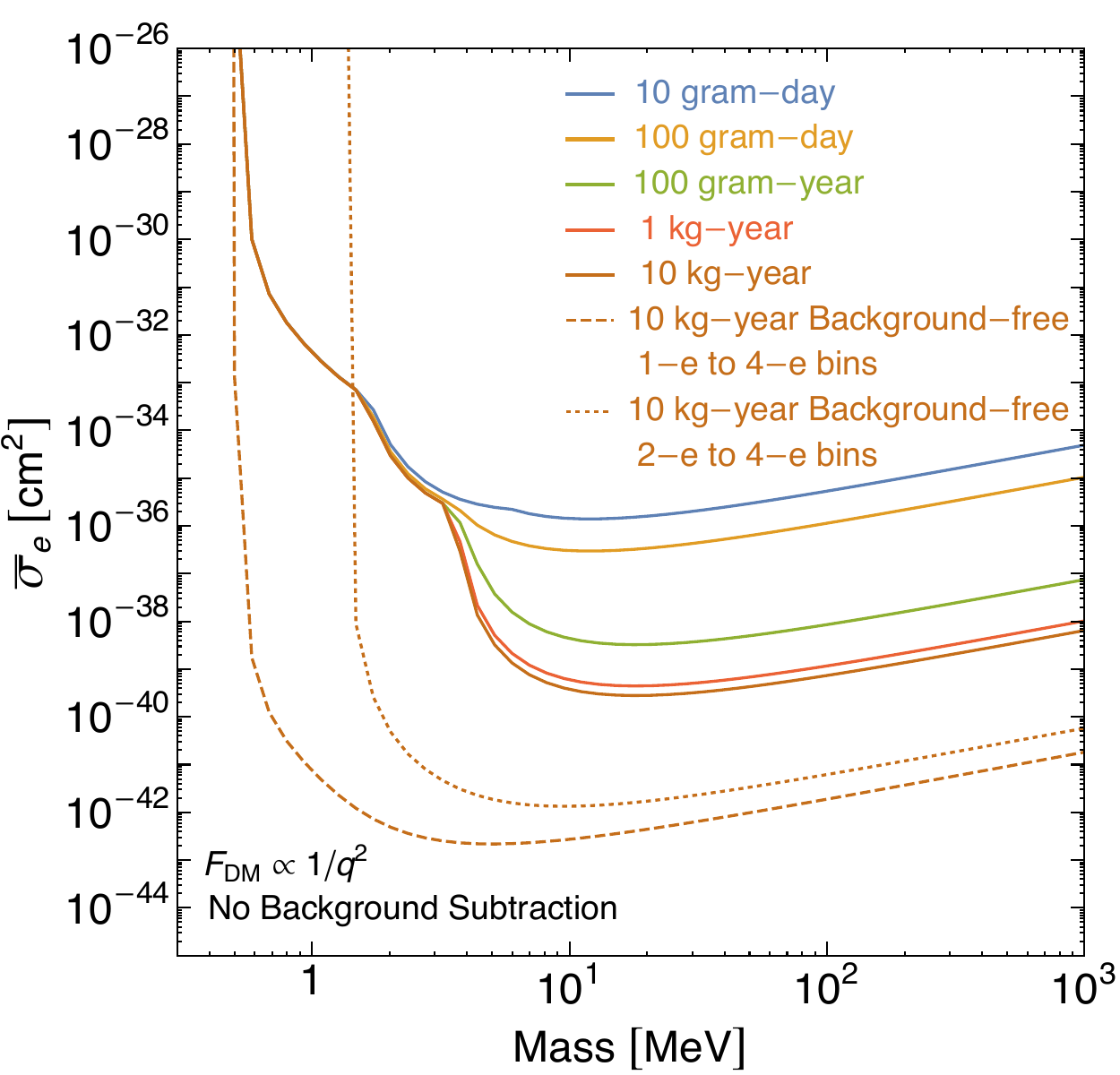}
	\caption{Ultraconservative sensitivity projections at 95\% c.l.~for DM-electron scattering for various exposures for heavy mediators (left) and light mediators (right), \textbf{assuming no reduction of backgrounds} compared to those observed in the SENSEI data~\cite{SENSEI:2020dpa} (i.e., also no improvement in shielding that would remove the radiative backgrounds). These projections also do not subtract the 2, 3, or 4-electron backgrounds that arise from coincident 1-electron events. Note that we assume the same exposure-plus-readout cycle as was done for taking the SENSEI data in~\cite{SENSEI:2020dpa}. } \label{fig:projections_withCherenkov_noSubtraction}
\end{figure} 

We assume that 2-, 3-, and 4-electron background events arise purely from 1-electron coincidences.
To estimate these coincidences, we also add a uniform 1-electron spurious charge component  of $1.6\times10^{-4}~\text{e}^{-}/\text{pixel}$ to the simulated images. We analyze the identified coincident events as described in Sec~\ref{sec:analysis}. Due to limited statistics we are unable to compute 4-electron coincidences from our simulations, so the 4-electron event rate is estimated by analytically computing the probability of coincidences of the 1-electron rate. 
Table~\ref{tab:coincidencewithcherenkov} shows the coincident-event rates at a halo radius of 20-pixels.  We do not add any true 2, 3, or 4-electron backgrounds as no sizable evidence for the existence of such events was observed in~\cite{SENSEI:2020dpa}, but we point out that 2-electron events may arise from energetic Cherenkov photons from the epoxy, $\mathrm{SiO}_2$ or $\mathrm{Si}_3\mathrm{N}_4$, or UV luminescence from these materials, as their large bandgap allows for energetic recombination photons that can lead to multiple-electron events~\cite{durmus2011optical,distefano1971band,vila2003mechanical}.  

For calculating projections for large-exposure experiments, we assume that the background rates of Table~\ref{tab:coincidencewithcherenkov} stay constant as the exposure increases, so the total number of background events grows in proportion to the exposure.

Fig.~\ref{fig:projections_withCherenkov_noSubtraction} shows the resulting projections for large exposures. The projections show the best sensitivity limit that can be obtained, and they are calculated by comparing signal ($S_{\mathrm{DM}}$) with background ($B$) in each bin. As a simplified prescription, in the low background regime ($B<1$) we set bounds when $S_{\mathrm{DM}}>3.09$ based on Poisson 95\%~C.L.~intervals ~\cite{Feldman_1998}, while in the background-limited regime ($B>1$) we use Gaussian statistics and set  $95\%$~C.L.~limits for $S_{\mathrm{DM}}>B + 1.64\sqrt{B}$.

\begin{figure}[t]
	\centering
	\includegraphics[width=0.48\textwidth]{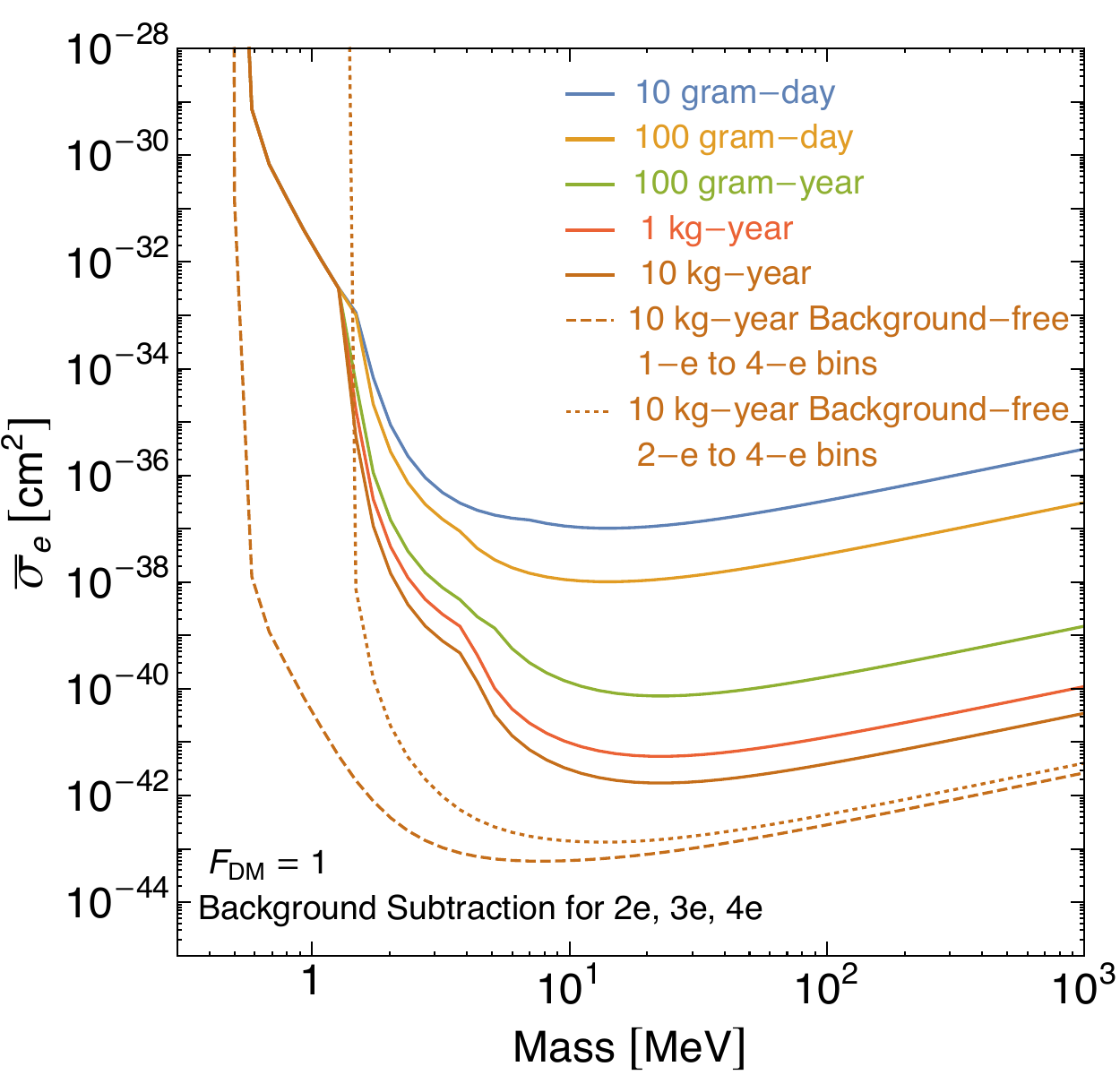}
	\includegraphics[width=0.48\textwidth]{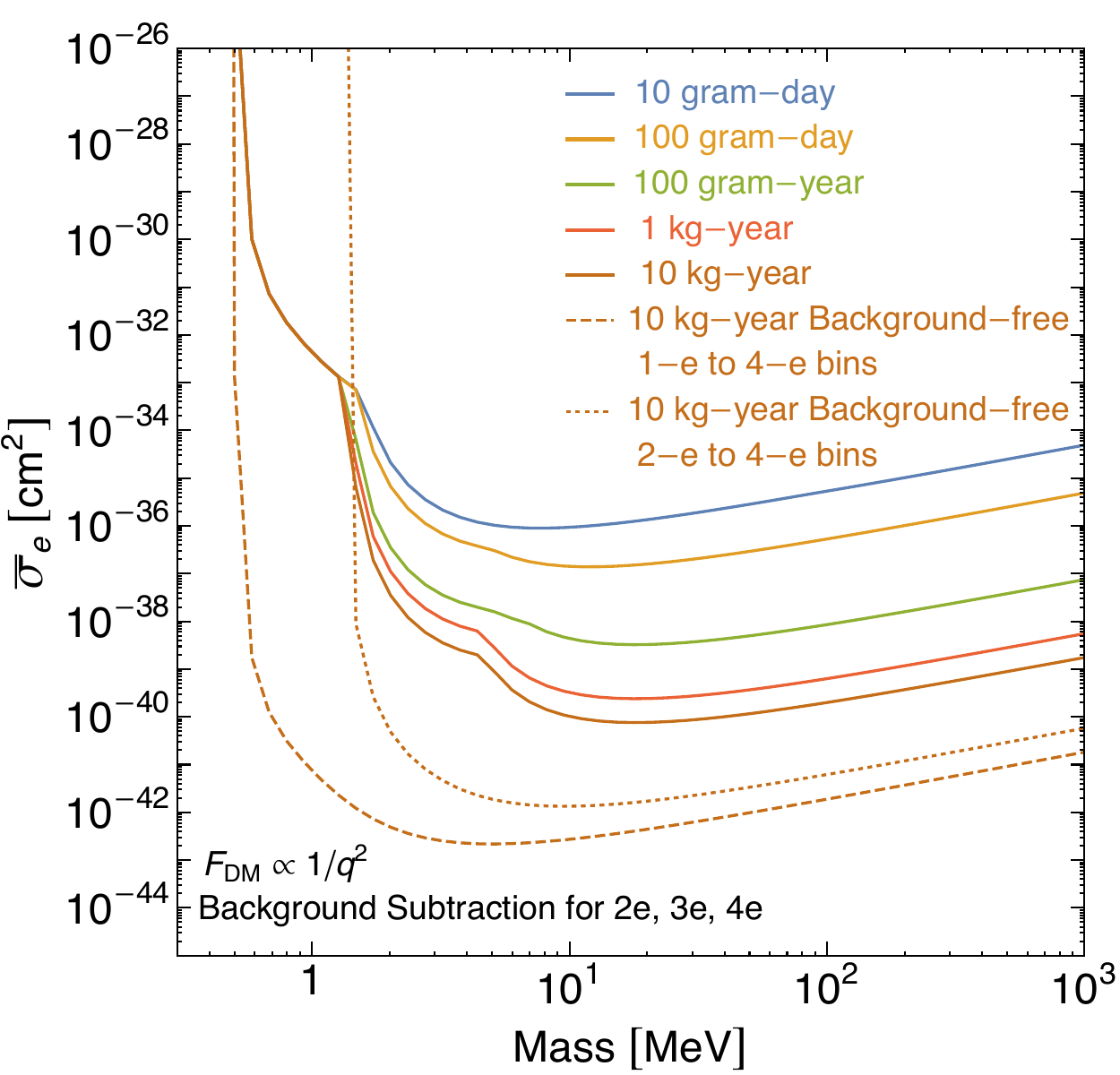}
	\caption{Conservative sensitivity projections at 95\% c.l.~for DM-electron scattering for various exposures for heavy mediators (left) and light mediators (right), \textbf{assuming no reduction of environmental backgrounds} compared to those observed in the SENSEI data~\cite{SENSEI:2020dpa} (i.e., no improvement in shielding that would remove the radiative backgrounds). In contrast to the projections shown in Fig.~\ref{fig:projections_withCherenkov_noSubtraction}, these projections here do include the subtraction of the 2, 3, or 4-electron backgrounds that arise from coincident 1-electron events. Note that we assume the same exposure-plus-readout cycle as was done for taking the SENSEI data in~\cite{SENSEI:2020dpa}. } \label{fig:projections_withCherenkov_WithSubtraction}
\end{figure} 

The DM event rate in each electron bin is obtained using \texttt{QCDark}~\cite{dreyer2023fully}, with an additional electron generated for every 3.8 eV of additional energy above the bandgap. In our projection, a multi-electron DM event corresponds only to a \textit{true} 2, 3, and 4-electron event, \textit{i.e.}, DM multi-electron events from coincidences of DM single-electron events are not counted as signal events. The left (right) panel in the figure shows the projections for a heavy (light) mediator model. We also show background-free limits with 1-e threshold and 2-e threshold for the exposure of 10 kg-year, obtained by equating the total signal in 1-e to 4-e bins for 1-e threshold and 2-e to 4-e bins for 2-e threshold to 3.09.  

Since we assume that 2-, 3-, and 4-electron events arise only through coincidences of uncorrelated 1-electron events, 
these coincident events can be subtracted to obtain an improved bound. In this case
we set a bound whenever the DM signal exceeds the  Poisson statistical uncertainty in the background at $95\%$~confidence-level (for a gaussian-distributed background this corresponds to setting a bound when $S_{\mathrm{DM}}\geq 1.64\sqrt{B}$). Fig.~\ref{fig:projections_withCherenkov_WithSubtraction} shows projections for large exposures assuming such a background subtraction. The left (right) panel shows the projections for a heavy (light) mediator model.

\subsection{Background hypothesis: uniform component only}
We expect the future experiments to be well-shielded and deep underground, in which case radiative backgrounds are likely to be subdominant. As a second background hypothesis we will then consider the possibility that only the uniform component, which may be unrelated to radiative processes as discussed above, is measured. 

We will consider three different rates for the uniformly distributed background and the spurious charge component. First, we will assume that the uniform component required to be added to the fiducial Cherenkov background to explain the SENSEI data (which corresponds to a 1-electron event rate of $\sim 300 ~\text{gram}^{-1}~\text{day}^{-1}$) does \textit{not} decrease in future detectors.  To estimate 2-, 3-, and 4-electron events with this background, we analytically compute the rates of coincidences of this uniform background along with a uniform spurious charge component. As alternative possibilities, we also consider that both the uniform 1-electron background rate and the spurious charge background are either one or two orders of magnitude smaller than what is described above, to account for the possibility that both components may improve with new mitigation strategies (such as~\cite{Tiffenberg:2023vil}). Reducing the 1-electron background by an order of magnitude diminishes the rate of coincidences to create $n$-electron events by $n$ orders of magnitude. The coincident event rate for 2-, 3-, and 4-electron events generated from 
the various uniform background assumptions  and assuming the same readout as in~\cite{SENSEI:2020dpa} are shown in Table~\ref{tab:coincidencewithoutcherenkov}. 

\begin{table}[t!]
\centering
\begin{tabular}{ |c|c|c|c| }
 \hline
& $R_{1e}$=$300~\text{gram}^{-1}\text{day}^{-1}$& $R_{1e}$=$30~\text{gram}^{-1}\text{day}^{-1}$ & $R_{1e}$=$3~\text{gram}^{-1}\text{day}^{-1}$ \\
\hline
$R_{2e}$& $0.096~\text{gram}^{-1}\text{day}^{-1}$& $9.6\times10^{-4}~\text{gram}^{-1}\text{day}^{-1}$ & $9.6\times10^{-6}~\text{gram}^{-1}\text{day}^{-1}$ \\
\hline
$R_{3e}$& $0.0012~\text{gram}^{-1}\text{day}^{-1}$& $1.2\times10^{-6}~\text{gram}^{-1}\text{day}^{-1}$& $1.2\times10^{-9}~\text{gram}^{-1}\text{day}^{-1}$ \\
\hline
$R_{4e}$& $1.84 \times10^{-6}~\text{gram}^{-1}\text{day}^{-1}$& $1.84 \times10^{-10}~\text{gram}^{-1}\text{day}^{-1}$& $1.84 \times10^{-14}~\text{gram}^{-1}\text{day}^{-1}$ \\
\hline
\end{tabular}
\caption{
The 2-, 3-, and 4-electron event rates from coincident 1-electron events, after applying all masks discussed in~\cite{SENSEI:2020dpa}, including a 20-pixel halo radius. We assume that the detector is well-shieled and that the 1-electron events generated radiatively are negligible; instead, the 1-electron rates arise only from an (unspecified) uniform 1-electron background of either $300~\text{gram}^{-1}~\text{day}^{-1}$, $30~\text{gram}^{-1}~\text{day}^{-1}$, or $3~\text{gram}^{-1}~\text{day}^{-1}$, along with a uniform spurious charge of either $1.6\times10^{-4}~\text{e}^{-}/\text{pixel}$, $1.6\times10^{-5}~\text{e}^{-}/\text{pixel}$, or 
$1.6\times10^{-6}~\text{e}^{-}/\text{pixel}$, respectively. Note that we assume the same exposure-plus-readout cycle as was done for taking the SENSEI data in~\cite{SENSEI:2020dpa}. \label{tab:coincidencewithoutcherenkov}}

\end{table}

\begin{figure}[t]
	\centering
	\includegraphics[width=0.48\textwidth]{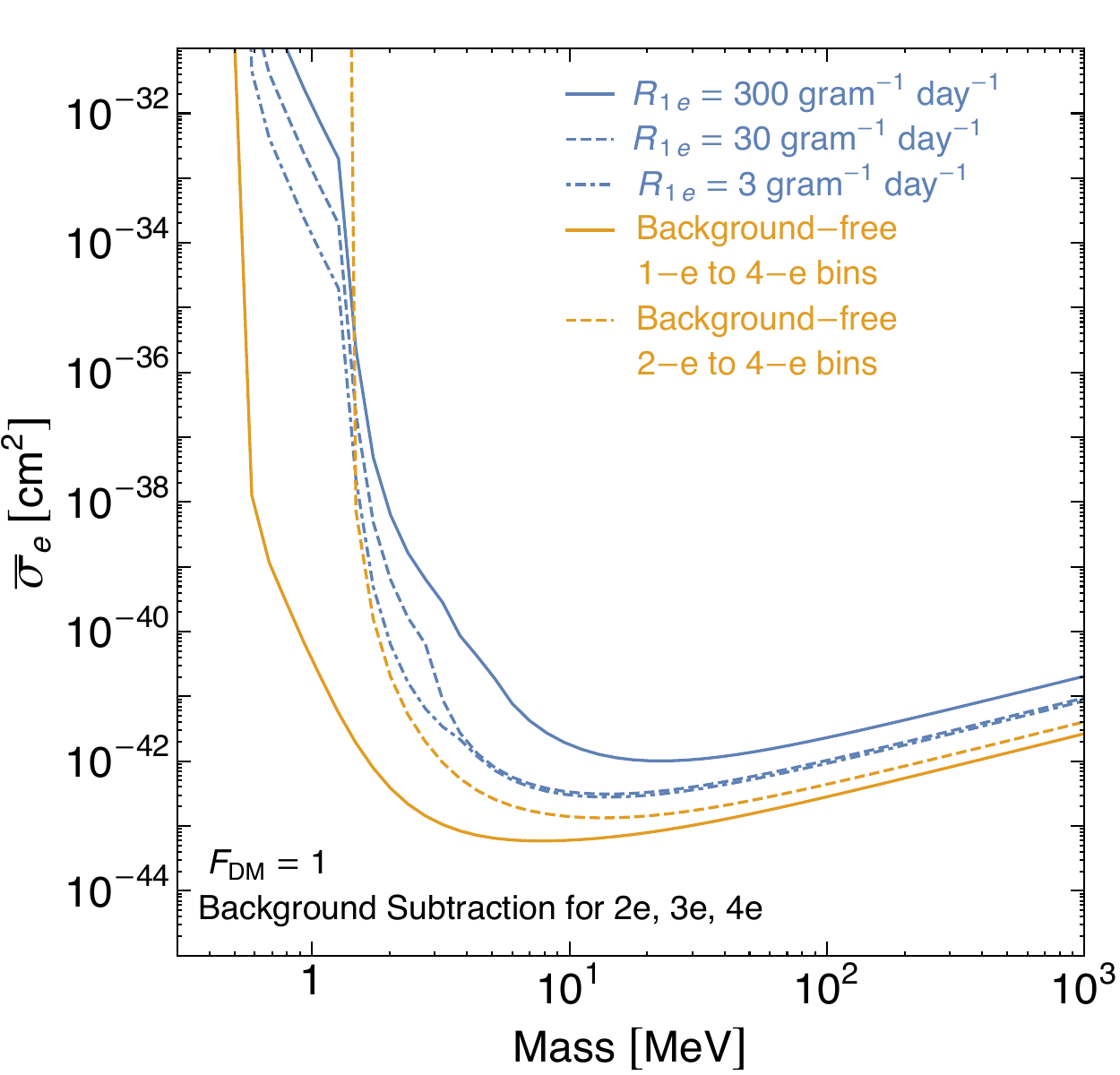}
	\includegraphics[width=0.48\textwidth]{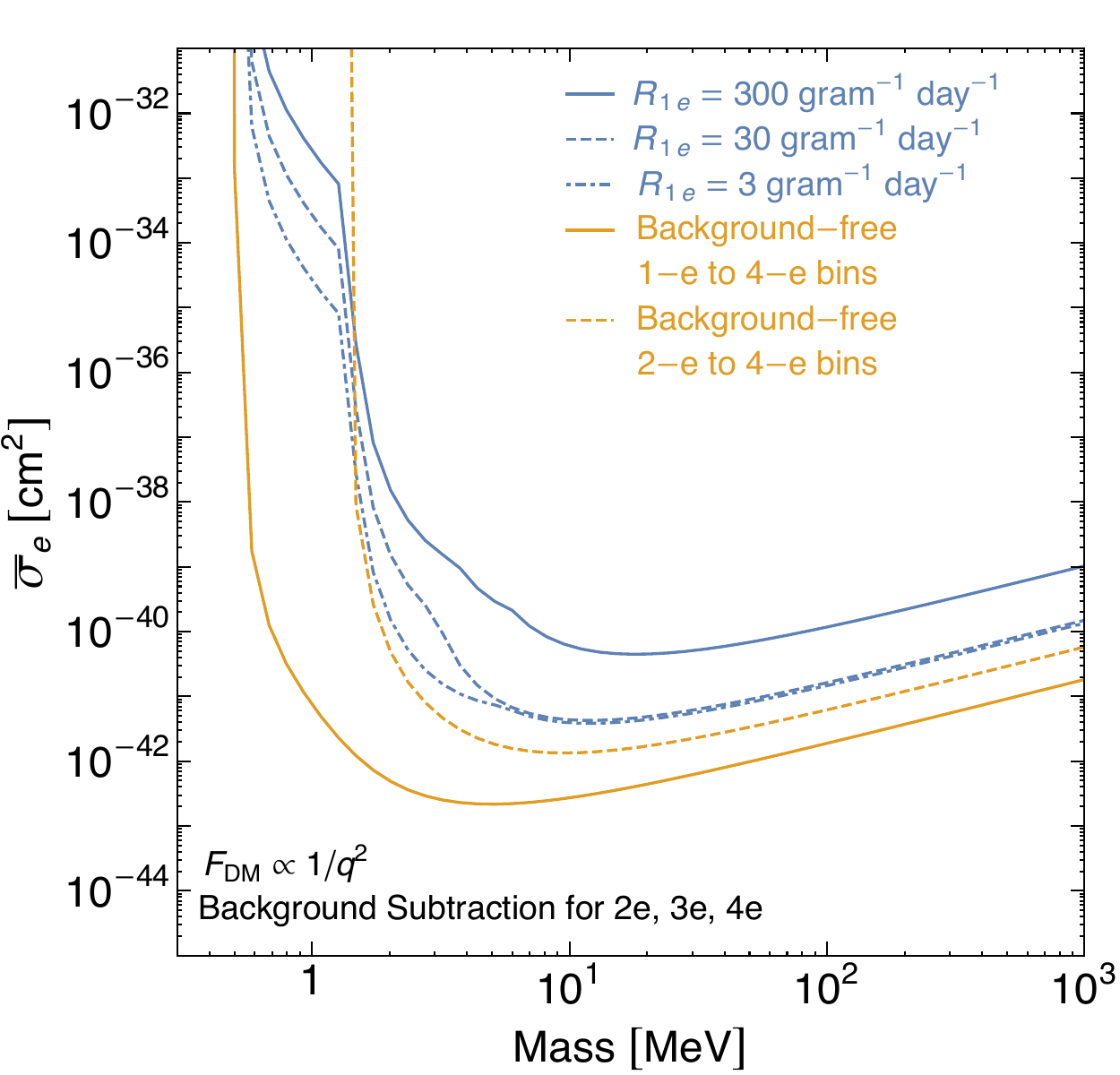}
	
	\caption{Sensitivity projections at 90\% c.l.~for DM-electron scattering for an exposure of 10~kg-years under various background assumptions (see Table~\ref{tab:coincidencewithoutcherenkov} and text), for heavy mediators (left) and light mediators (right).  Different blue lines correspond to different assumptions about the 1-electron rates, namely 300/gram-day (solid), 30/gram-day (dashed), and 3/gram-day (dot-dashed), along with different assumptions about the uniform spurious charge, namely  $1.6\times10^{-4}~\text{e}^{-}/\text{pixel}$,  $1.6\times10^{-5}~\text{e}^{-}/\text{pixel}$, and 
$1.6\times10^{-6}~\text{e}^{-}/\text{pixel}$, respectively.  The 2-, 3-, and 4-electron events from coincident 1-electron events are subtracted.  The orange curve assumes a perfect experiment with no 1-electron backgrounds. Note that we assume the same exposure-plus-readout cycle as was done for taking the SENSEI data in~\cite{SENSEI:2020dpa}. \label{fig:projections_withoutCherenkov_WithSubtraction}
 } 
\end{figure}

In Fig.~\ref{fig:projections_withoutCherenkov_WithSubtraction}, we show the projections for an experiment with an exposure of 10 kg-year with uniform 1-electron background rates as described above. We assume that the 2-, 3-, and 4-electron coincident background rates can be subtracted to look for true multi-electron events from DM. We set bounds as in the previous section, 
comparing the signal and background events in each electronic bin.
The DM event rate in each electron bin is obtained with \texttt{QCDark}~\cite{dreyer2023fully}. 

\section{Conclusions}\label{sec:conclusions}

In this work, we performed the first detailed simulations of secondary photon backgrounds at CCD-based experiments, with a focus on the SENSEI experiment at Fermilab near the MINOS cavern~\cite{SENSEI:2020dpa}. 
We found that secondary radiative backgrounds likely constitute $\approx 40\%$ of the observed 1-electron event rates at SENSEI-MINOS, with  most of these events coming from Cherenkov photons, while events from luminescent electron-hole radiative recombination are subleading in our fiducial simulation. Our fiducial simulations also show that the observed 2-electron event rates are likely dominated by coincidences of 1-electron events from secondary radiation and spurious charge generated during detector readout.

We also found, however, that there are sizable systematic uncertainties in the luminescent 1-electron rate due to insufficient knowledge of material properties on the CCD's backside, where the majority of electron-hole radiative recombination is expected to occur due to high levels of phosphorus doping. Such levels of doping lead to complications in the charge diffusion modeling, which itself affects the radiative recombination rates. To account for these systematics we consider alternative diffusion models, from which we obtain radiative recombination rates and also develop first-principles models of the charge-collection efficiency in the CCD's backside. We show that in extreme scenarios where diffusion is quenched by doping-induced bandgap narrowing, radiative recombination on the detector backside can lead to a dominant contribution to the event rates, to a level that  together with the aforementioned Cherenkov events could fully explain the excess events observed by SENSEI. While this suggest the possibility that all the observed backgrounds could be due to secondary radiation, we find that such extreme diffusion scenarios are to a certain extent disfavored as they lead to charge-collection efficiency profiles that are in tension with the  measurements performed in~\cite{Moroni_2021}. 

By subtracting the secondary radiation background events obtained from our simulations from the observed data, we then conclude that there is still likely a major ($\sim70\%$) component of the  1-electron backgrounds that remains unaccounted for and that is spatially homogeneous across the device. While the origin of these events remains unkown, our simulations suggest that they are not radiative in origin. Plausible explanations are charge leakage through gate insulators or to surface traps~\cite{lenzlinger1969fowler,ranuarez2006review,janesick1987scientific}, in which case the background rates are not expected to be mitigated by improvements in detector shielding or the environmental radio-purity, but could be mitigated with new technologies that suppress surface dark counts, such as~\cite{Tiffenberg:2023vil}.

While no definitive statements regarding the remaining backgrounds can be made at this point, our results allow us to make informed projections regarding the future expected sensitivity of detectors as they collect more data. We perform projections under two background hypotheses. The first is representative of a device that is run under similar operating conditions as SENSEI-MINOS, and considers the backgrounds to be composed by the $40\%$ secondary Cherenkov component plus the $60\%$ spatially homogeneous background of unspecified origin, with a total rate compatible with the one observed at SENSEI-MINOS scaling up linearly with exposure. The second hypothesis assumes improved shielding and radiopurity, which lowers the Cherenkov backgrounds but does not mitigate the unspecified homogenous events that are assumed to be non-radiative in origin. In both cases, we find that such backgrounds would have a significant impact on the sensitivity of planned experiments, even if in specific scenarios this impact could be mitigated by performing background substraction. 

Our simulations also enable to vary the detector's operating conditions to evaluate possible strategies for background mitigation. In this regard we evaluate the impact of thinning the backside of the CCD to remove radiative recombination backgrounds. For a CCD detectors with operating conditions  as for the SENSEI experiment at MINOS, we  find that this would likely be detrimental as it would lead to an outsized increase of secondary Cherenkov events, since the doped backside acts accidentally as an absorber of Cherenkov photons that would otherwise constitute backgrounds. These conclusions may change, however, for a detector that is operated differently from the one at SENSEI-MINOS. 

We also perform simulations under lower high-energy radiation conditions, and find a consistent decrease of secondary radiative events; this result is important for understanding the relevance of Chenrekov and recombination backgrounds for future experiments such as Oscura \cite{Oscura:2022vmi}. Oscura aims to reduce the high-energy background rate by approximately five-orders of magnitude with respect to the SENSEI setup at MINOS down to $10^{-2}$ events/kg/keV/day, and will have a seven-order of magnitude larger exposure. Our results then indicate that assuming a  similar data analysis as in SENSEI-MINOS, Oscura could see approximately $10^4$ one-electron Cherenkov backgrounds. Determining if this will or will not be the leading 1-e background component at Oscura depends on how the remaining unidentified background components will scale as the detectors are improved (if these backgrounds do not improve, then Cherenkov will be a subdominant component to the event rate). A similar conclusion holds for the 2-e events. In this case, 
Cherenkov backgrounds will contribute to the event rate via coincidences of 1-e events. The rate for 1-e Cherenkov-Cherenkov coincidences will be negligible, as it scales down quadraticaly with the lower high-energy event rate at Oscura. The rate of 1-e Cherenkov coincidences with spurious charge and with the remaining 1-e backgrounds (likely non-radiative in origin), however, scales only linearly with the improvements in the high-energy event rate, and can be a sizable contribution to the 2-e bin. Which component will dominate the 2-e bin will then depend on the rates for spurious charge and the remaining 1-e backgrounds at Oscura.

Finally, in our simulations we find an interesting population of 1-electron events due to Cherenkov radiation from the epoxy layer located close to the front-side of the device (see Appendix~\ref{sec:epoxy}). While this population is mostly vetoed by analysis masks, these events should be observable in unmasked data. We also speculate that in the future, additional multiple-electron events could be observed from luminescence from the epoxy or gate insulating layers.

Our work motivates new efforts for studying alternative hypotheses for the remaining unexplained events at CCD-based experiments and developing new techniques to suppress detector backgrounds. As detectors collect more data, these advancements will be critical for reaching the scientific goals of sub-GeV DM experiments.

\section*{Acknowledgements}

 The authors wish to thank Jerry Vavra for useful discussions on transmission and absorption measurements of the  Epotek 301-2 epoxy resin, and Steve Holland for many discussions about the detailed properties of CCDs.
We also thank the SENSEI Collaboration for many useful discussions, in particular Mariano Cababie, Javier Tiffenberg, and Sho Uemura. 
The work of P.D. is supported by the US Department of Energy under grant DE-SC0010008.
 R.E.~acknowledges support from DoE Grant DE-SC0009854, Simons Investigator in Physics Award 623940, the US-Israel Binational Science Foundation Grant No.~2016153, and the Heising-Simons Foundation Award No.~79921 (for SENSEI). DEU is supported by Perimeter Institute for Theoretical Physics and
by the Simons Foundation. Research at Perimeter Institute is supported in part by the Government of Canada
through the Department of Innovation, Science and Economic Development Canada and by the Province of Ontario through the Ministry of Colleges and Universities. The work of M.S. is supported by Department of Energy Grants DE-SC0009919 and DE- SC0022104.

\appendix 

\section{Cherenkov emission in epoxy}\label{sec:epoxy}

\begin{figure}[t]
	\centering
	\includegraphics[width=0.42\textwidth]{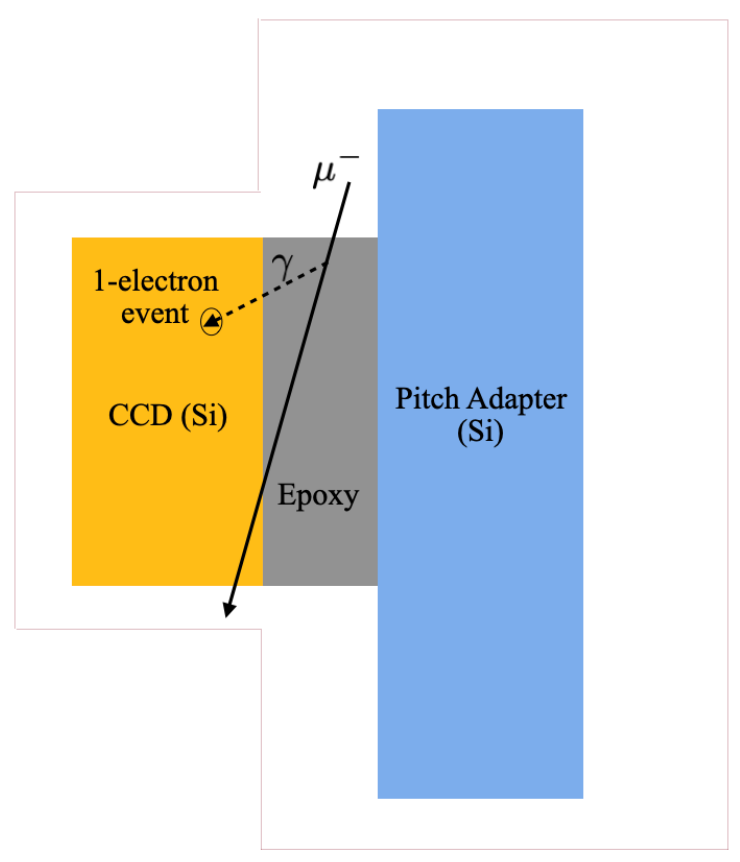}
	\includegraphics[width=0.42\textwidth]{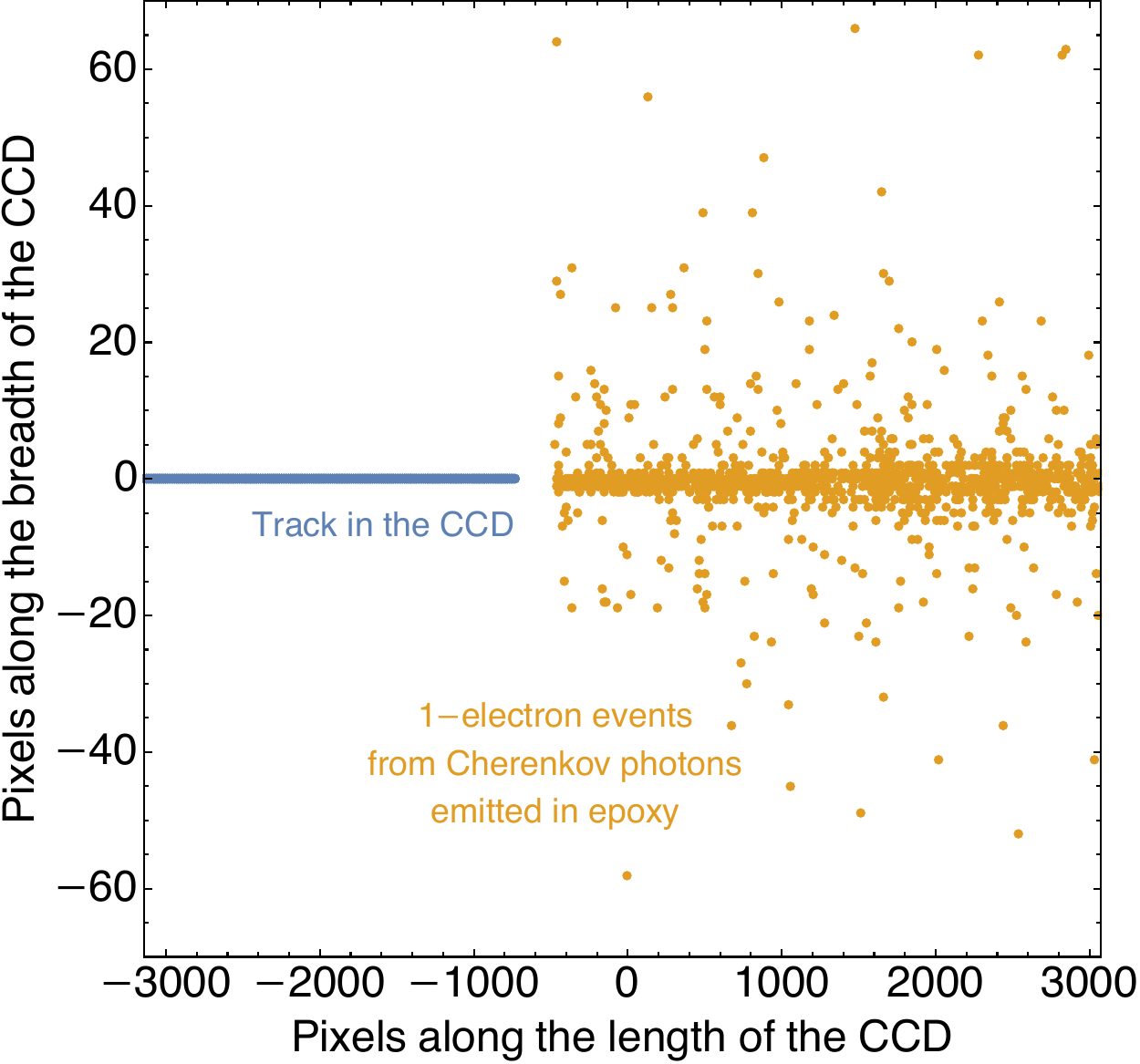}
	\caption{\textbf{Left:} Schematic of a high-energy muon track entering the SENSEI 2020 setup through the epoxy at an angle close to the vertical. \textbf{Right:} The positions of 1-electron events created by photons generated in the epoxy, and the pixels containing the track of the muon.} \label{fig:epoxy_track}
\end{figure}

In the simulations described in this paper, we do not consider photons emitted in the epoxy layer between the CCD and the pitch adapter. However, as the epoxy is a dielectric, high-energy tracks passing through the epoxy can emit Cherenkov photons. A fraction of these photons can enter the CCD and get absorbed to produce 1-electron events. While we do not consider this process in detail in this work, we perform a simulation of a single high-energy muon crossing through the epoxy, and track the Cherenkov photons emitted in the epoxy. We note that the epoxy could also emit scintillation light after being excited by a charged particle, but we do not include this~\cite{ALLEN198297}. 

As shown in the left panel of Fig.~\ref{fig:epoxy_track}, we simulate a high-energy muon track at an angle very close to the vertical, which enters through the epoxy from the top and leaves through the CCD at the bottom. Please note that the left panel of Fig.~\ref{fig:epoxy_track} is not to scale. The width of the epoxy layer is only 80~$\mu$m whereas its length is 9.2~cm. Thus, the orientation for such a muon track needs to be almost vertical. For simplicity, we place the track exactly in the middle of the breadth of the epoxy and the CCD. We use the formalism in Sec.~\ref{sec:Cherenkov_generation} to simulate the generation of Cherenkov photons in the epoxy from this high-energy track. We assume that the real part of the dielectric function of epoxy in the photon energy range of 1.07~eV to 2.2~eV is just the square of the real part of the refractive index of epoxy ($\sim 1.6$). For practical purposes, we do not change the photon energy range from the one used in the rest of the simulations described in this paper (i.e., 1.07~eV to 2.2~eV). We then use the photon propagation module described in Sec.~\ref{sec:photon} to identify the photons that get absorbed in the CCD. We show the positions of the 1-electron events created by these in the right panel of Fig.~\ref{fig:epoxy_track}. We also show the pixels containing the track of the high-energy muon. We see that such a track could generate a large number of 1-electron events that are spread out along a straight path. 

Even though this observation is interesting, we note again that the orientation of such a track has to be near-vertical for it to travel large enough distance in the epoxy and emit a significant number of Cherenkov photons. A very large number of Cherenkov photons needs to be emitted to see any significant effect, as only a very small fraction penetrate the CCD from the epoxy, and are able to evade the masks associated with the track. Considering the mean free path of Cherenkov emission in epoxy ($\sim 22~\mu\text{m}$) and the width of the epoxy ($\sim 80~\mu\text{m}$), and requiring that the track emits at least 100 Cherenkov photons in the epoxy, the maximum polar angle of the track with the vertical is forced to be $\sim$0.036~rad. With the muon flux discussed in Sec.~\ref{sec:muon-flux}, we find that such tracks are expected at a rate of $\sim$0.003 per day per CCD. Tracks at larger polar angles can also emit large number of Cherenkov photons in a narrower range of azimuthal angles. The contribution of such tracks is suppressed though for near-horizontal tracks, as the muon-flux is mostly vertical. Thus, with the SENSEI 2020 setup, it is a good approximation to neglect the contribution of such tracks. However, if a similar setup is exposed for timescales larger than $\sim$year, then the effect of such tracks could be observable.    
\bibliographystyle{apsrev4-1.bst}
\bibliography{cherenkov.bib}
		
\end{document}